\documentclass[aps,prab,twocolumn,groupedaddress,showkeys]{revtex4-2}

\usepackage{textcomp}
\usepackage[T1]{fontenc}
\usepackage{lmodern}
\usepackage{tgtermes}
\usepackage{newtxmath}

\usepackage[utf8x]{inputenc}
\usepackage[english]{babel}
\usepackage{amsmath}
\usepackage{amsfonts}
\usepackage{xcolor}
\usepackage{graphicx}
\usepackage{enumitem}

\begin{document}

\title{A method for accurate and efficient simulations of slow beam degradation due to electron clouds}

\author{Konstantinos Paraschou}
\email[]{konstantinos.paraschou@cern.ch}
\affiliation{Aristotle University of Thessaloniki, 54124 Thessaloniki, Greece}
\affiliation{CERN, 1211 Geneva, Switzerland}
\author{Giovanni Iadarola}
\affiliation{CERN, 1211 Geneva, Switzerland}

\date{\today}

\begin{abstract}
In the Large Hadron Collider, electron clouds have been observed to cause slow beam degradation in the form of beam lifetime reduction and slow emittance growth.
We present a method for the simulation of such slow effects with arbitrarily complex electron clouds and in the presence of
non-linearities in the lattice of the accelerator.
The application of the electron cloud forces, as obtained from a particle-in-cell simulation of the electron cloud dynamics, is kept symplectic by using an appropriate tricubic interpolation scheme.
The properties of such an interpolation scheme are studied in detail and numerical artifacts that can be introduced by the scheme are identified and corrected through a suitable refinement procedure.
The method is applied to the case of the Large Hadron Collider for protons at injection energy to compute Dynamic Aperture, to perform Frequency Map Analysis, and to simulate beam losses and emittance growth by tracking particle distributions for long time scales.
\end{abstract}


\maketitle

\section{Introduction}\label{sec:intro}

The formation of electron clouds (e-clouds) is often observed in synchrotrons that are operating with closely spaced bunches of positively charged particles.
This is the result of an exponential multiplication of electrons in the beam pipe driven by secondary
electron emission and photo-emission from the surface of the beam pipe~\cite{ecloud-general,cimino,gianni-benevento}.

The electromagnetic fields generated by the e-cloud can
affect the dynamics of the particle beam posing significant limitations
to the performance of accelerators.
In particular, the interaction of the beam with the e-cloud
can result in coherent transverse instabilities, which need 
to be mitigated with the use of a feedback system and by operating with large chromaticity and tune spread from 
octupole magnets~\cite{PhysRevLett.85.3821,PhysRevSTAB.5.121002,annalisa,PhysRevAccelBeams.23.081002}. 

Even when instabilities are successfully controlled, the
e-cloud can still induce significant beam degradation 
through incoherent effects, resulting in slow beam losses
and transverse emittance growth.
Notably, at the Large Hadron Collider (LHC)~\cite{LHC-design} e-clouds induce
a significant beam lifetime degradation both at injection energy and in collision~\cite{iadarola-ipac2021,paraschouAnalysisBunchbyBunchBeam2019,paraschouIncoherentElectronCloud2020}.
The modelling of such beam degradation is particularly
difficult since these effects are the result of an interplay between the non-linear e-cloud forces and the non-linearities of the accelerator lattice.
Furthermore, such effects are often visible only on very long time scales corresponding to several millions of beam revolutions.
Therefore, numerical models and computer programs used for this purpose need to be sufficiently accurate to correctly describe the phenomena
while at the same time being sufficiently fast to allow the simulation of such long
time scales.

\begin{figure*}
    \includegraphics[width=\textwidth]{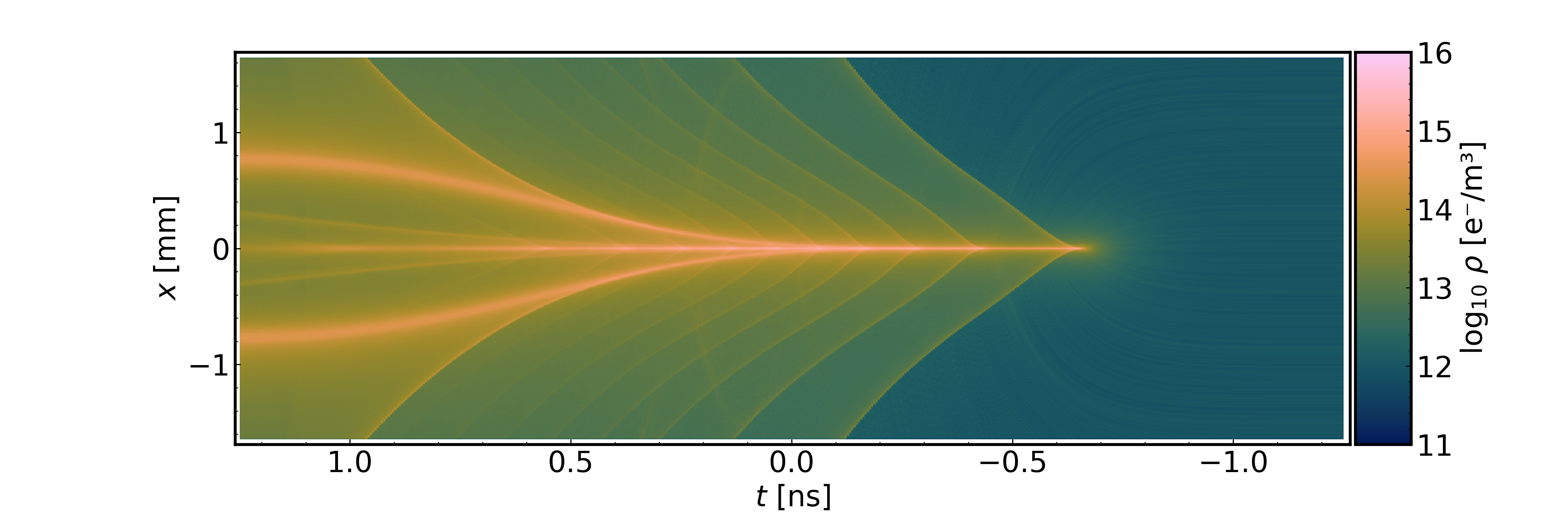}
    \caption{\label{fig:fig-intro}Time evolution of the $y=0$ slice of the electron density distribution during an e-cloud pinch. The head of the bunch is at $t < 0$}.
    \centering
    \begin{minipage}[h]{0.45\textwidth}
      \includegraphics[width=\textwidth]{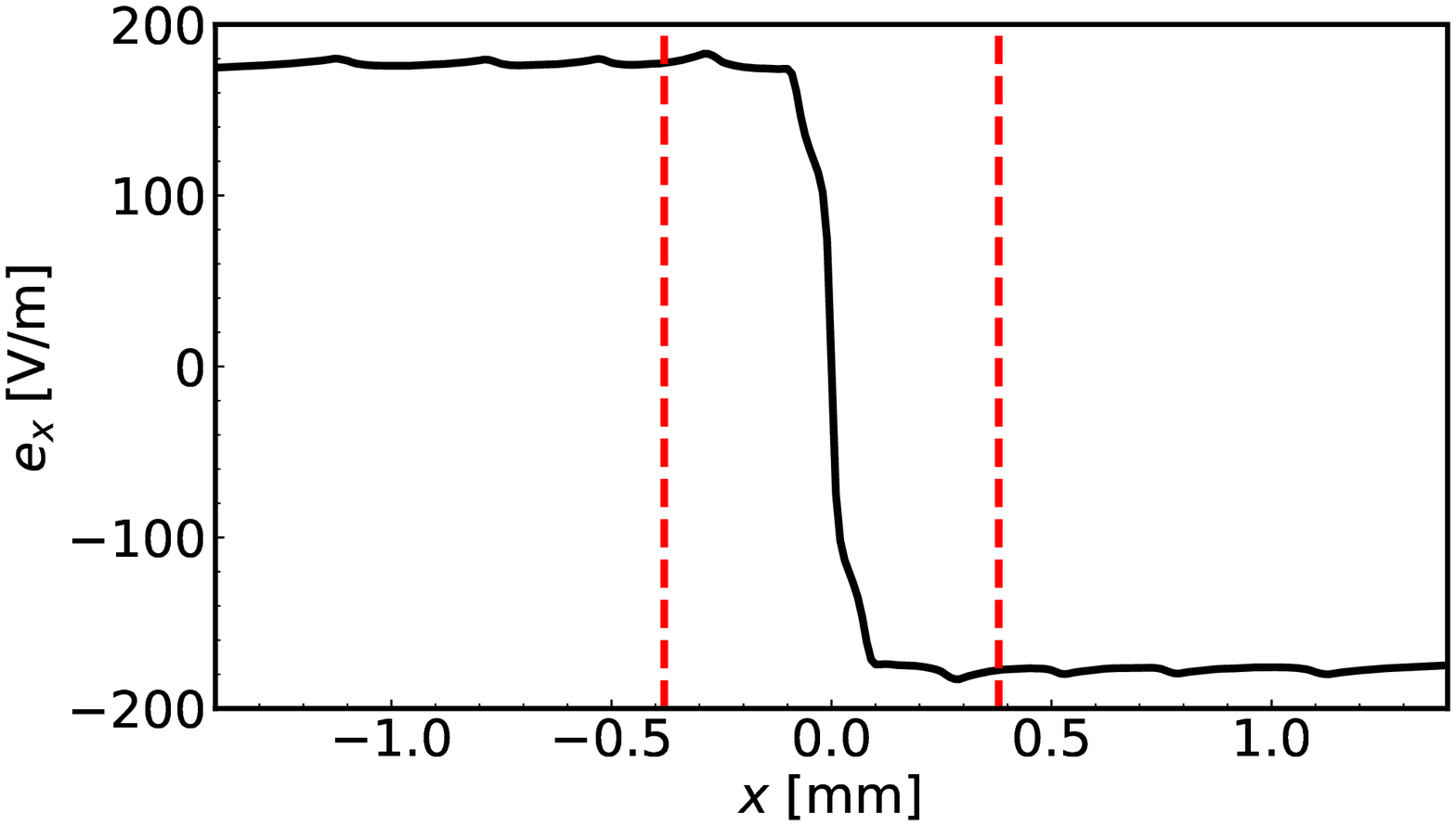}
     \\
    (a)
    \end{minipage}
    \begin{minipage}[h]{0.45\textwidth}
      \includegraphics[width=\textwidth]{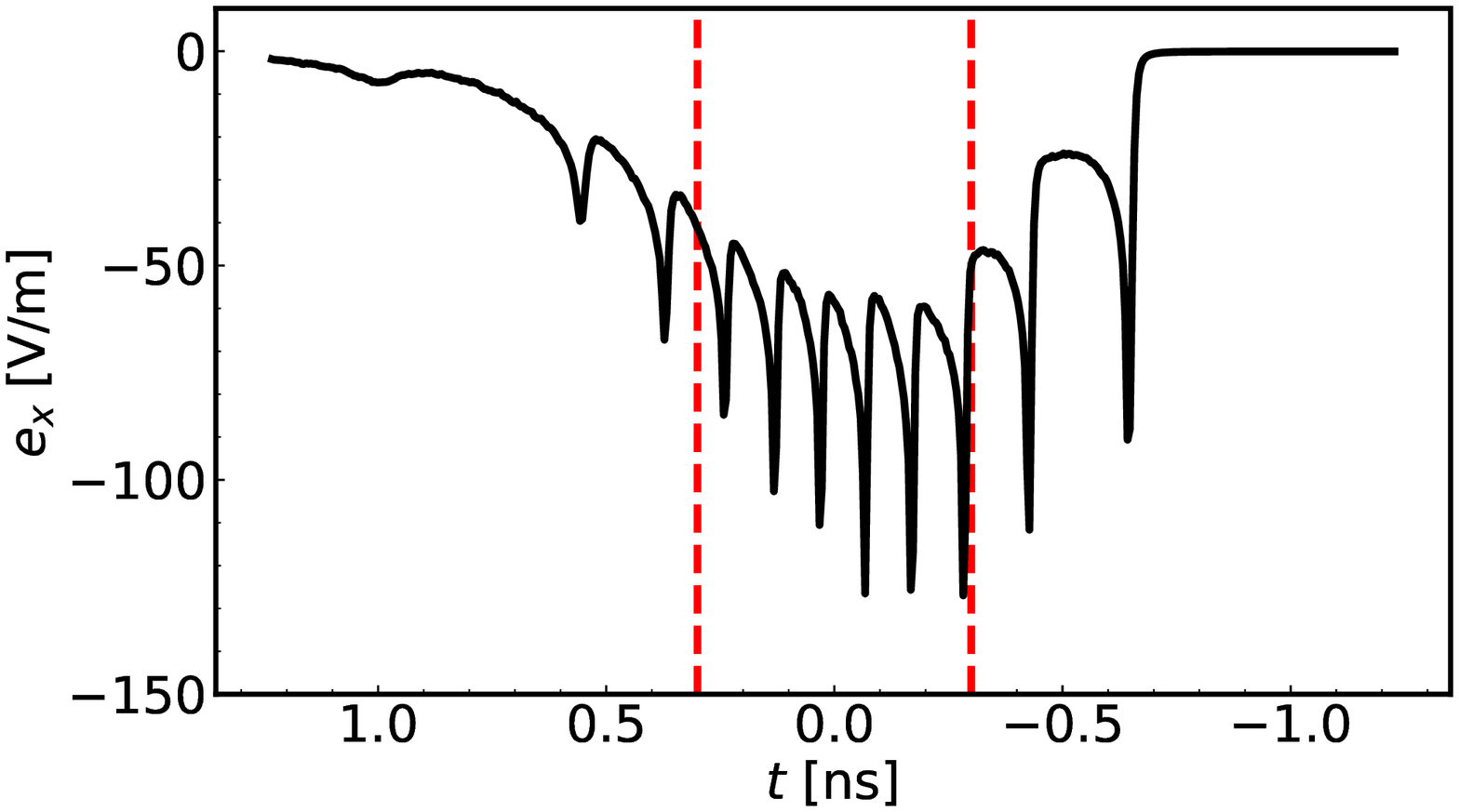}
    \\
    (b) 
    \end{minipage}
    \caption{\label{fig:fig-intro2} (a) Horizontal field as a function of horizontal position $x$ at $y=0, t=0.14$ ns. (b) Horizontal field as a function of time $t$ at $y=0, x=5$ \textmu m.}
\end{figure*}

Incoherent effects from e-cloud were addressed by different studies over the last two decades.
Furman~\textit{et~al}.~\cite{furman-incoherent} showed that e-clouds can cause a tune spread, distort the betatron and dispersion functions,
 as well as creating a synchro-betatron coupling.
The incoherent tune shift was also studied by Romano~\textit{et~al.}~\cite{romano-ipac} and 
Petrov~\textit{et~al.}~\cite{petrov-prstab}.
Franchetti~\textit{et~al.}~\cite{franchetti-incoherent} used simplified cloud distributions to express analytically the forces induced on the beam particles by the e-cloud.
Such an approach is very convenient in terms of computation time
but lacks the capability of accurately describing realistic
e-cloud distributions, especially in the presence of
magnetic field gradients.
Ohmi and Oide~\cite{ohmi-chaos} studied the incoherent emittance growth driven by e-cloud with self-consistent 
Particle-In-Cell (PIC) simulations of the coupled dynamics between the e-cloud and the beam particles.
Such an approach is extremely demanding in terms of computation time and
ultimately does not allow the simulation of the long time scales required for the study of these effects in realistic machine conditions.

Incoherent modifications of the beam distribution driven by e-cloud effects are typically very slow processes.
Hence, it is possible to assume that over a relatively large number of turns the impact of the changes in the beam distribution on the e-cloud dynamics can be neglected.
Based on this consideration, Benedetto~\textit{et~al}.~\cite{benedetto-paper,benedetto-thesis} introduced the approach of pre-recording the e-cloud forces on a discrete grid and computing the forces on the beam particle location using an interpolation scheme.
In this work the authors do not address the issue of the symplecticity of the simulated interaction.
As will be discussed in more detail in Sec.~\ref{sec:tricubic}, the use of an interpolation scheme does not guarantee the preservation of the symplecticity of
the particle interaction with the e-cloud.
If the numerical model is not symplectic, artificial growth or damping can be introduced in the amplitude of the particle motion, leading to unacceptable modifications on observables of interest like beam lifetime and emittance evolution~\cite{wolski}.
For example, in studies of space-charge effects, Qiang~\cite{qiang1,qiang2,qiang3} found that by using symplectic models, it is possible to suppress numerical emittance growth, which was observed instead when using non-symplectic models.

In the following, we propose a more advanced method that 
allows the e-cloud field map of arbitrary
complexity to be modelled in a way that preserves the symplecticity of the interaction.
In order to illustrate the capability of such an approach we
apply the method to tracking simulations together with
a realistic non-linear model of the LHC lattice.
The resulting non-linear beam dynamics in the presence of the e-cloud is investigated through Dynamic Aperture (DA) evaluations, Frequency Map Analysis (FMA) and direct predictions of lifetime and emittance evolution obtained through long-term
tracking simulations of particle distributions.
Graphics Processing Units (GPU)
are used to boost the simulation speed in order to directly simulate the time
scales over which such effects are observed in the LHC.

The article is structured as follows:
in Sec.~\ref{complex}, we discuss the main features and e-cloud distribution and forces during the passage of the bunch;
in Sec.~\ref{sec:map}, a robust method is developed to model an arbitrarily complex e-cloud interaction within 
the non-linear model of an accelerator lattice;
and in Sec.~\ref{sec:application}, applications to the LHC case are illustrated.

\section{The e-cloud pinch}\label{complex}

The dynamics of the e-cloud depends on a large set of parameters, including spacing between bunches, 
bunch charge, transverse beam sizes and bunch length.
It also depends on properties of the accelerator, for example the magnetic field configuration, the geometry of the vacuum chamber and the material properties of the vacuum chamber's walls, in particular their Secondary Electron Yield (SEY).
Figure~\ref{fig:fig-intro} shows a typical evolution of the e-cloud charge density during the passage of a bunch in the
absence of externally applied magnetic fields.
The illustrated distribution is obtained for typical parameters of the proton bunch and the e-cloud in the LHC.
In particular, 
the bunch is defined by a 3D Gaussian distribution with horizontal and vertical r.m.s. beam size equal to $0.4$\,mm, r.m.s. bunch length equal to $0.3$\,ns and a bunch intensity of $1.2 \cdot 10^{11}$\,p/bunch.
The initial electron density of the e-cloud is uniform and equal to $10^{12}$\,e/m.
The boundary conditions are defined by the beam pipe geometry of the LHC arcs.

While the bunch passes through the cloud, the electron density increases significantly at the bunch location and 
complex structures appear in the density profile.
This is commonly referred to as the e-cloud ``pinch''~\cite{franchetti-incoherent,pinch}.
The field generated by such an evolving charge distribution has very specific features illustrated in Fig.~\ref{fig:fig-intro2}.
In the transverse plane, the field changes sign very abruptly within the core of the bunch, as visible in Fig.~\ref{fig:fig-intro2}a.
Moreover, strong oscillations of the fields are observed as a function of time as illustrated in Fig.~\ref{fig:fig-intro2}b.
It is evident that describing such a field behaviour with analytical expressions is practically unfeasible.

Furthermore, in the presence of a magnetic field the electrons are confined by the field lines and 
the electron dynamics can get even more complicated.
Both the non-linear behavior of the transverse field distribution and its time-varying nature have a strong
impact on the dynamics of the beam particles that are subject to these fields.
It is therefore important to correctly model these features when simulating the beam dynamics in the presence of the e-cloud.

\section{Weak-strong modelling of e-clouds}\label{sec:ws}

In the current state-of-the-art self-consistent PIC simulations, which are used to study e-cloud driven coherent instabilities\cite{pyparis}, the computation time required to accurately resolve the coupled dynamics between beam particles and e-cloud particles is in the order of hundreds of seconds per revolution\cite{luca-note}. It is clear that such an approach becomes prohibitively time consuming for the simulation of slow effects that become evident only after millions of revolutions.

In the absence of coherent beam instabilities, e-cloud effects at the LHC cause only very minor changes in the beam distribution, consisting mostly in the loss of large amplitude particles.  
It is therefore appropriate to assume a weak-strong modelling of the e-cloud forces, as done by 
Benedetto~\textit{et~al}.~\cite{benedetto-paper,benedetto-thesis}.
In the weak-strong modeling approach, the electron dynamics and the electromagnetic fields generated by the electron distribution during the bunch passage are computed only once using the unperturbed bunch distribution. After that, the electromagnetic fields can be applied turn after turn without being recomputed, as long as changes in the beam distribution remain small enough not to alter significantly the electron dynamics.

In this study, the aforementioned weak-strong approximation is employed within a symplectic six-dimensional thin-lens model of the LHC non-linear lattice\cite{sixtrack-origin}.
The simulation method is therefore separated in two parts:
\begin{enumerate}[label=\Alph*)]
    \item In the first part, which can be considered to be the pre-processing part of the simulation, the e-cloud map is constructed following the considerations of Sec.~\ref{sec:map}. 
    This stage includes simulating the e-cloud dynamics with a PIC simulation using the unperturbed bunch distribution in order to calculate the potential and the e-cloud forces to be applied in the beam-particle tracking simulations. 
    At this stage the PIC simulation of the single bunch passage can  be executed multiple times in parallel to reduce discrete macroparticle noise (shot noise). 
    \item The second part is the actual simulation of the beam dynamics in the presence of electron cloud, during which the beam particles are tracked around the accelerator under the action of the accelerator lattice elements and of the e-cloud interactions. In this part, the e-cloud forces acting on the bunch are obtained from the map recorded at the part A through a 3D interpolation scheme. 
\end{enumerate}

When employing the weak-strong modelling, the computational time is not limited anymore by the e-cloud interactions, but rather by the rest of the elements in the accelerator lattice.
For this reason, it is possible to afford a high-order interpolation scheme (described in Sec.~\ref{sec:tricubic}) in order to ensure the symplecticity of the map.

Additionally, in the weak-strong modelling, the computation time required for part A does not depend at all on the number of beam particles nor on the number of turns that are to be simulated in part B.
Instead, the computation time of part B  scales linearly with both the number of simulated particles and the number of turns.
However, an advantage of this modelling is that different particles can be simulated independently of each other and hence the workload can be easily parallelized on GPU architectures, by tracking the beam particles concurrently.


\section{Symplectic implementation of the e-cloud map}\label{sec:map}

The forces acting on a beam particle due to the effect 
of an e-cloud can be conveniently calculated in the frame of the reference particle. 
The forces acting on the bunch are proportional to the gradient of a scalar potential $\phi$.
For a very relativistic beam such a potential can be calculated with 
good approximation as the solution of the 2D Poisson equation~\cite{iadarola-note-interaction}:
\begin{equation}\label{eq:laplace-2D}
    \frac{\partial^2 \phi\left(x,y,\tau\right)}{\partial x^2} + 
    \frac{\partial^2 \phi\left(x,y,\tau\right)}{\partial y^2} = -\frac{\rho\left(x,y,\tau\right)}{\varepsilon_0} ,
\end{equation}
with $\varepsilon_0$ being the permittivity of free space and $\rho\left(x,y,\tau\right)$ being the charge density 
of the e-cloud.
This approximation is used in most simulation codes used for this purpose, for example PyECLOUD~\cite{pyparis}, POSINST~\cite{posinst}, HEADTAIL~\cite{headtail}, openECLOUD~\cite{openecloud}.
With good approximation, the integrated forces from the e-cloud in a short portion of the accelerator of length $L$ centered around the longitudinal position $s_0$, 
can be generated from the following Hamiltonian as a function of the position coordinates

\begin{equation}\label{eq:hamilton}
    H = \frac{qL}{\beta_0 P_0 c} \phi\left(x,y,\tau\right) \delta \left( s - s_0 \right) ,
\end{equation}
where $\beta_0$ is the ratio between the speed of the reference particle and the speed of light $c$, and $P_0$ is the total momentum of the reference particle.
Here $x,y$ are the horizontal and vertical coordinates respectively and $p_x, p_y$ are their canonical conjugate momenta. The longitudinal variables 
$\tau, p_\tau$ are defined as:
\begin{align}
    \tau = \frac{s}{\beta_0} - ct , \\ 
    p_\tau = \frac{E - E_0}{P_0 c} ,
\end{align}
where $t$ is time, $E$ is the energy of the particle and $E_0$ is the energy of the reference particle.
The map obtained from this Hamiltonian is:
\begin{align}
    x &\mapsto x , \label{eq:map1}\\
    p_x &\mapsto p_x +\frac{qL}{\beta_0 P_0 c} e_x\left(x,y,\tau\right) ,\label{eq:map2}\\
    y &\mapsto y ,\label{eq:map3}\\
    p_y &\mapsto p_y +\frac{qL}{\beta_0 P_0 c} e_y\left(x,y,\tau\right) ,\label{eq:map4}\\
    \tau &\mapsto \tau ,\label{eq:map5}\\
    p_\tau &\mapsto p_\tau +\frac{qL}{\beta_0 P_0 c} e_\tau\left(x,y,\tau\right) ,\label{eq:map6}
\end{align}
where 
\begin{equation}
    e_x = -\frac{\partial \phi}{\partial x},\;\;    
    e_y = -\frac{\partial \phi}{\partial y},\;\;    
    e_\tau = -\frac{\partial \phi}{\partial \tau}     .
\end{equation}

As discussed in Sec.~\ref{sec:intro}, incoherent modifications of the beam distribution driven by e-cloud effects are typically slow processes for situations 
of practical interest.
Hence, it is possible to assume that, over a relatively large number of turns, the impact on 
the e-cloud dynamics from the changes in the beam distribution can be neglected.
Therefore, the e-cloud dynamics can be simulated once and the corresponding potential can be stored to apply the 
e-cloud forces to the beam over multiple turns.
Moreover, following this approach, each recorded map of the e-cloud forces can be used for multiple locations of the accelerator where the beam size and the geometry of the chamber are the same.
This is convenient especially in lattices with a strong degree of periodicity such as those of high energy accelerators.
This approach is similar to the ones typically used for the
simulation of beam-beam effects (with a rigid strong beam) or
of space-charge effects with a frozen potential that can also be
adiabatically adapted over time.
Furthermore, the resulting map will not exhibit turn-by-turn fluctuations from the limited number of macroparticles used to 
describe the electron distributions.

\subsection{Tricubic Interpolation}\label{sec:tricubic}

Particle-In Cell codes are the standard tool for the simulations of the e-cloud dynamics.
The PyECLOUD~\cite{pyparis} code is used for this purpose in LHC studies and has been extensively benchmarked against experiments~\cite{galina-ipac,iadarola-arcs} and other simulation codes~\cite{eric}.
PyECLOUD provides the charge density $\rho(x,y,\tau)$ and the scalar potential $\phi(x, y, \tau)$ on a regular 
three-dimensional grid with cell sizes $\Delta x, \Delta y, \Delta \tau$:
\begin{align}
    \phi^{ijk} &= \phi(x_i, y_j, \tau_k) , \\
    \rho^{ijk} &= \rho(x_i, y_j, \tau_k) ,
\end{align}
where $x_i = x_0 + i\Delta x$, $y_j = y_0 + j\Delta y$, $\tau_k = \tau_0 + k\Delta \tau$ and $x_0, y_0, \tau_0$ define the position of the grid in the three-dimensional space.
Within PyECLOUD, the fields are calculated by solving Eq.~\eqref{eq:laplace-2D} by means of a finite difference method,
evaluating the gradient of the potential at the grid nodes using central differences and using linear interpolations
to obtain the field at the location of the electrons.

\begin{figure}
      \centering
    \begin{minipage}[h]{0.45\textwidth}
        \centering
    \includegraphics[width=\textwidth]{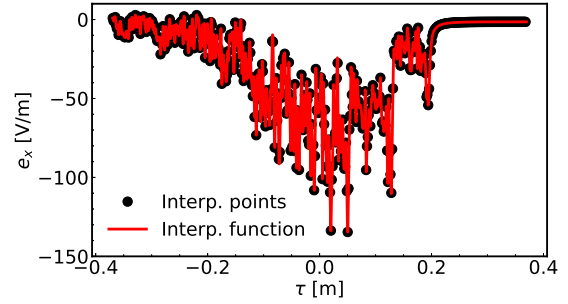}
      \\
        (a) 
    \end{minipage}
    \begin{minipage}[h]{0.45\textwidth}
      \centering
    \includegraphics[width=\textwidth]{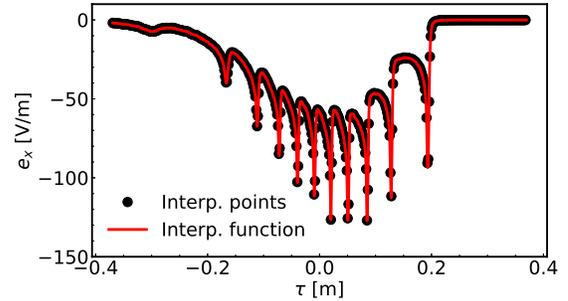}\\
        (b) 
    \end{minipage}
    \caption{\label{fig:fig01-redo}Horizontal field of the e-cloud as a function of $\tau$ in a single simulation (a) and the average of 4000 simulations (b).}
\end{figure}

The field map obtained by such simulations with typical numerical settings exhibit significant macroparticle noise as visible
in Fig.~\ref{fig:fig01-redo}a.
This does not affect the motion of electrons, which is largely dominated by the beam field but is not acceptable when
evaluating its effect on the beam repeated over many turns,
as the noise can artificially enhance the diffusion of
the beam particles.
As the potential table is computed and stored before starting the beam particle tracking simulations, 
such a problem can be effectively solved by simulating the e-cloud dynamics multiple times with different random seeds
in the generation of the initial electron distribution, and averaging the resulting charge density and scalar potential.
The result of such an averaging process over 4000 simulations is shown by the black points of Fig.~\ref{fig:fig01-redo}b. 

A map in the form given by Eqs.~\eqref{eq:map1}-\eqref{eq:map6} is symplectic if the following conditions are satisfied:
\begin{equation}\label{eq:symplecticity}
    \frac{\partial}{\partial x} \left(\frac{\partial\phi}{\partial y} \right) = 
    \frac{\partial}{\partial y} \left(\frac{\partial\phi}{\partial x} \right) ,
\end{equation}
\begin{equation}
    \frac{\partial}{\partial x} \left(\frac{\partial\phi}{\partial \tau} \right) = 
    \frac{\partial}{\partial \tau} \left(\frac{\partial\phi}{\partial x} \right) ,
\end{equation}
\begin{equation}
    \frac{\partial}{\partial y} \left(\frac{\partial\phi}{\partial \tau} \right) = 
    \frac{\partial}{\partial \tau} \left(\frac{\partial\phi}{\partial y} \right) .
\end{equation}
If the potential $\phi$ is a smooth function, for example if it is an analytic function, these conditions are
automatically satisfied.
However, as from the PyECLOUD simulation, the potential $\phi^{ijk}$ is known only on a discrete grid, 
an interpolation scheme needs to be used to obtain the potential or the field in an arbitrary point in space.
The conventional interpolation scheme used in PIC simulations,
based on linear interpolation and finite differences for evaluation of the fields, 
does not preserve these conditions as discussed in Appendix~\ref{sec:app-linear}.

The symplecticity condition will  hold, however, if the derivatives are computed analytically from an interpolating function $\phi^\text{int}(x,y,\tau)$, which has continuous mixed derivatives: 
\begin{equation}\label{eq:set1}
    \frac{\partial ^2 \phi^\text{int}}{\partial x \partial y},\;\;\frac{\partial ^2 \phi^\text{int}}{\partial x \partial \tau}, \;\;\frac{\partial ^2 \phi^\text{int}}{\partial y \partial \tau}.
\end{equation} 
One could create such a function by fitting globally a high-order polynomial in three dimensions to the discrete samples $\phi^{ijk}$.
The order of such a polynomial would need to be prohibitively high to accurately 
reproduce complex features of the fields such as those visible in Fig.~\ref{fig:fig-intro2}.
For this reason, we decided instead to investigate the use of a local interpolation scheme, as done in PIC simulations, while still preserving symplecticity.

This can be done following the approach introduced by Lekien and Marsden~\cite{lekien}. 
They identified a local tricubic interpolation scheme that is able to preserve the global continuity of the functions:
\begin{equation}\label{eq:set2}
\phi^\text{int}, \frac{\partial \phi^\text{int}}{\partial x}, \frac{\partial \phi^\text{int}}{\partial y}, \frac{\partial \phi^\text{int}}{\partial \tau}, 
\frac{\partial^2 \phi^\text{int}}{\partial x\partial y},\frac{\partial^2 \phi^\text{int}}{\partial y\partial \tau},\frac{\partial^2 \phi^\text{int}}{\partial x\partial \tau},\frac{\partial^3 \phi^\text{int}}{\partial x\partial y\partial \tau}
 .
\end{equation}
Such a list of globally continuous quantities includes those in Eq.~\eqref{eq:set1}, guaranteeing the symplecticity
of the scheme.
In such a scheme, the interpolating function is represented as a piece-wise polynomial:
\begin{equation}\label{eq:interp}
    \phi^\text{int} \left(x,y,\tau\right) = \sum_{i,j,k=0}^3 \mathrm{a}_{ijk} x^i y^j \tau^k ,
\end{equation}
where the coefficients $\mathrm{a}_{ijk}$ are different for each hexahedral cell of the grid and are calculated by imposing the quantities listed in Eq.~\eqref{eq:set2}
at the 8 nodes of the corresponding cell.
The coefficients $\mathrm{a}_{ijk}$ can be obtained from the matrix equation:
\begin{equation}\label{eq:acoef}
    \boldsymbol{\alpha} = \mathbf{B}^{-1} \mathbf{b} ,
\end{equation}
where the $64$ elements of the vector $\boldsymbol{\alpha}$ are defined such that $\mathrm{\alpha}_{1 + i + 4j + 16k} = \mathrm{a}_{ijk}$. 
The $64$ elements $b_i$ of the vector $\mathbf{b}$ represent the imposed constraints and are equal to:
\begin{equation}\label{eq:bvector}
b_i = 
    \begin{cases}
        \phi \left(p_i\right) &  1 \leq i \leq 8,\\ 
        \frac{\partial \phi}{\partial x} \left(p_{i-8}\right) &  9 \leq i \leq 16,\\ 
        \frac{\partial \phi}{\partial y} \left(p_{i-16}\right) &  17 \leq i \leq 24,\\ 
        \frac{\partial \phi}{\partial \tau} \left(p_{i-24}\right) &  25 \leq i \leq 32,\\ 
        \frac{\partial^2 \phi}{\partial x \partial y} \left(p_{i-32}\right) &  33 \leq i \leq 40,\\ 
        \frac{\partial^2 \phi}{\partial y \partial \tau} \left(p_{i-40}\right) &  41 \leq i \leq 48,\\ 
        \frac{\partial^2 \phi}{\partial x \partial \tau} \left(p_{i-48}\right) &  49 \leq i \leq 56,\\ 
        \frac{\partial^3 \phi}{\partial x \partial y \partial \tau} \left(p_{i-56}\right) &  57 \leq i \leq 64 ,\\ 
    \end{cases}
\end{equation}
where the indices $p_1, p_2, \dots, p_8$ correspond to the 8 nodes of the grid cell.
The $64\times64$ matrix $\mathbf{B}^{-1}$ is the core of the tricubic interpolation scheme, which is the same for all cells and can be computed
by combining Eqs.~\eqref{eq:interp},~\eqref{eq:acoef} and \eqref{eq:bvector}, as illustrated in Ref.~\cite{lekien}.
The imposed constraint allows all functions in Eq.~\eqref{eq:set2}  to be globally continuous across the domain covered by the grid.

Since in the case of the e-clouds the exact derivatives of the potential are not known, central differences are used to evaluate them from the discrete samples $\phi^{ijk}$, for example:
\begin{equation}\label{eq:central}
    \frac{\partial \phi}{\partial x}^{ijk} \approx \frac{\phi^{i+1,j,k} - \phi^{i-1,j,k}}{2\Delta x} .
\end{equation}

\begin{figure}[t]
      \centering
    \begin{minipage}[h]{0.45\textwidth}
        \centering
    \includegraphics[width=\textwidth]{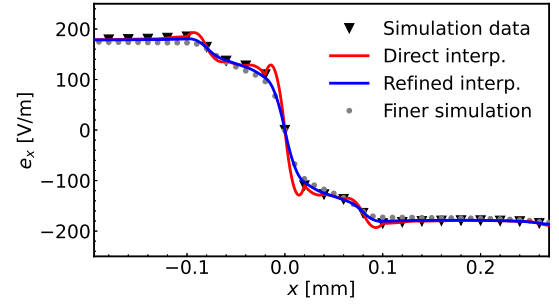}
    \\
        (a)
    \end{minipage}
    \begin{minipage}[h]{0.45\textwidth}
        \centering
    \includegraphics[width=\textwidth]{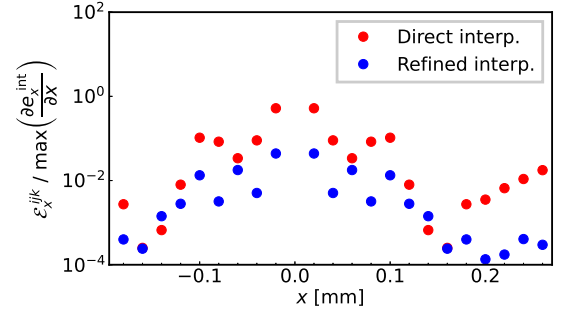}
      \\
        (b)
    \end{minipage}
    \caption{\label{fig:fig03}(a) Horizontal forces from the e-cloud interaction as a function of the transverse coordinate $x$ in the vicinity of 
    the closed orbit of the beam for $x=0,\ t=0.14$\,ns.
    The interpolating function obtained by the direct application of the tricubic method is shown by the red curve alongside the result obtained with the refinement procedure, visible by the blue line. The black triangles represent the simulation data used for the interpolation and the grey dots represent a finer simulation used for comparison.
    (b) Discontinuities on the first derivatives obtained with the two compared methods.}
\end{figure}

\subsection{Refinement of the potential}

We found that for a typical e-cloud distribution, a direct application of the scheme described in the previous section leads to unacceptable artifacts on the interpolated force.
This is shown by the red line in Fig.~\ref{fig:fig03}, which represents the result of the direct application of tricubic interpolation to the test e-cloud depicted in Fig.~\ref{fig:fig-intro}.
It is evident that the interpolating function is irregular and in particular shows discontinuous derivatives of the fields at the cell boundaries.
Specifically, the quantities 
\begin{equation}
\frac{\partial e_x^\text{int}}{\partial x} = -\frac{\partial^2 \phi^\text{int}}{\partial x^2},\;\;
\frac{\partial e_y^\text{int}}{\partial y} = -\frac{\partial^2 \phi^\text{int}}{\partial y^2},\;\;
\frac{\partial e_\tau^\text{int}}{\partial \tau} = -\frac{\partial^2 \phi^\text{int}}{\partial \tau^2} ,
\end{equation}
 are not globally continuous 
and, in fact are discontinuous at the cell boundaries across the $x,y,\tau$ directions, respectively, as discussed in Ref.~\cite{lekien}.
Expressions for the first derivative of $e_x^\text{int}$ at the grid points are derived in Ref.~\cite{lekien}
\begin{align}
\left.\frac{\partial e_x^{\text{int}}}{\partial x} \right|_{x\to x_i^+}
&=
    - 6\frac{ \phi^{i+1,j,k} - \phi^{i,j,k}}{\Delta x^2}  
    - 2 \frac{ 
      e_x^{i+1,j,k} 
    + 2  e_x^{i,j,k} 
    }{\Delta x} ,\\
\left.\frac{\partial e_x^{\text{int}}}{\partial x} \right|_{x\to x_i^-}
&=
    - 6 \frac{ 
    \phi^{i-1,j,k} 
    - \phi^{i,j,k}
    }{\Delta x^2}  
    + 2 \frac{ 
      e_x^{i-1,j,k} 
     + 2  e_x^{i,j,k} 
    }{\Delta x} ,
\end{align}
from which we can compute the discontinuity at the grid nodes as
\begin{align}
    \mathcal{E}_{x}^{ijk}
          &=\left.\frac{\partial e_x^\text{int}}{\partial x}\right|_{x\rightarrow x_i^+} - 
          \left.\frac{\partial e_x^\text{int}}{\partial x}\right|_{x\rightarrow x_i^-} \nonumber \\
    &= - 6\frac{ \phi^{i+1,j,k} - \phi^{i-1,j,k}}{\Delta x^2}  
    - 2 \frac{ 
      e_x^{i+1,j,k} 
    + 4  e_x^{i,j,k} 
    + e_x^{i-1,j,k} 
    }{\Delta x} .
\end{align}
The discontinuity of the derivative of red line of Fig.~\ref{fig:fig03}a is shown with red dots in Fig.~\ref{fig:fig03}b where it can be seen that large ripples correspond to large discontinuities.
Taking into account that $e^{ijk}_x$ 
are evaluated by directly applying central differences (Eq.~\eqref{eq:central}) on the discrete samples $\phi^{ijk}$, we obtain:
\begin{equation}
    \mathcal{E}_{x,\text{direct}}^{ijk}
    =  
     \frac{ 
      \phi^{i+2,j,k} 
    - 2  \phi^{i+1,j,k} 
    + 2  \phi^{i-1,j,k} 
    - \phi^{i-2,j,k} 
    }{\Delta x^2} .
\end{equation}
By Taylor expanding $\phi(x,y,\tau)$ with respect to x around $x_i$, we obtain:
\begin{equation}\label{eq:direct}
    \mathcal{E}_{x,\text{direct}}^{ijk}
 = -2 \frac{\partial^3 \phi}{ \partial x^3}\!\left(x_i,y_j,\tau_k\right) \Delta x + O\!\left(\Delta x^3\right) .
\end{equation}
From this expression it is clear that the discontinuity could in principle be minimized by reducing the grid spacing $\Delta x$ in the PIC simulations of the electron dynamics.
Unfortunately, when reducing the grid spacing by a factor of $h$ in all three dimensions, the number of macroparticles must be increased by a factor $h^3$ in order to avoid introducing numerical shot noise.
For a realistic e-cloud simulation, this approach quickly becomes prohibitive both in terms of memory consumption and computation time.

It is important to remark that the initially chosen $\Delta x$ is sufficient to properly resolve
the electron dynamics and the introduced forces on the grid while
it is only at the interpolation stage that the artifact is introduced.
Hence, it is possible to leave $\Delta x$ unchanged in the PIC simulation and 
instead apply a refinement scheme directly on the result of such a simulation.
In particular, we can increase locally the resolution of $\phi$ by linearly interpolating $\rho$ on a grid with spacing reduced by a factor of $h$
\begin{equation}
 \Delta x_\text{refined}  =   \frac{\Delta x}{h} ,
\end{equation}
and solving the Poisson equation on this finer grid.
This provides the potential in additional grid points.
For example along the x direction of the cell with indices $i,j,k$, we have:
\begin{equation}
    \phi^{i,j,k},\phi^{i+1/h,j,k},\phi^{i+2/h,j,k},\ \dots\ ,\phi^{i+(h-1)/h,j,k}, \phi^{i+1,j,k} .
\end{equation}

Working on such a refined grid allows for a more local estimate of all quantities in Eq.~\eqref{eq:set2} on the grid nodes using central differences, for example:
\begin{equation}
    \frac{\partial \phi}{\partial x}^{ijk} \approx h\frac{\phi^{i+1/h,j,k} - \phi^{i-1/h,j,k}}{2\Delta x} .
\end{equation}
The memory required to store such quantities in all nodes of the finer grid
scales proportionally to $h^3$, which in practical cases severely limits the choice of the factor $h$.
To avoid such a limitation, the quantities in Eq.~\eqref{eq:set2} as obtained from the refined grid are stored only on the nodes of the original grid and used to perform tricubic interpolation on this coarse grid.
It is possible to evaluate the discontinuities introduced by such a scheme by following the same approach as before:
\begin{widetext}
\begin{equation}
\begin{aligned}
    \mathcal{E}_{x,\text{refined}}^{ijk}
    =&  
     -6\frac{ 
        \phi^{i+1,j,k} 
      -  \phi^{i-1,j,k} 
    }{\Delta x^2} 
    +h\frac{
      -  \phi^{i+1+1/h,j,k} 
      +  \phi^{i+1-1/h,j,k} 
    }{\Delta x^2}\\
    &+h\frac{
      - 4 \phi^{i+1/h,j,k} 
      + 4 \phi^{i-1/h,j,k} 
      -  \phi^{i-1+1/h,j,k} 
      +  \phi^{i-1-1/h,j,k} 
    }{\Delta x^2} .
\end{aligned}
\end{equation}
\end{widetext}
Expanding $\phi(x,y,\tau)$ as a Taylor series with respect to x around $x_i$, we obtain:
\begin{equation}\label{eq:refined}
    \mathcal{E}_{x,\text{refined}}^{ijk}
 = -2 \frac{\partial^3 \phi}{ \partial x^3}\!\left(x_i,y_j,\tau_k\right) \frac{\Delta x}{h^2} + O\!\left(h^{-4}\Delta x^3\right).
\end{equation}
Comparing Eqs.~\eqref{eq:direct} and~\eqref{eq:refined} shows that the proposed refinement scheme is able to arbitrarily reduce 
the discontinuities while keeping the memory required to store the interpolation coefficients independent of the choice of $h$.
The result of applying such a scheme to the samples in Fig.~\ref{fig:fig03}a is shown with the blue line in the same figure.
It is evident that the artifacts are practically suppressed.
Figure~\ref{fig:fig03}b provides a quantitative comparison of the observed discontinuities, which shows that the artifacts are reduced by one order of magnitude in the most affected points.

We underline that the refinement method presented in this section is applied during the pre-processing stage of the simulation (called part A in Sec.\,\ref{sec:ws}) and therefore does not at all affect the computation time of the actual beam dynamics simulation (part B in Sec.\,\ref{sec:ws}).

\begin{figure}
    \centering
    \begin{minipage}[h]{0.45\textwidth}
    \centering
    \includegraphics[width=\textwidth]{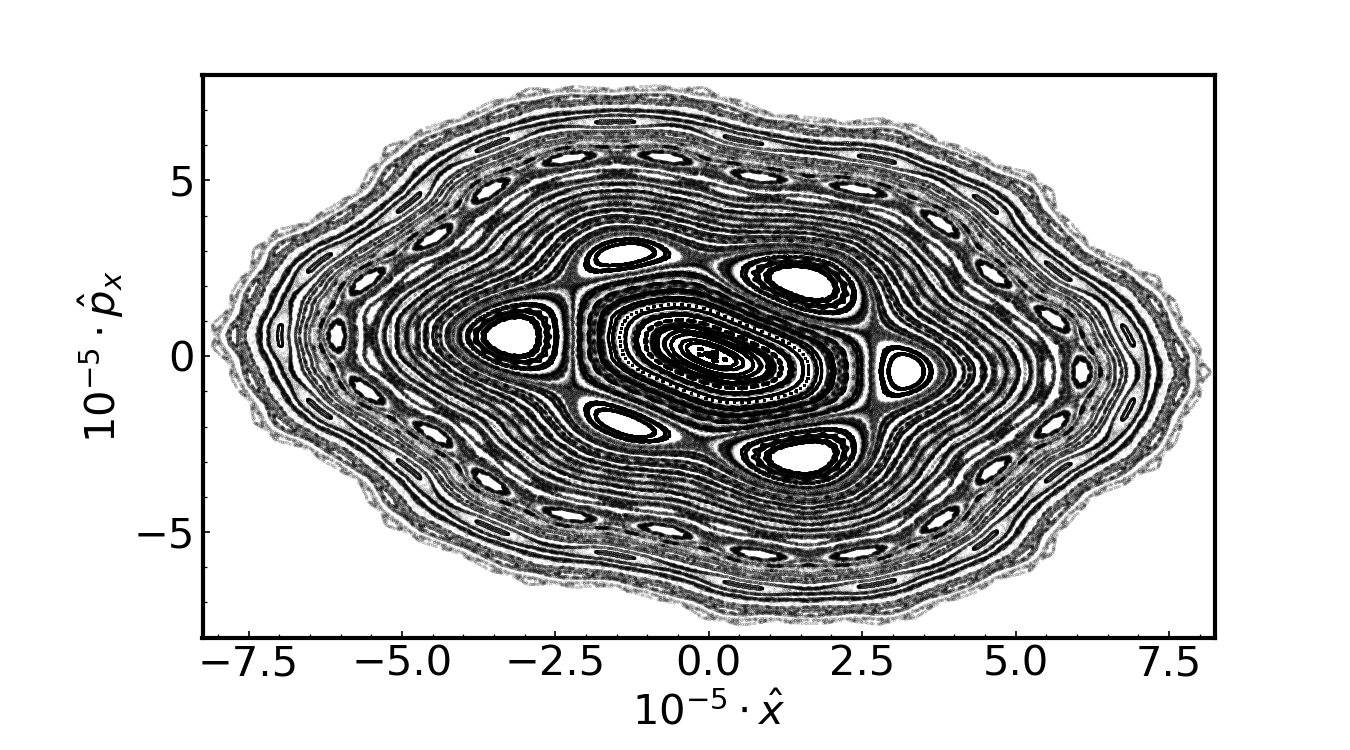}
     \\ (a)
    \end{minipage}
    \begin{minipage}[h]{0.45\textwidth}
    \centering
    \includegraphics[width=\textwidth]{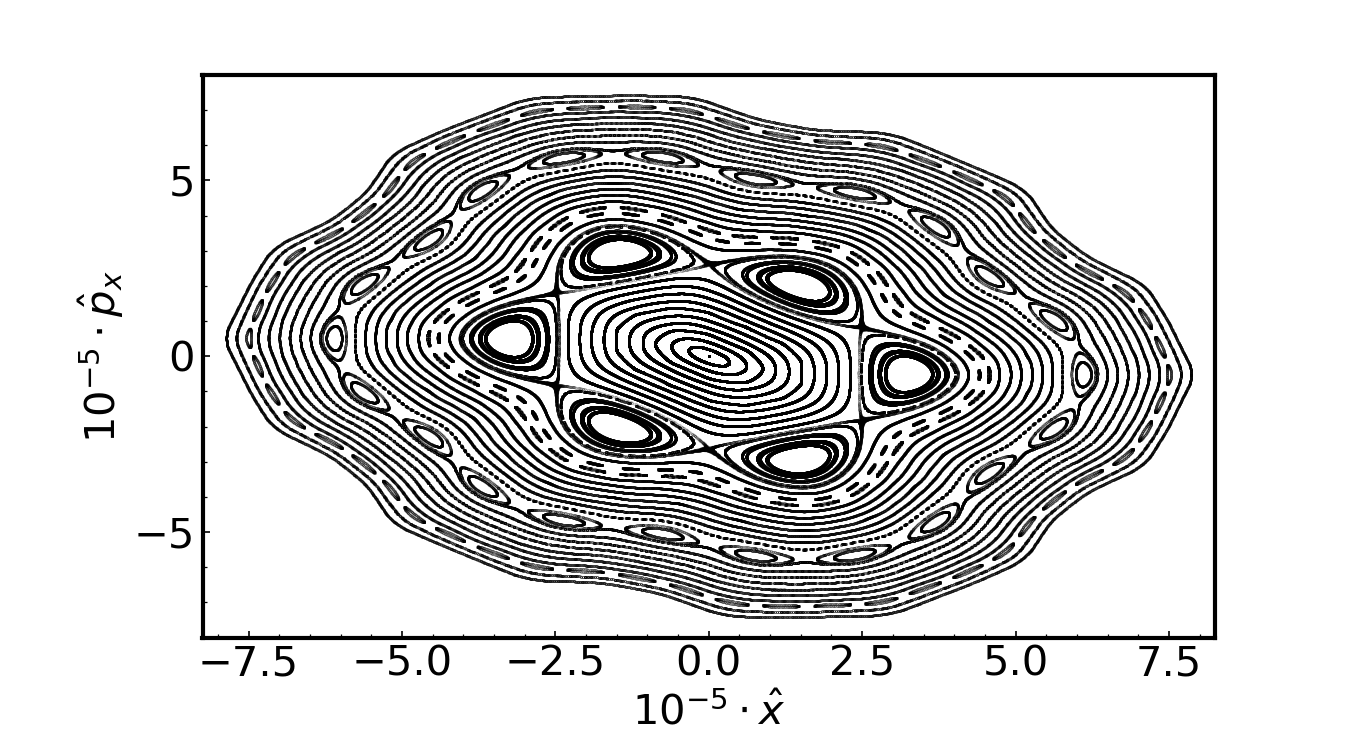}
     \\ (b)
    \end{minipage}
    \caption{\label{fig:fig04}Poincar\'e plot in normalized phase space with direct interpolation (a) and refined interpolation (b) of the e-cloud scalar potential.}
\end{figure}

The importance of applying this refinement scheme is visible in Fig.~\ref{fig:fig04}, where Poincar\'e plots of the non-linear motion of the beam particles over several turns are shown in the normalized phase space.
These were produced by tracking in the two degrees of freedom $x, p_x$ through the successive interaction of the e-cloud shown in Fig.~\ref{fig:fig03} interleaved with a linear one-turn-map.
The artifacts introduced by the direct interpolation method result
into modifications of the particle dynamics, which are mitigated by the refined interpolation method. 
This is visible through the presence of both higher-order
resonances at large amplitudes and
chaotic motion in Fig.~\ref{fig:fig04}a, which are absent in Fig.~\ref{fig:fig04}b where
the refined interpolation was used instead.

\begin{figure}
    \begin{minipage}[h]{0.45\textwidth}
    \includegraphics[width=\textwidth]{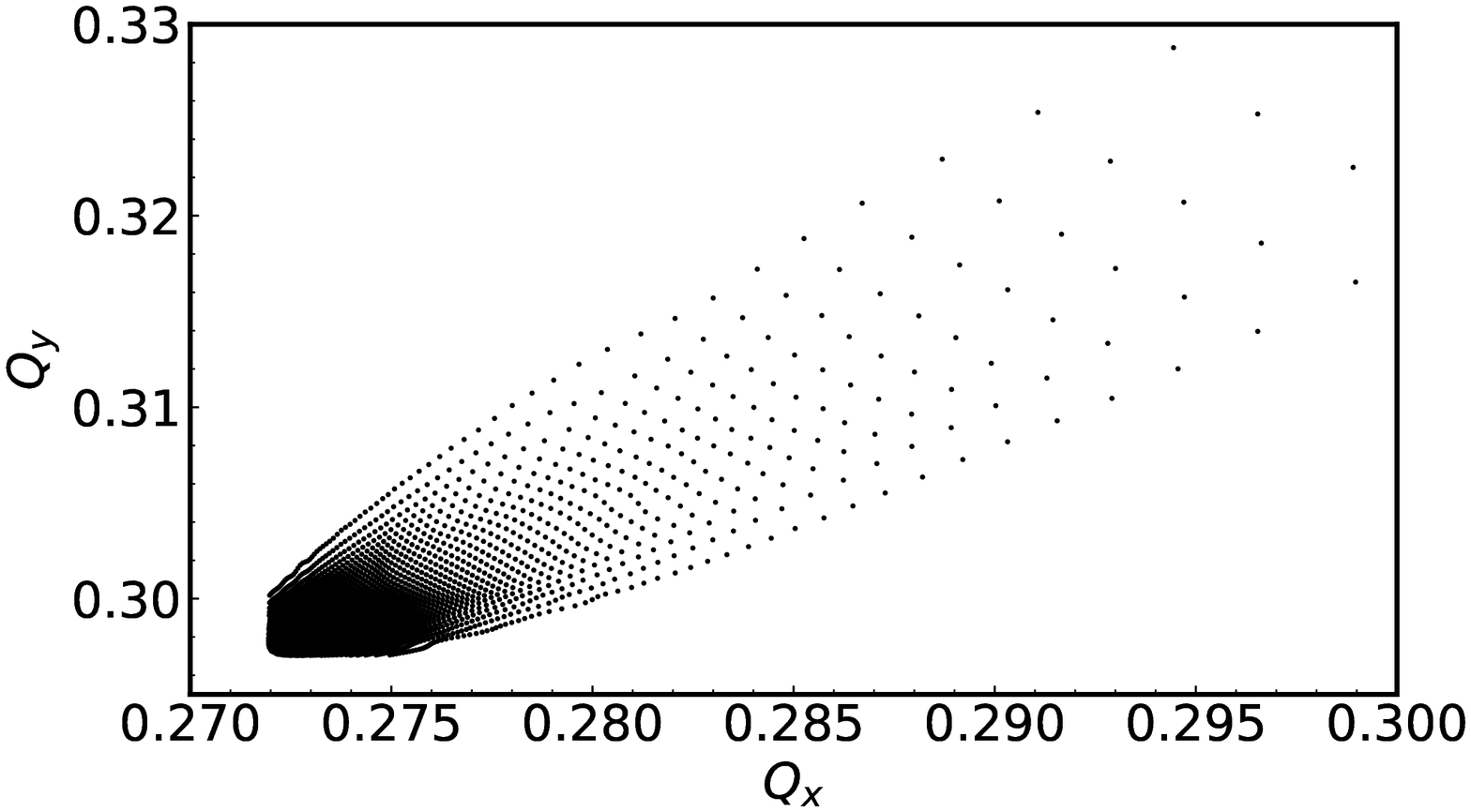}
     \\ (a)
    \end{minipage}
    \begin{minipage}[h]{0.45\textwidth}
    \includegraphics[width=\textwidth]{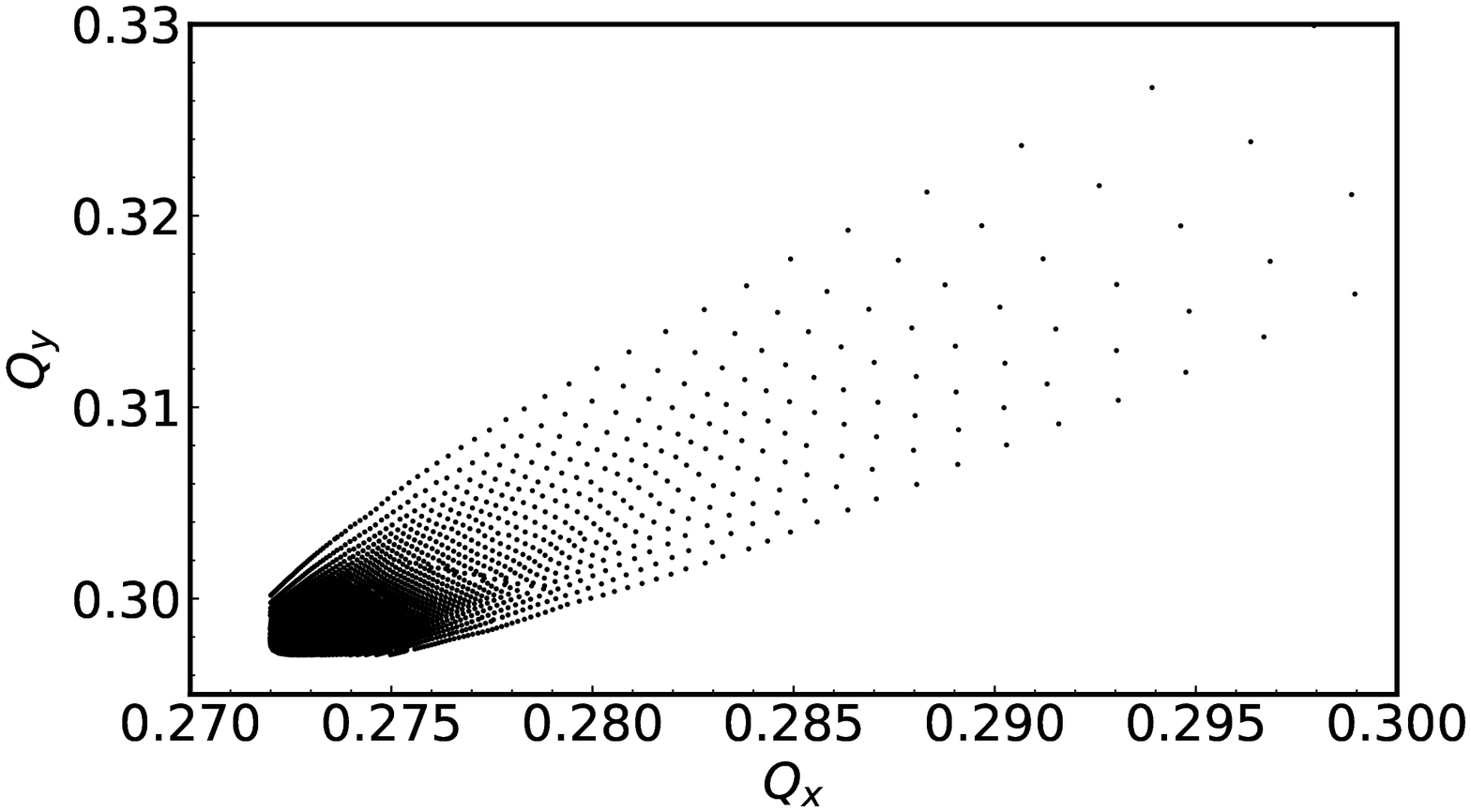}
     \\ (b)
    \end{minipage}
    \caption{\label{fig:foot}Betatron tunes calculated with the refined interpolation method in a non-linear lattice model (a) and with PyHEADTAIL simulations (b).}
\end{figure}

For the purpose of validating the fields computed with the refinement method described above, we simulated with the PyECLOUD code the same e-cloud pinch on a grid having smaller cell size.
The results of such simulation are shown by the grey points of Fig.~\ref{fig:fig03}a showing how
the refinement method is able to accurately reproduce the e-cloud field between the provided grid points. 
To validate the new method also in terms of its impact on the simulated beam dynamics, we compare  the obtained betatron tunes for different transverse amplitudes against an independent simulation performed with the PyECLOUD-PyHEADTAIL suite. The result of such a comparison is illustrated in Fig.~\ref{fig:foot}. The tunes are computed for a distribution of on-momentum test particles whose horizontal and vertical amplitudes of oscillation are uniformly distributed between $0.03$ and $1.1$ mm.
In Fig.~\ref{fig:foot}a, the newly introduced method was used in combination with the non-linear lattice model of the LHC and the e-cloud interaction is modeled by 368 interactions uniformly distributed over the LHC arcs.
In Fig.~\ref{fig:foot}\,b a PyECLOUD-PyHEADTAIL Particle-In-Cell simulation is used. In this case 
the e-cloud effect is modeled by only 23 interactions and the particles are tranported between interactions using a linear transfer matrix.
The test particles are tracked for 1000 turns and the betatron tunes are calculated from the horizontal and vertical turn-by-turn positions.
Excellent agreement between the two methods can be observed comparing the two plots.

\section{Application to the LHC with protons at 450 GeV}\label{sec:application}

\begin{figure*}[p]
    \centering
    \begin{minipage}[h]{0.40\textwidth}
    \centering
    \includegraphics[width=\textwidth]{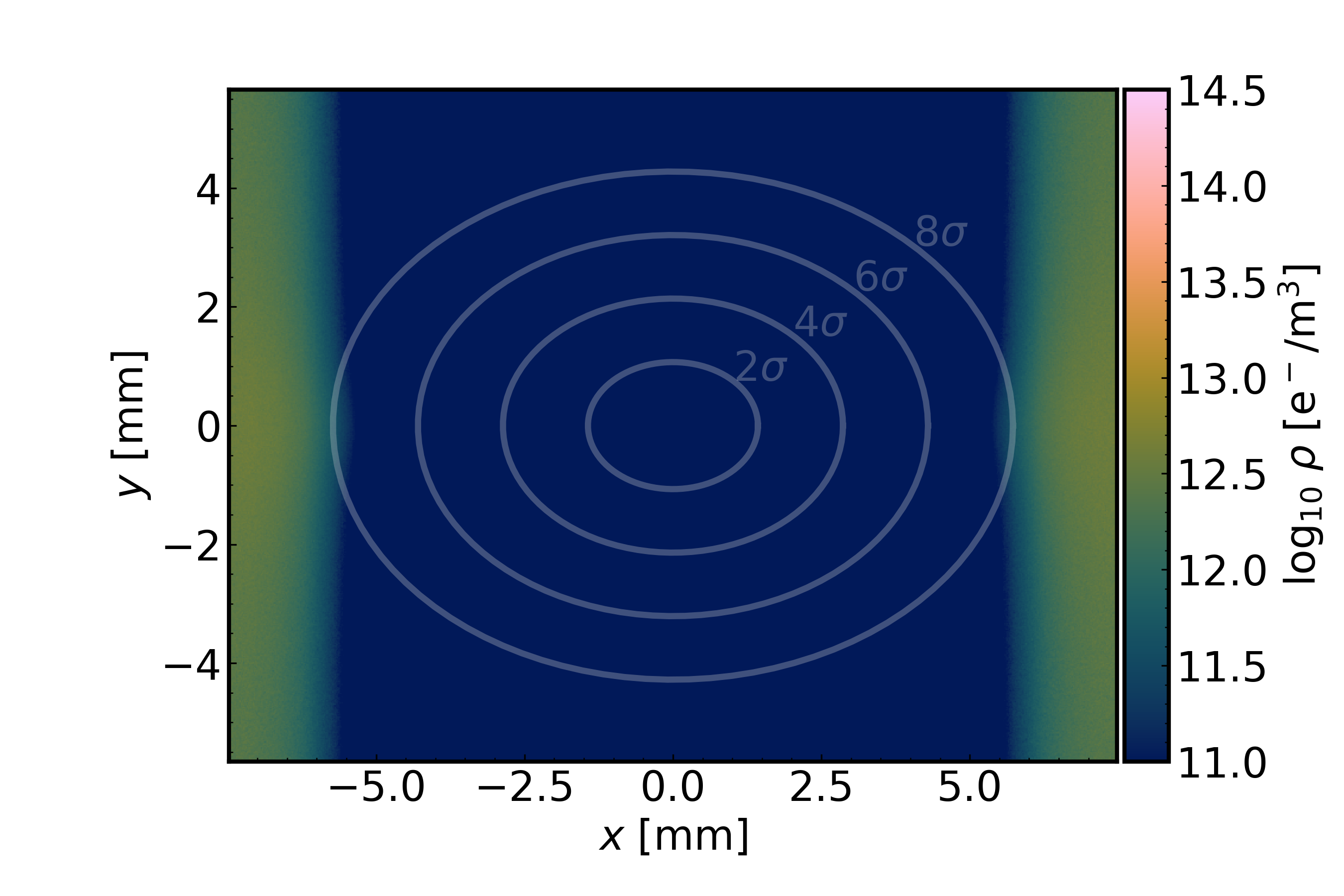}
     \\ (a) 
    \end{minipage}
    \begin{minipage}[h]{0.40\textwidth}
    \centering
    \includegraphics[width=\textwidth]{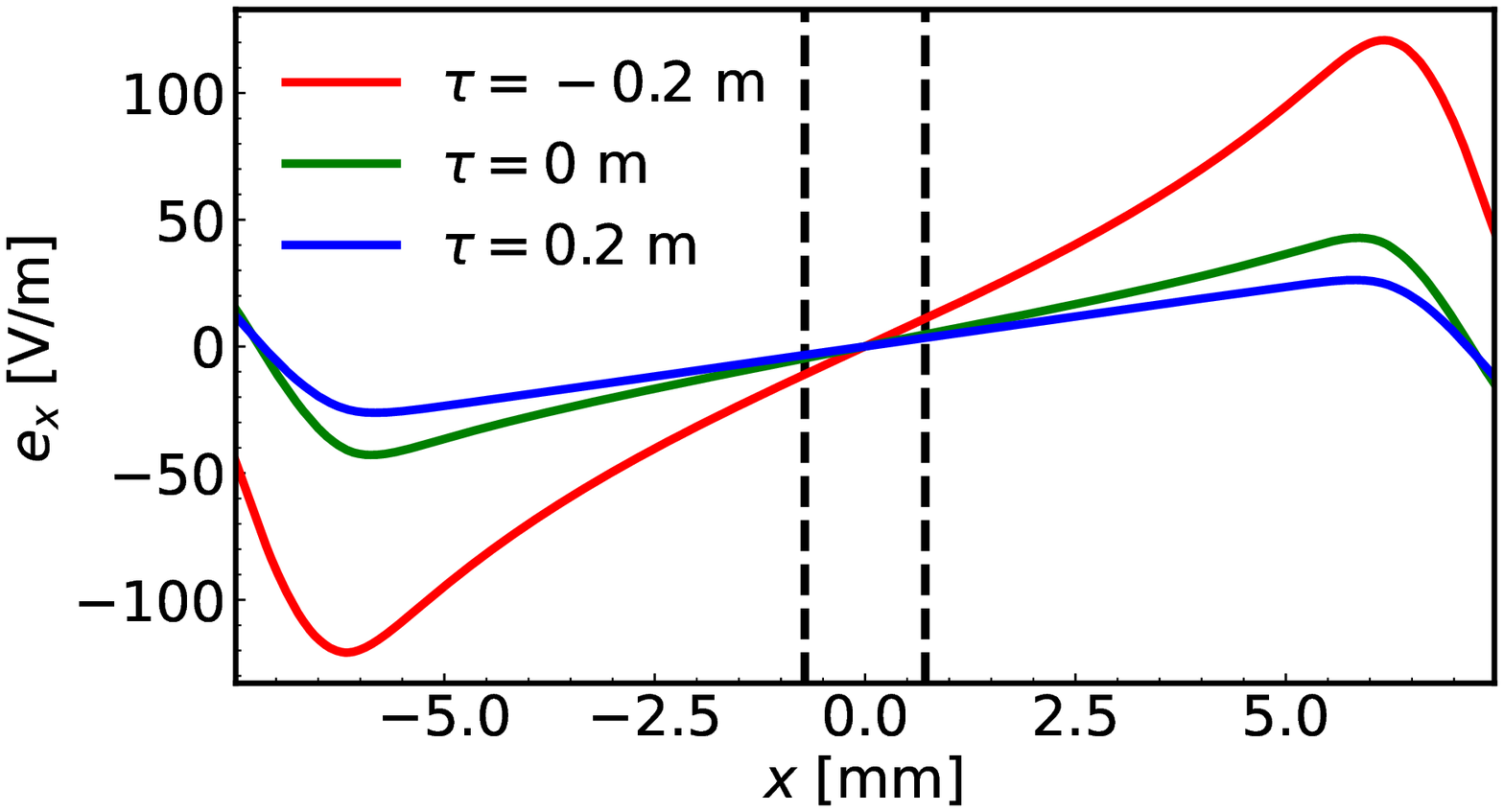}
     \\ (b) 
    \end{minipage}
    \caption{\label{fig:fig06}Snapshot of the electron cloud density in an MB magnet (a) and horizontal electric field in the plane $y=0$ at different moments during the bunch passage (b) for the nominal bunch intensity of $1.2\cdot 10^{11}$\,p/bunch
    and $\mathrm{SEY}=1.3$.
    }

    \begin{minipage}[h]{0.40\textwidth}
    \centering
    \includegraphics[width=\textwidth]{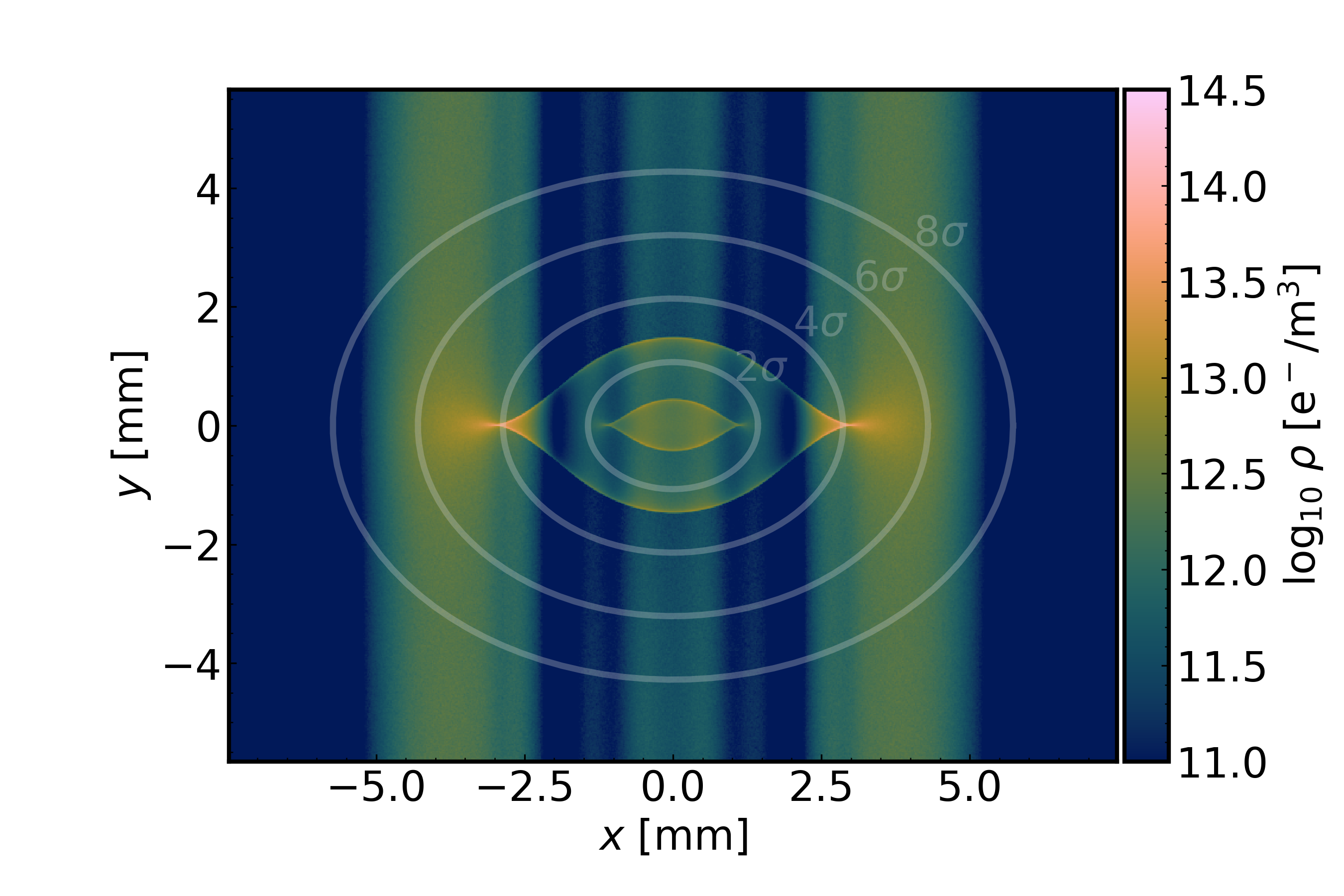}
     \\ (a) 
    \end{minipage}
    \begin{minipage}[h]{0.40\textwidth}
    \centering
    \includegraphics[width=\textwidth]{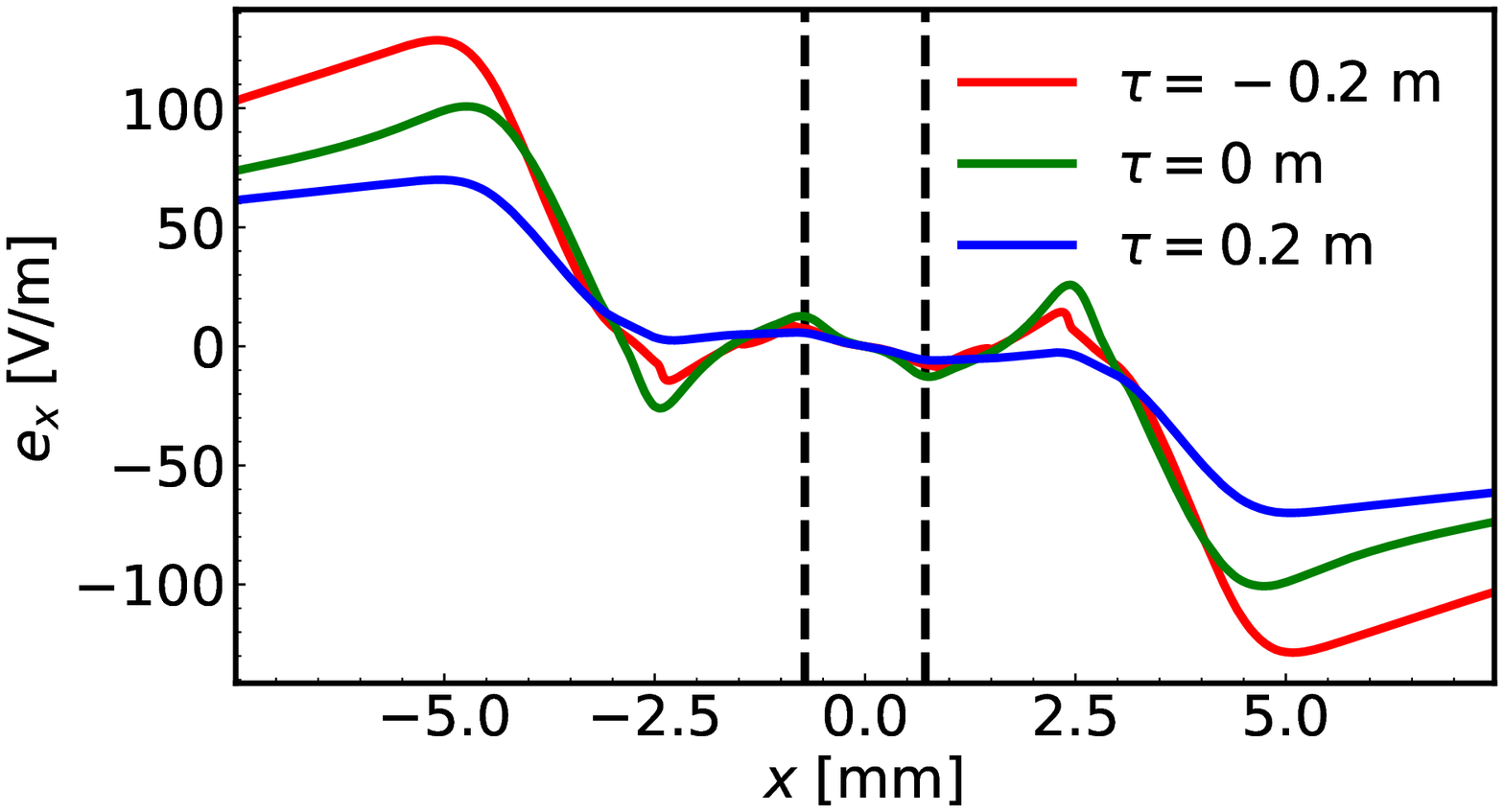}
     \\ (b) 
    \end{minipage}
    \caption{Snapshot of the electron cloud density in an MB magnet (a) and horizontal electric field in the plane $y=0$ at different moments during the bunch passage (b) for the reduced bunch intensity of $0.6\cdot 10^{11}$\,p/bunch
    and $\mathrm{SEY}=1.3$}
    \label{fig:fig05}

    \begin{minipage}[h]{0.40\textwidth}
    \centering
    \includegraphics[width=\textwidth]{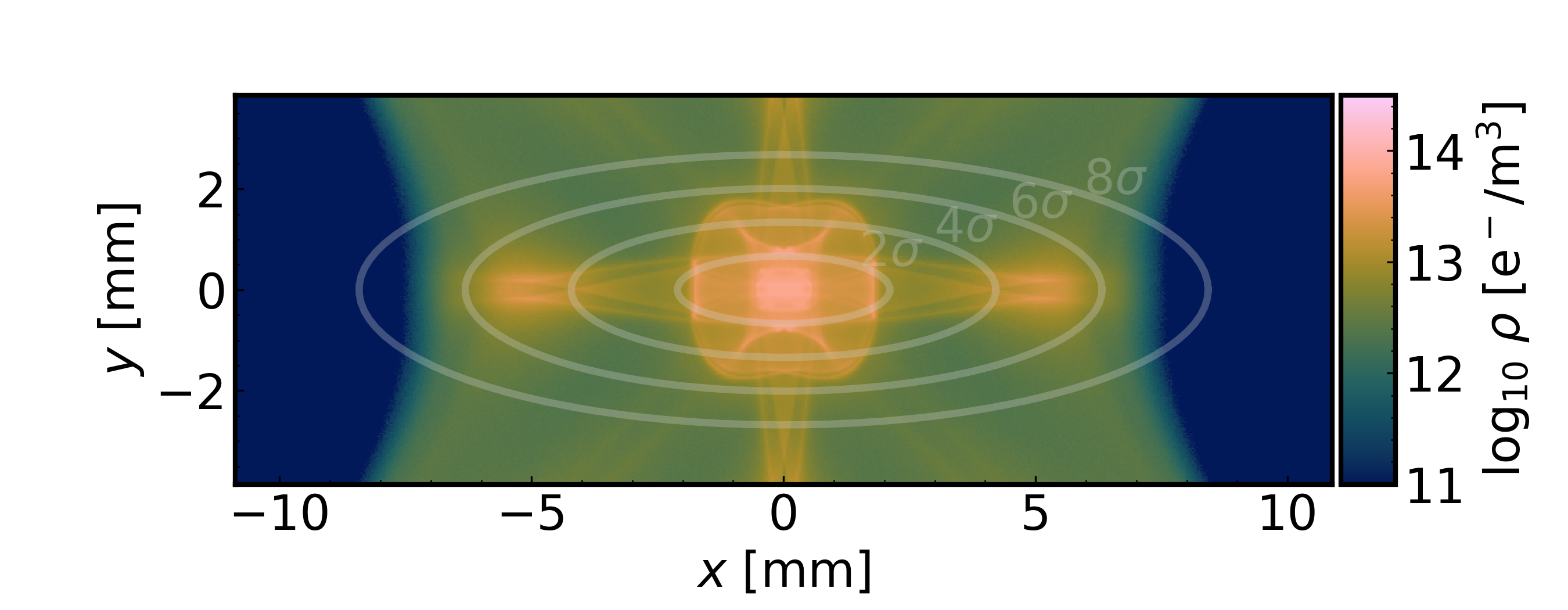}
     \\ (a) 
    \end{minipage}
    \begin{minipage}[h]{0.40\textwidth}
        \centering
    \includegraphics[width=\textwidth]{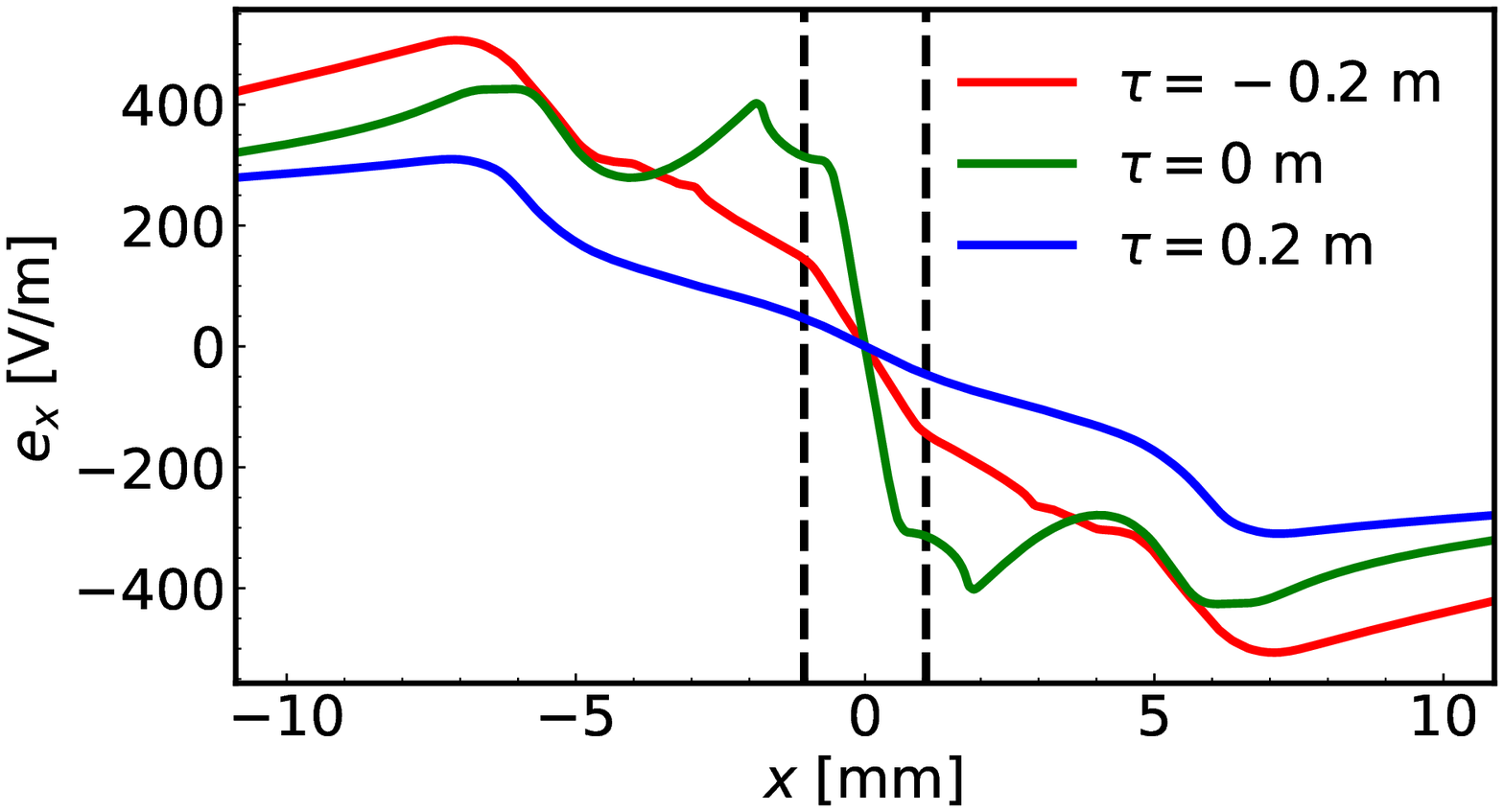}
     \\ (b) 
    \end{minipage}
    \caption{Snapshot of the electron cloud density in a focusing MQ magnet (a) and horizontal electric field in the plane $y=0$ at different moments during the bunch passage (b) for the nominal bunch intensity of $1.2\cdot 10^{11}$\,p/bunch
    and $\mathrm{SEY}=1.3$}
    \label{fig:fig08}
    \begin{minipage}[h]{0.40\textwidth}
    \centering
    \includegraphics[width=\textwidth]{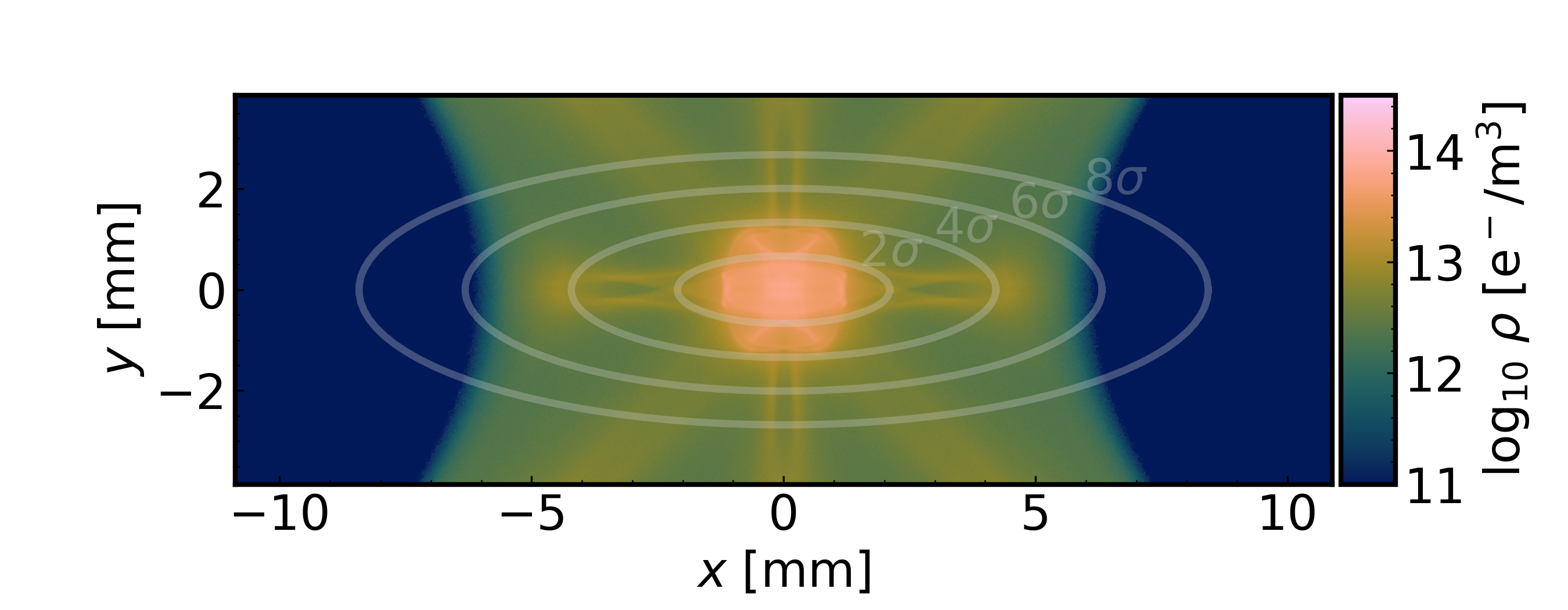}
     \\ (a) 
    \end{minipage}
    \begin{minipage}[h]{0.40\textwidth}
    \centering
    \includegraphics[width=\textwidth]{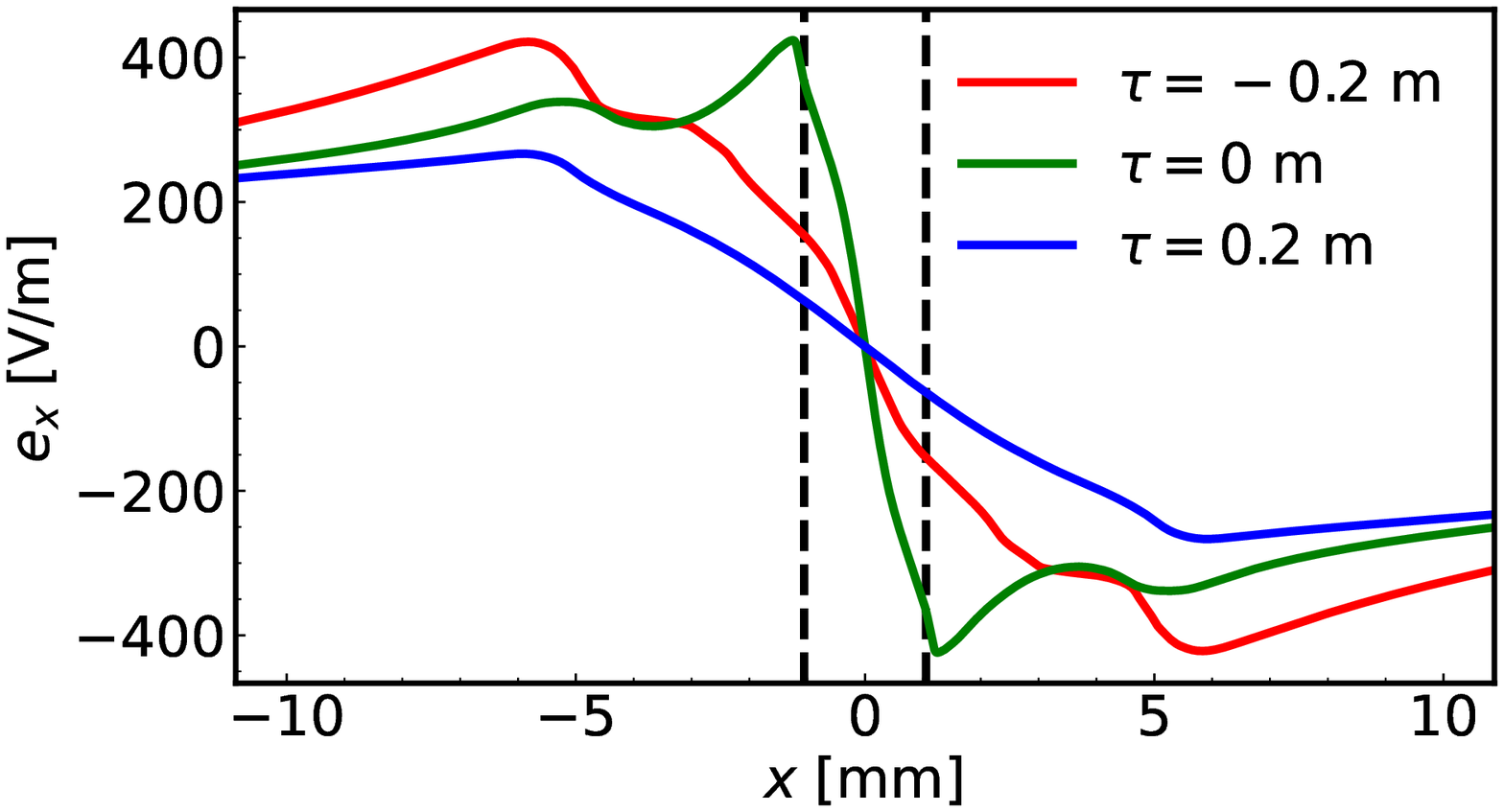}
     \\ (b) 
    \end{minipage}
    \caption{Snapshot of the electron cloud density in a focusing MQ magnet (a) and horizontal electric field in the plane $y=0$ at different moments during the bunch passage (b) for the reduced bunch intensity of $0.6\cdot 10^{11}$\,p/bunch and $\mathrm{SEY}=1.3$}
    \label{fig:fig07}
\end{figure*}

To illustrate the potential of the method presented in the previous sections, we consider the case of the LHC.
In particular, we focus on operation with protons at injection energy with the typical configuration used during the second physics run (Run 2, 2015-2018) of the LHC.
The main parameters defining the considered scenario are reported in Table~\ref{tab:tab1}.
In this configuration, the effect of the beam-beam interaction is negligible, the optics do not have particularly large 
betatron functions around the interaction points (as compared to when the beams are put in collision) and the strongest non-linear effects apart from the e-cloud are due to the large current 
in the magnetic octupoles and to the large chromaticity, both of which are required to mitigate coherent beam instabilities from e-cloud effects.
We consider only the e-clouds in the LHC arcs, 
which constitute a large fraction of the ring circumference.
In particular we include in our model the effect of e-cloud in the main dipole magnets (MB) ($\sim$66\% of the ring circumference) and the main quadrupole magnets (MQ) ($\sim$7\% of the ring circumference).
The e-cloud buildup and the electron dynamics are simulated with the PyECLOUD code~\cite{pyparis}. 
More information on such a simulation model and its comparison against 
experimental data can be found in Refs.~\cite{galina-ipac,iadarola-arcs}.

\begin{table}
\caption{\label{tab:tab1} Typical operational parameters of the LHC in Run 2 used in the simulations}
\begin{ruledtabular}
\begin{tabular}{lc}
    Bunch population [p/bunch] & $1.2 \cdot 10^{11}$\\
    Reference energy [GeV] & 450 \\
    R.m.s. bunch length [cm] & 9 \\
    R.m.s. horizontal emittance (normalized) [\textmu m] & 2\\
    R.m.s. vertical emittance (normalized) [\textmu m] & 2\\
    Horizontal betatron tune & 62.27 \\
    Vertical betatron tune & 60.295 \\
    Synchrotron tune & $5.1\cdot 10^{-3}$ \\
    Horizontal chromaticity & 15 \\
    Vertical chromaticity & 15 \\
    Octupole magnets' current [A] & 40 \\
    Amplitude detuning coefficient~\cite{nicolas-thesis}, $\alpha_{xx}$ [\textmu m\textsuperscript{-1}] &  0.31 \\
    Amplitude detuning coefficient~\cite{nicolas-thesis}, $\alpha_{yy}$ [\textmu m\textsuperscript{-1}] &  0.32 \\
    Amplitude detuning coefficient~\cite{nicolas-thesis}, $\alpha_{xy}$ [\textmu m\textsuperscript{-1}] &  -0.22 \\
    RF voltage [MV] & 6 \\ 
    Bunch spacing [ns] & 25 \\
    MB magnet's length [m] & 14.3 \\
    MB magnet's field [T] & 0.535 \\
    MQ magnet's length [m] & 3.1 \\
    MQ magnet's field gradient [T m\textsuperscript{-1}] & 12.1\\
    Primary collimators' set distance [$\sigma$] & 7.5 \\
\end{tabular}
\end{ruledtabular}
\end{table}

The numerical parameters of the simulations concerning the buildup and the dynamics of the electrons are summarized in Table~\ref{tab:tab2} and have been chosen in accordance to previous studies in Refs.~\cite{PhysRevAccelBeams.23.081002,luca-note}. 
A significant effort was devoted in those studies in order to ensure the numerical convergence of the simulations.

\begin{table}
\caption{\label{tab:tab2} Numerical parameters used in PyECLOUD to simulate the electron dynamics in the e-cloud}
\begin{ruledtabular}
\begin{tabular}{lc}
    Time step [ps] & 5 \\
    Transverse grid resolution at the beam location [\textmu m] & 20 \\
    Number of macroparticles for the e-cloud & $5\cdot 10^{5}$ \\
    Number of macroparticles for the proton bunch & $25 \cdot 10^{7}$ \\
    Number of longitudinal slices along the proton bunch & 500 \\
\end{tabular}
\end{ruledtabular}
\end{table}

Figure~\ref{fig:fig06}a shows the simulated electron distribution in an arc dipole magnet (MB) for the nominal bunch intensity of $1.2\cdot10^{11}$\,p/bunch and $\mathrm{SEY}=1.3$.
It is a characteristic of the e-cloud developing in the dipolar field of the LHC MB magnets that two vertical stripes form left and right of the proton beam's closed orbit position~\cite{annalisa}.
For the nominal bunch intensity, such stripes are located far away from the beam location and the force exerted within the bunch is rather linear as shown in Fig.~\ref{fig:fig06}b.
The situation is significantly different for a reduced bunch intensity as shown
in Fig.~\ref{fig:fig05}a for $0.6 \cdot 10^{11}$\,p/bunch.
In this case, the electron stripes overlap with the beam distribution introducing significant non-linearities in the forces within the bunch as shown in Fig.~\ref{fig:fig05}b.

In the main quadrupole magnets (MQ), the e-cloud overlaps with the beam distribution regardless of the considered bunch intensity, generating non-linearities
in the forces within the bunch, as shown in Figs.~\ref{fig:fig08} and~\ref{fig:fig07} for the nominal and reduced bunch intensities, respectively.

\begin{figure}
    \centering
    \includegraphics[width=\columnwidth]{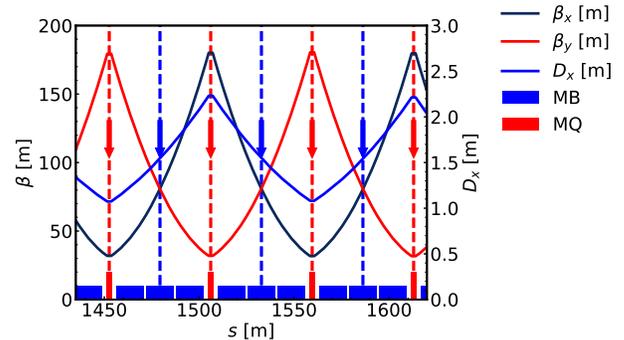}
    \caption{\label{fig:fig09}Beta functions and horizontal dispersion in the FODO cells of the LHC arcs. The arrows indicate the places where the e-cloud interactions are applied.}
\end{figure}

\begin{figure*}[t]
    \centering
    \begin{minipage}[t]{0.49\textwidth}
    \centering
        \includegraphics[width=\textwidth]{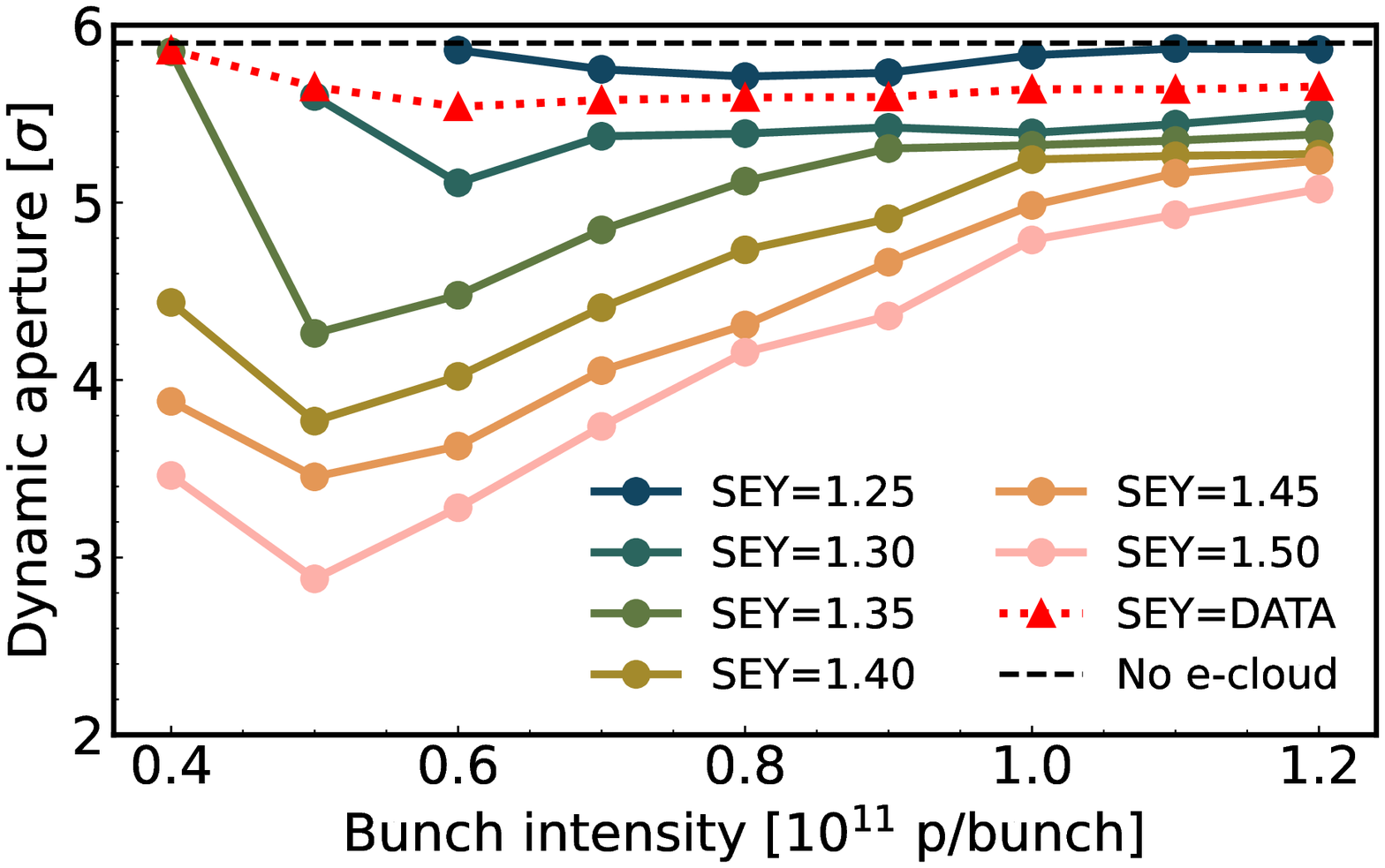}
     \\ (a)
    \end{minipage}
    \begin{minipage}[t]{0.49\textwidth}
    \centering
        \includegraphics[width=\textwidth]{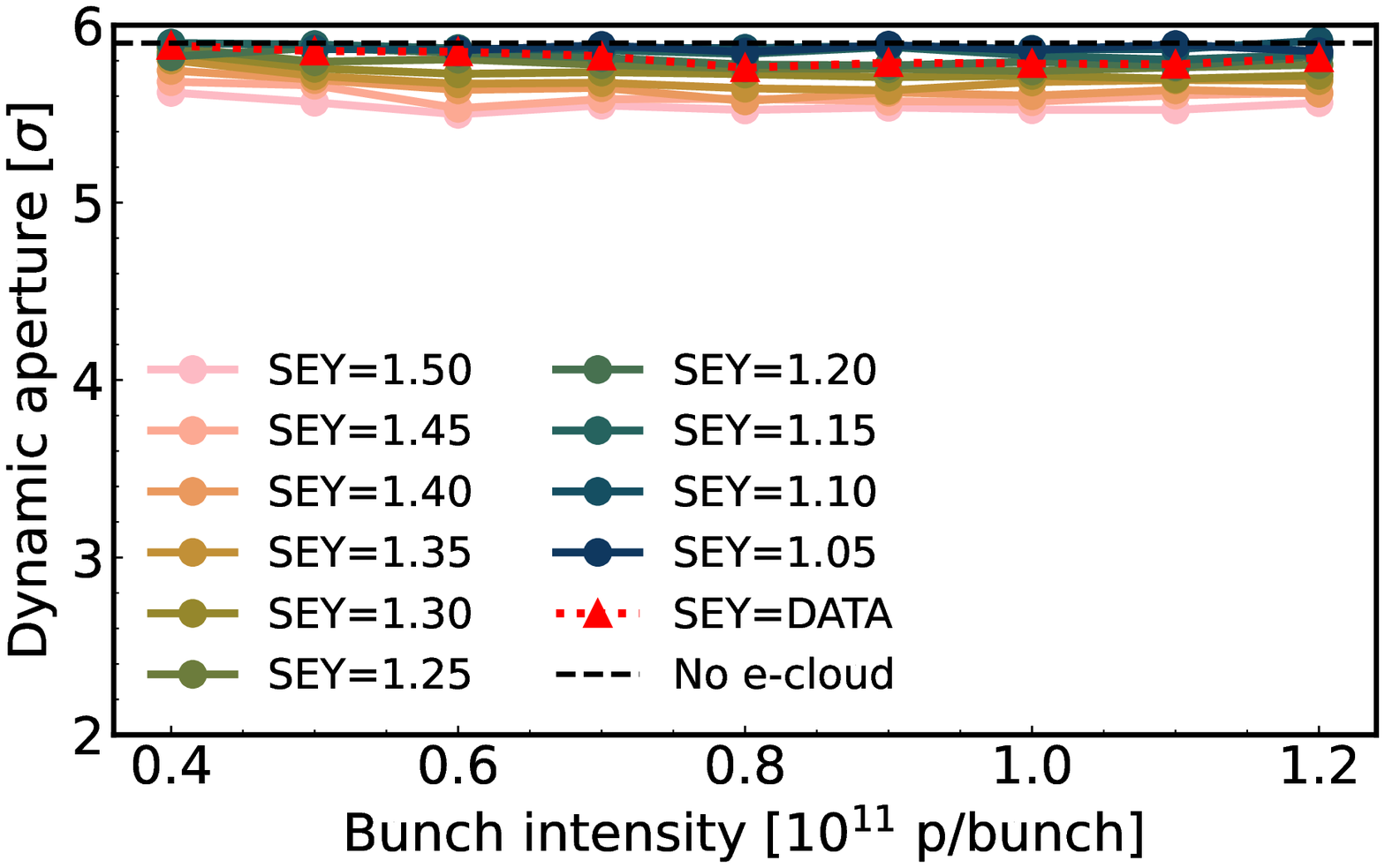}
     \\ (b)
    \end{minipage}
    \vspace{0.6cm}
    \begin{minipage}[t]{0.49\textwidth}
    \centering
        \includegraphics[width=\textwidth]{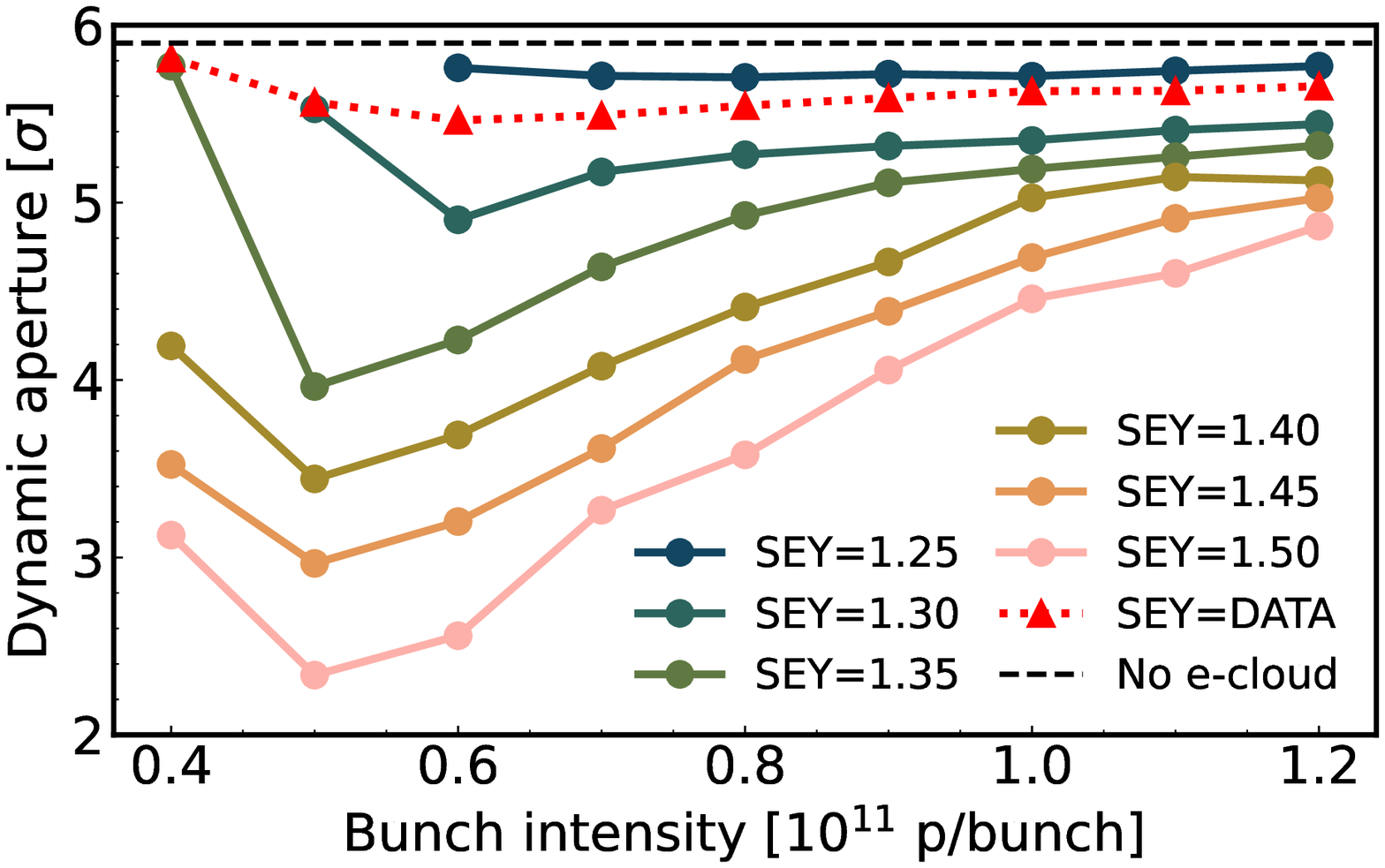}
     \\ (c)
    \end{minipage}
    \caption{\label{fig:fig10}Dynamic aperture as a function of the bunch intensity and SEY with simulations including e-clouds in the MB magnets only (a), in the MQ magnets only (b) and in the MB and MQ magnets (c).}
\end{figure*}

Each arc of the LHC consists of 23 regular FODO cells each including 6 MB magnets and two MQ magnets~\cite{LHC-design}.
A small part of an arc is shown in Fig.~\ref{fig:fig09} where the beta functions and the horizontal dispersion are plotted alongside an illustration of the position and extent of the main magnets.

The beam dynamics is simulated using the SixTrackLib code~\cite{sixtracklib}.
Each element of the lattice is modelled with an appropriate number of thin lenses. 
To model the e-cloud, we include in each cell four e-cloud interactions, two modelling the e-cloud in the six MB magnets and two modelling the e-cloud in the two MQ magnets.
In total, 368 interactions are used to model the e-cloud in the MB magnets, 180 interactions are used to model the e-cloud in the focusing MQ magnets while another 180 interactions are used for the defocusing MQ magnets.
Their locations inside the LHC cells are marked by the arrows in Fig.~\ref{fig:fig09}.
Particles are intercepted by the LHC primary collimators~\cite{LHC-design} which, in the simulation, limit the amplitude of the betatron oscillations to $7.5$ times the r.m.s. beam size.

To study incoherent effects from e-cloud a large number of turns needs to be simulated, which results in a significant computation time.
For this reason, it is convenient to perform the simulations using GPUs instead of 
conventional CPUs, which is possible within the SixTrackLib framework.
For the simulations illustrated in the following sections, the use of high-end GPUs provided a speedup in computation time of about 100 with respect to a single-thread simulation on a typical CPU.

\subsection{Characterization of non-linear dynamics}
\label{sec:characterization}

\begin{figure*}[t]
    \centering
    \begin{minipage}[t]{0.49\textwidth}
    \centering
        \includegraphics[width=\textwidth]{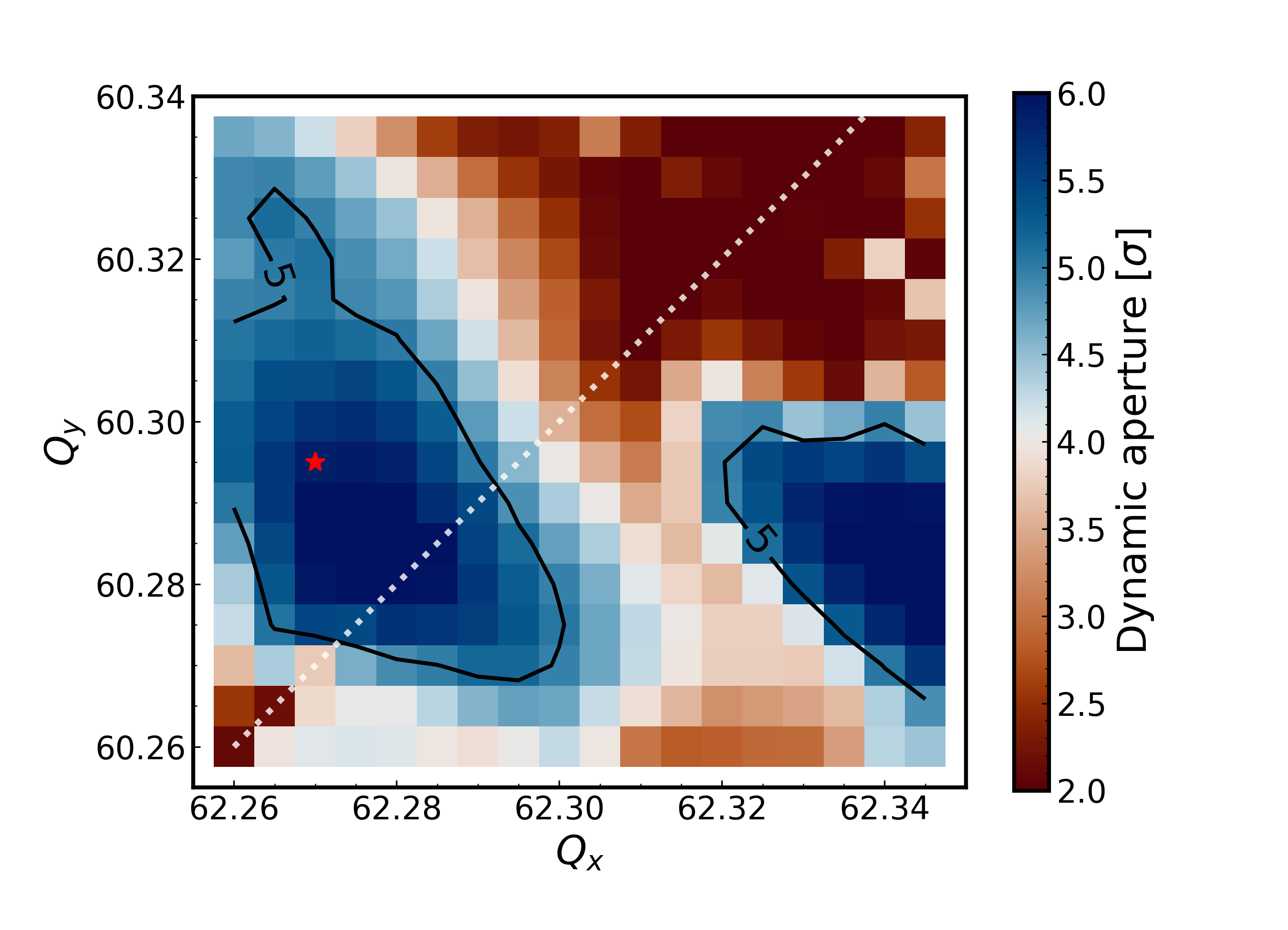}
     \\ (a)
    \end{minipage}
    \begin{minipage}[t]{0.49\textwidth}
    \centering
        \includegraphics[width=\textwidth]{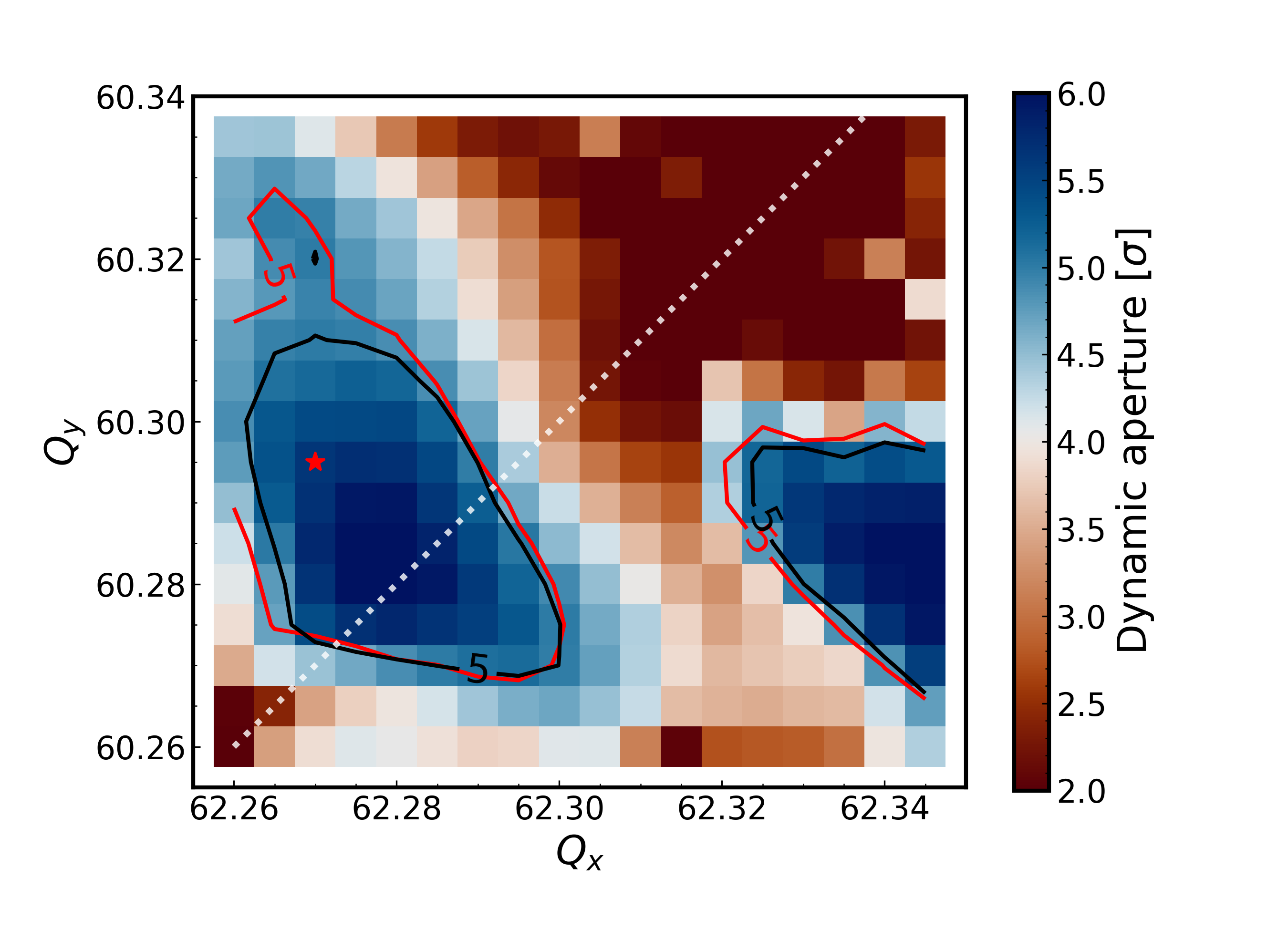}
     \\ (b)
    \end{minipage}
    \begin{minipage}[t]{0.49\textwidth}
    \centering
        \includegraphics[width=\textwidth]{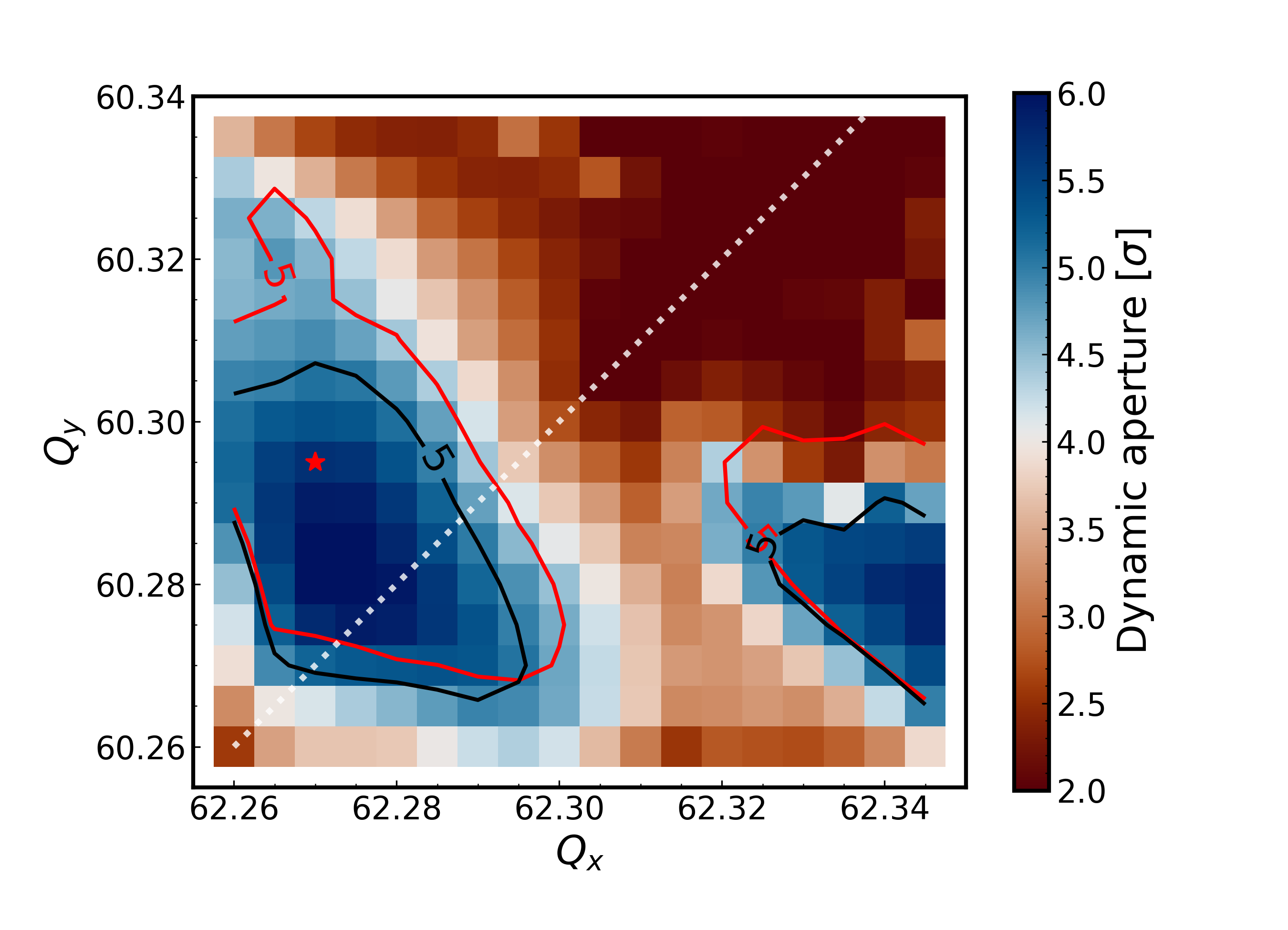}
     \\ (c)
    \end{minipage}
    \begin{minipage}[t]{0.49\textwidth}
    \centering
        \includegraphics[width=\textwidth]{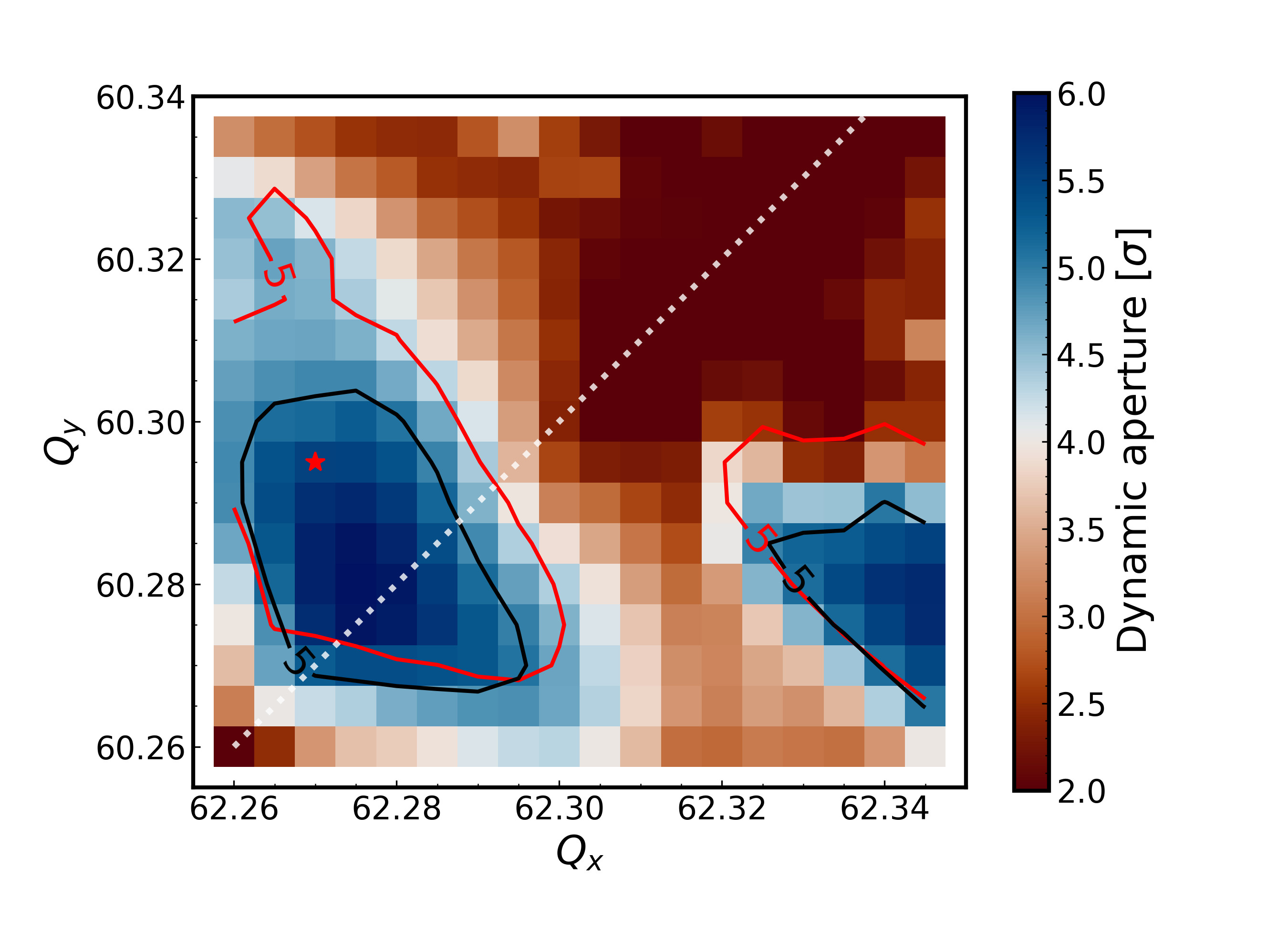}
     \\ (d)
    \end{minipage}
    \caption{\label{fig:fig11}Dynamic aperture as a function of the horizontal and vertical betatron tunes in the absence of e-cloud (a) and with e-clouds in the MB magnets only (b), in the MQ magnets only (c), and in the MB and MQ magnets (d). The red marker corresponds to the working point set during operation. Contours of DA equal to $5\sigma$ are shown with the black lines while the red lines in (b), (c) and (d) report the contours obtained without e-cloud.}
\end{figure*}

For a first characterization of the single-particle stability under the influence of the different e-cloud interactions, we use tracking simulations to evaluate the DA~\cite{dynap} of the machine. 
As usually done for LHC studies, the DA is defined as the normalized transverse oscillation amplitude above which particles are lost within $10^6$ turns.
The particles are initialized with an energy deviation of $p_\tau = 5.5 \cdot 10^{-4}$, to include effects from synchrotron oscillations in the evaluation of DA.
The DA is evaluated by simulating particles with initial coordinates on a polar grid in the (x, y) plane covering 
amplitudes 
between 0 and 7.5 times the r.m.s. beam size. This domain is discretized using 200 amplitude values and 100 angle values in the range (0, 90 degrees).
The DA estimations presented in the following refer to the average evaluated over the simulated angles.

Simulation results are illustrated in Fig.~\ref{fig:fig10} as a function of the bunch intensity for different values of the SEY.
The dashed black line marks the DA obtained in the absence of e-cloud effects.
Figure~\ref{fig:fig10}a shows the DA when only the e-cloud in the MB magnets is included in the simulation.
Increasing the SEY makes the e-cloud stronger, which results in a stronger degradation
of the DA.
The e-cloud in the MB magnets causes a degradation of the single-particle stability that is more severe for lower bunch intensities, for which the stripes in the electron distribution overlap with the bunch, resulting in stronger non-linear forces
(see Fig.~\ref{fig:fig05} and Appendix \ref{sec:app-ecloud}).
For comparison, we performed simulations approximating the actual SEY in the LHC chambers (SEY=DATA in the plots). In fact, the beam pipes in different half-cells of the LHC arcs show different SEY as a result of
different oxidation states of the
surface~\cite{petitBeaminducedSurfaceModifications2021}.
The red dotted curve shows the DA computed with a special e-cloud potential which has been constructed by averaging over the potentials obtained from simulations performed with the appropriate SEY for each half-cell (as estimated through cryogenic heat-load measurements).
The measurement and estimation of the SEY distribution is carried out and explained in Ref.~\cite{iadarola-arcs}.

Figure~\ref{fig:fig10}b shows the DA when only the e-cloud in the MQ magnets is included in the simulation.
The effect of the e-cloud in the MQ magnets on the DA is significantly weaker than in the MB case with only a very small dependence on the bunch intensity.
This is expected from the comparison of the electron distributions for the different intensities, as shown for example in Figs.~\ref{fig:fig07} and~\ref{fig:fig08}.
The combined effect of e-clouds in both MB and MQ magnets is shown in Fig.\ref{fig:fig10}c.

\begin{figure*}[p]
    \centering
    \begin{minipage}[t]{0.49\textwidth}
    \centering
        \includegraphics[width=\textwidth]{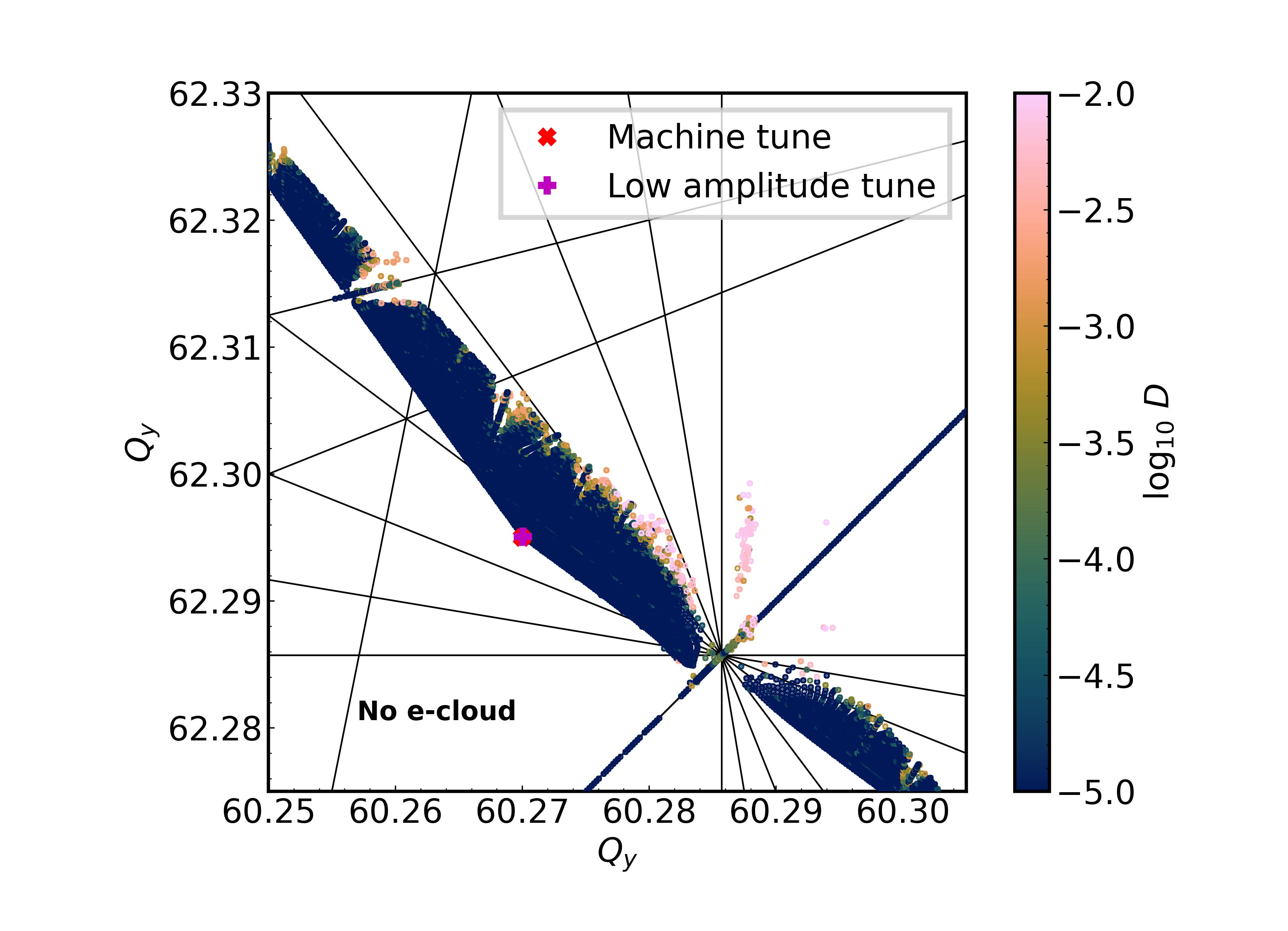}
    \\ (a) 
    \end{minipage}\linebreak
    \begin{minipage}[t]{0.49\textwidth}
    \centering
        \includegraphics[width=\textwidth]{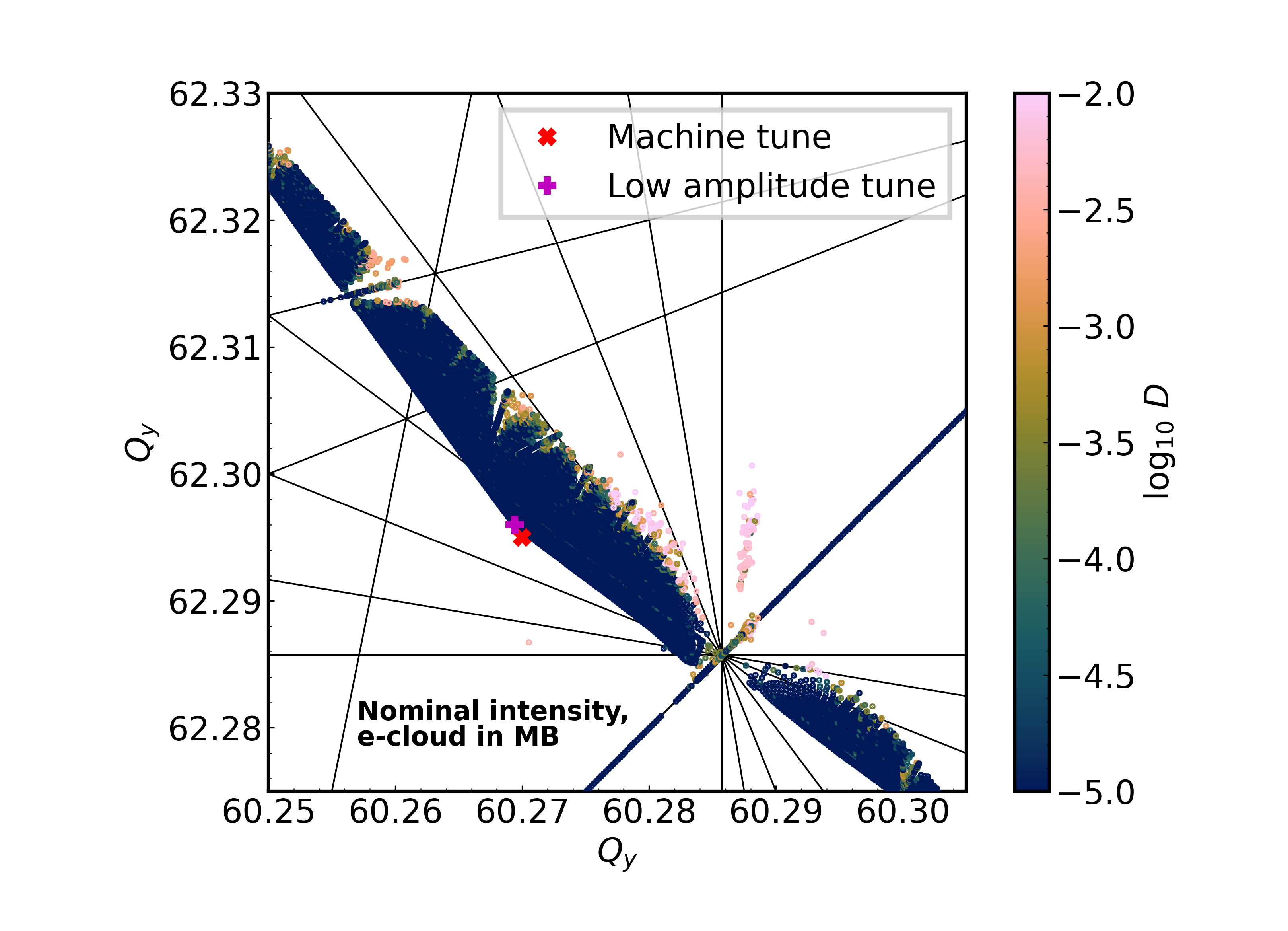}
    \\ (b) 
    \end{minipage}
    \begin{minipage}[t]{0.49\textwidth}
    \centering
        \includegraphics[width=\textwidth]{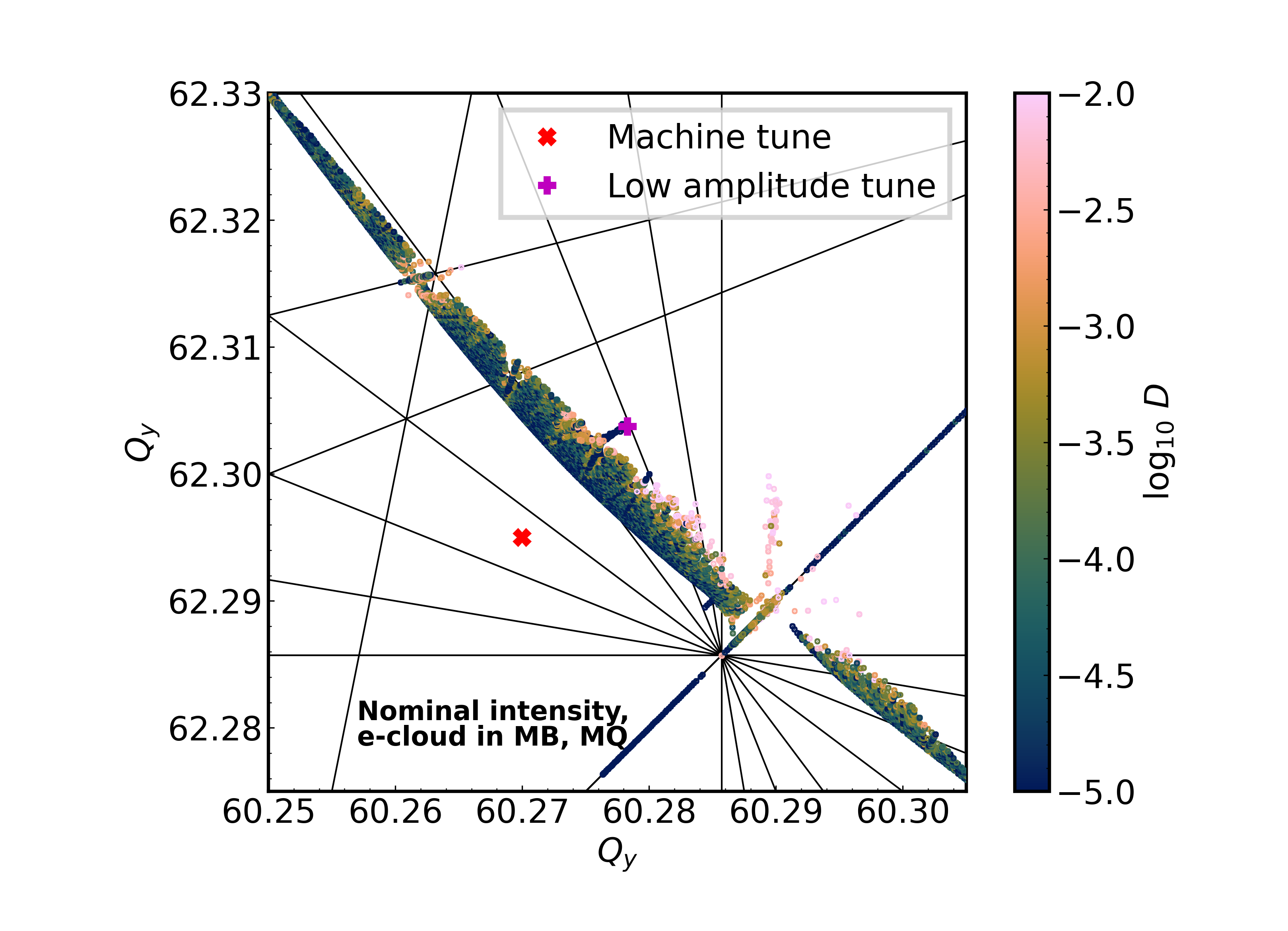}
    \\ (c) 
    \end{minipage}
    \begin{minipage}[t]{0.49\textwidth}
    \centering
        \includegraphics[width=\textwidth]{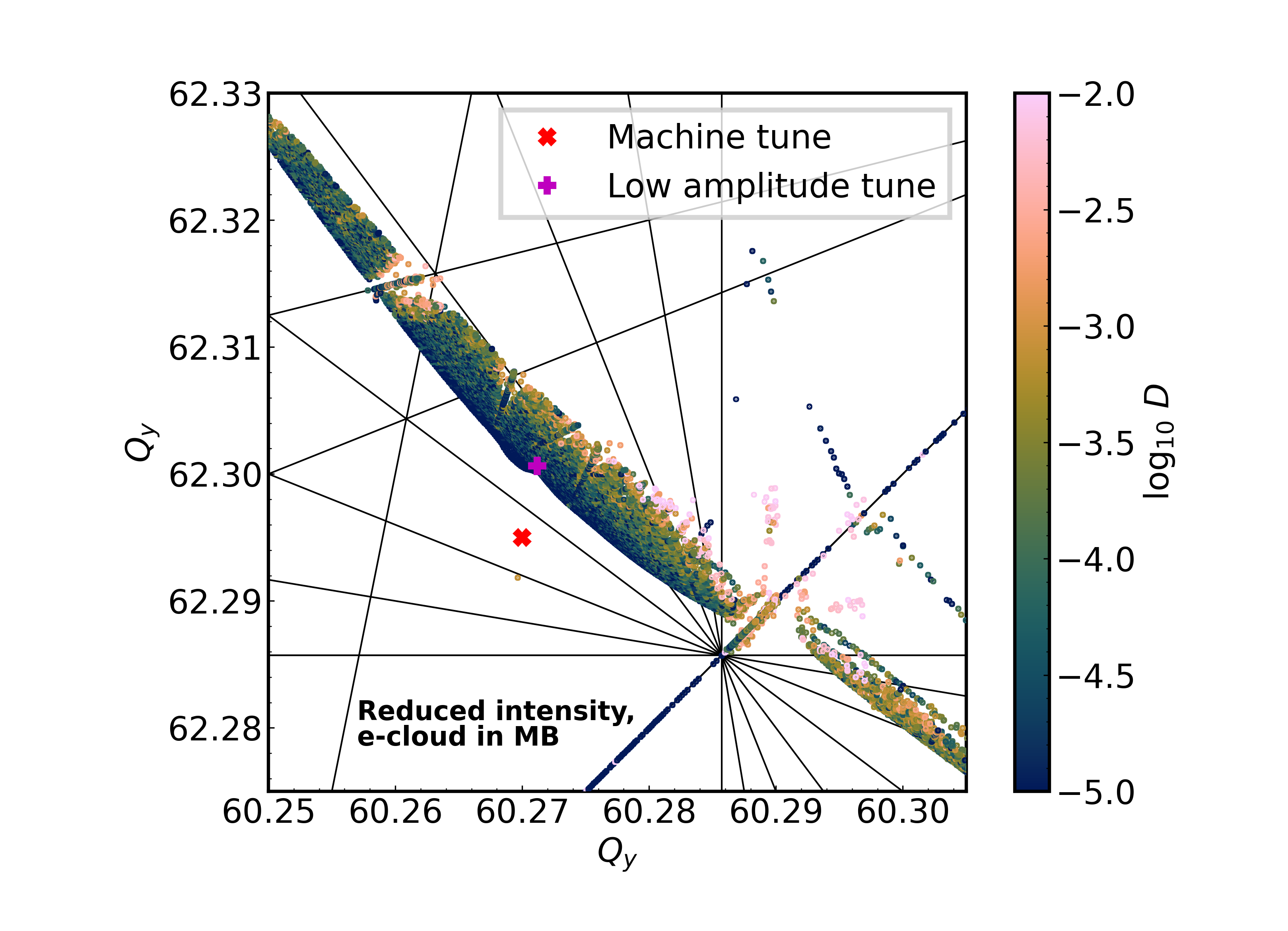}
    \\ (d) 
    \end{minipage}
    \begin{minipage}[t]{0.49\textwidth}
    \centering
        \includegraphics[width=\textwidth]{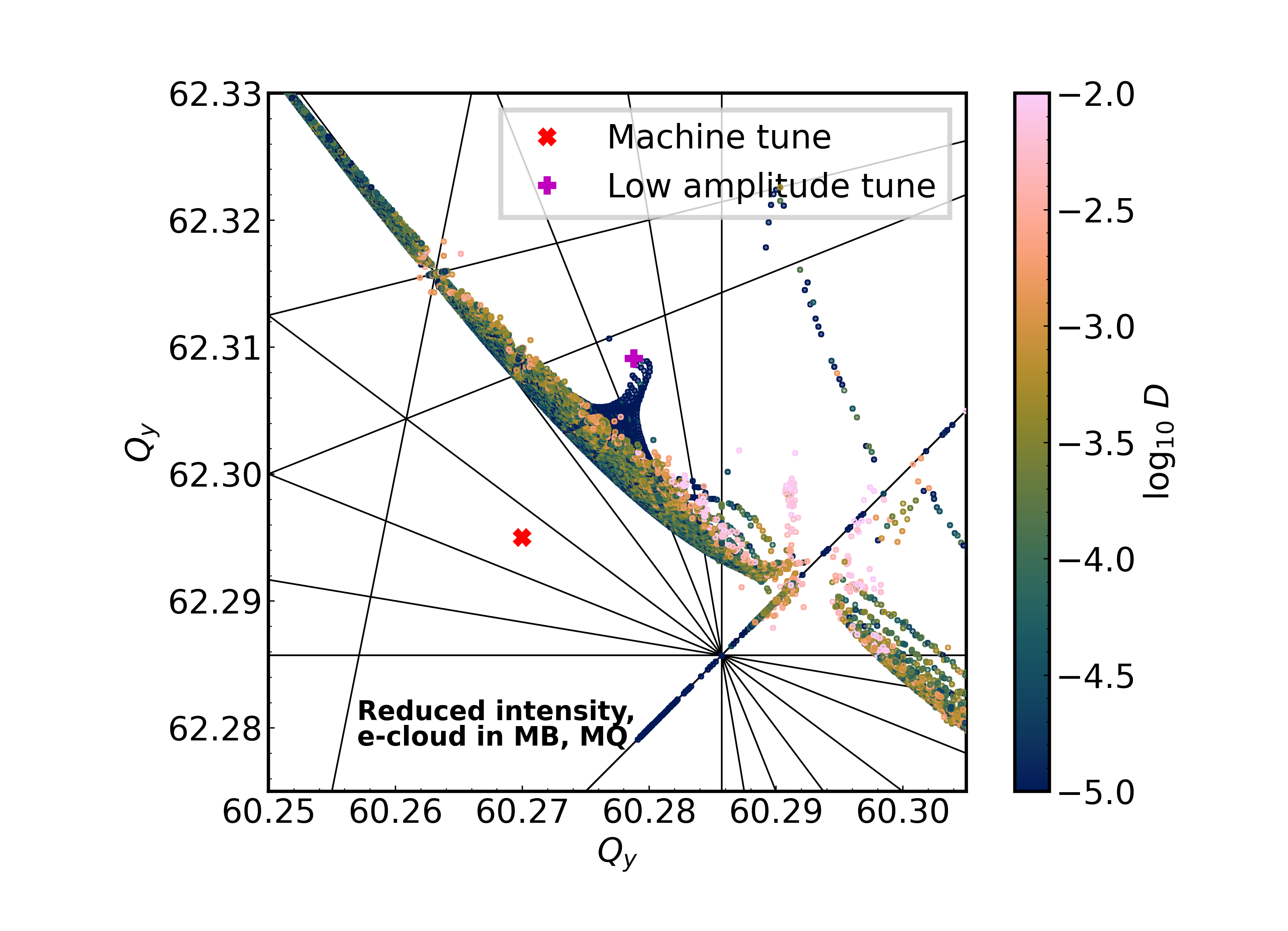}
    \\ (e) 
    \end{minipage}
    \caption{\label{fig:fig12}Frequency Map Analysis for on-momentum particles without e-clouds (a) and with e-clouds in the MB magnets at nominal intensity ($1.2\cdot 10^{11}$ p/bunch) (b), in the MB and MQ magnets at nominal intensity (c), in the MB magnets at reduced intensity ($0.6\cdot 10^{11}$ p/bunch) and in the MB and MQ magnets at reduced intensity (e), with $\mathrm{SEY}=1.3$. Transverse resonance lines up to order 7 are shown.}
\end{figure*}

The dependence of the DA on the betatron tune settings has also been studied through simulations with the nominal bunch intensity
and $\mathrm{SEY}=1.3$.
As shown in Fig.~\ref{fig:fig11}, it is possible to observe how the effect of the e-cloud reduces the region of the tune diagram available for operation, defined such
that $\textrm{DA} > 5\sigma$.
To ensure coherent beam stability, a minimum difference of at least $0.01$ is recommended between the fractional part of the horizontal and vertical tunes~\cite{lee-coupling}.
For this reason, the tune settings used in the operation of the LHC are $Q_x=62.27$ and $Q_y = 60.295$, which is marked by the red star in Fig.~\ref{fig:fig11}.

Additional insight on the beam dynamics in the presence of e-cloud can be gained through the FMA technique~\cite{FMA-chaos}.
In this study, the FMA is realized by simulating a particle distribution that is uniform in the horizontal and vertical normalized amplitude space, which is tracked for $2\cdot 10^4$ turns.
The turn-by-turn positions are used to estimate the betatron tunes in the first $10^4$ turns $\left(Q_{x,1}, Q_{y,1}\right)$ and the last $10^4$ turns $\left(Q_{x,2}, Q_{y,2}\right)$.
The numerical analysis of fundamental frequencies algorithm (NAFF)~\cite{naff} is used for the identification of the tunes.
The distance in the betatron tunes computed over the two time intervals can be used as an indicator of the chaoticity of the orbit:

\begin{equation}
    D = \sqrt{\left( Q_{x,2} - Q_{x,1} \right)^2 + \left( Q_{y,2} - Q_{y,1} \right)^2} .
\end{equation}

We carry out the FMA simulations on-momentum, \textit{i.e.}~initializing the particles in the absence of synchrotron oscillations, as the estimation of the betatron tunes can become elusive in the presence of strong tune modulation effects~\cite{sofia-modulation}.
The results of the FMA simulations are presented in Fig.~\ref{fig:fig12} for the two bunch intensities considered and with $\mathrm{SEY}=1.3$. 
Figure~\ref{fig:fig12}a shows the FMA without e-cloud while Figs.~\ref{fig:fig12}b, d show the FMAs including only MB-type e-clouds and Figs.~\ref{fig:fig12}c, e show the FMAs with both MB and MQ-type e-clouds. 
In all figures, the red marker indicates the betatron tune settings of the machine and the magenta marker shows the tune-shift found for low-amplitude particles.

For the nominal intensity, with e-clouds only in the MB magnets, there is no significant impact of the e-cloud on the on-momentum FMA, as observed comparing Figs.~\ref{fig:fig12}a and~b.
When including the e-cloud in the MQ magnets, as presented in Fig.~\ref{fig:fig12}c, we notice a much stronger low-amplitude tune-shift, the excitation of several resonances and significantly larger tune variation over the entire simulated particle distribution.
The observations can seem counter-intuitive when compared to the simulations of DA in Fig.~\ref{fig:fig10}, where it was observed that the effect of the e-cloud in the MB magnets was stronger than that of the e-cloud in the MQ magnets.
This is indeed explained by the fact that the losses from e-cloud are mostly driven from off-momentum particles, as we will discuss in Sec.~\ref{sec:long}, while only
on-momentum particles are shown in Fig.~\ref{fig:fig12}.
For this reason, these studies would strongly profit from the development of advanced FMA techniques which would work robustly in the presence of strong tune modulations driven by synchrotron motion (recent attempts in this direction were made in Ref.~\cite{sofia-modulation}).

A similar analysis is presented in Figs.~\ref{fig:fig12}d, e for the reduced bunch intensity. In this case, an even stronger tune-shift and resonance excitation are observed.
We notice that these effects are visible even when the e-cloud is present only in
the MB magnets, due to the increased electron density at the beam location (see Fig.~\ref{fig:fig05}) compared to the case with nominal bunch intensity.

\subsection{Direct simulation of the beam evolution}\label{sec:long}

Dynamic Aperture and FMA simulations provide important input to understand non-linear beam dynamics.
Recent research has identified promising approaches to link the variation of dynamic aperture with time and the evolution of intensity, or equivalently, the beam lifetime~\cite{da-scaling,da-measurement,latest-da-scaling-laws}.
Nevertheless, the accuracy of such methods in the presence of the complex forces from the e-cloud pinch has never been investigated.
For this reason we use direct simulations in order to study
beam lifetime and beam profile evolution.
At the LHC, such effects are visible only on very long timescales, in the order of several minutes.
For this reasons, we simulate $10^7$ turns, corresponding to approximately 15 minutes of beam time.

The choice of the initial particle coordinates plays an important role in the simulation.
One possibility would be to initialize the particles according to the realistic beam distribution, for example a Gaussian distribution.
However, such a choice is not optimal due to the fact that very few particles would be
generated with large oscillation amplitudes, although it is typically such large-amplitude particles that determine observables like the beam lifetime.

Instead, we found it convenient to generate particles randomly following a uniform distribution over the normalized transverse phase space $(\hat{x},\hat{p}_x,\hat{y},\hat{p}_y)$ and to repeat simulations for different synchrotron oscillation amplitudes in order to cover the entire 6D phase space.
The distributions are matched to the optics by using the normalizing transformation $\mathbf{W}$ obtained by an eigenvector analysis of the linear 6D one-turn map of the lattice $\mathbf{M}$ with the Courant-Snyder parameterization~\cite{wolski}:
\begin{equation}
    \begin{pmatrix}
     \hat{x} \\
     \hat{p}_x \\
     \hat{y} \\
     \hat{p}_y \\
     \hat{\tau} \\
     \hat{p}_\tau 
    \end{pmatrix}
    =
    \mathbf{W}^{-1}\cdot
    \begin{pmatrix}
     x \\
     p_x \\
     y \\
     p_y \\
     \tau \\
     p_\tau 
    \end{pmatrix} .
\end{equation}
$\mathbf{W}$ is obtained from the following factorization of $\mathbf{M}$:
\begin{equation}
\mathbf{M}=\mathbf{W} \mathbf{R} \mathbf{W}^{-1} ,
\end{equation}
$\mathbf{R}$ being the rotation matrix:
\begin{equation}
\mathbf{R}=\left(\begin{array}{ccccc}
\cos \mu_{1} & \sin \mu_{1} & & & \\
-\sin \mu_{1} & \cos \mu_{1} & & & \\
& & \ddots & & \\
& & & \cos \mu_{3} & \sin \mu_{3} \\
& & & -\sin \mu_{3} & \cos \mu_{3}
\end{array}\right) .
\end{equation}

The loss rate and the evolution of the beam profile for the real non-uniform beam distribution can be calculated by assigning a weight to each particle according to the local phase density of the assumed particle distribution.
Such an approach has the advantage that the same tracking data can be used to estimate the evolution for different initial particle distributions extending over the same phase space area by simply changing the particle weights.

\begin{figure}[t]
    \includegraphics[width=\columnwidth]{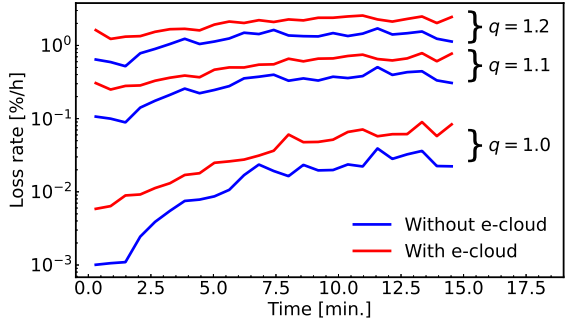}
    \caption{\label{fig:fig13_5}Loss rate as a function of time for three different values of the $q$ parameter that defines the population of particles at large amplitudes. The red lines correspond to simulations with e-clouds in both MB and MQ magnets for nominal intensity ($1.2\cdot 10^{11}$~p/bunch) and the blue lines to simulations without e-clouds.
    The loss rate is averaged over 30 seconds.}
\end{figure}

For the LHC, experimental measurements of the transverse beam profiles show that
their core can be described very well with Gaussian distributions.
However, measuring the tails of the beam profiles is more difficult.
The available data suggests that the tails of the profiles are slightly overpopulated
with respect to a purely Gaussian distribution~\cite{gorzawski}.
The beam loss rate is directly related to the population of particles
in the tails due to their large oscillation amplitude.
In order to study the sensitivity of our results to the tail population, we use 4D q-Gaussian distributions.
The population in the tails is controlled by the $q$ parameter.
For $q=1$, the distribution is a Gaussian distribution, 
for $q>1$ the distribution has overpopulated tails, 
and for $q<1$ the tails are underpopulated~\cite{tsallis,stefania-non-gaussian}.
The q-Gaussian distribution is written as:
\begin{equation}
    f(x) = \frac{\sqrt{\beta}}{C_q} e_q\left( - \beta x^2\right),
\end{equation}
where $\beta$ is a scale parameter, $e_q$ is the q-exponential defined as:
\begin{equation}
    e_q(x) =
    \begin{cases}
     \exp(x)&\textrm{if}\ q = 1,\\
     (1 + (1-q)x)^\frac{1}{1-q}& \textrm{if}\ q \neq 1\text{ and }1 + (1 - q)x > 0,\\
     0& \textrm{if}\ q \neq 1\text{ and }1 + (1 - q)x \leq 0,
    \end{cases}
\end{equation}
and $C_q$ is the normalization factor that is given by the expression:
\begin{equation}
    C_q =
    \begin{cases}
    \dfrac{2\sqrt{\pi} \Gamma\left(\frac{1}{1-q}\right)}{\left(3-q\right) \sqrt{1-q}\Gamma\left(\frac{3-q}{2\left(1-q\right)}\right)} &\textrm{if}\ q < 1,\\
    \sqrt{\pi} & \textrm{if}\ q = 1,\\
    \dfrac{\sqrt{\pi} \Gamma\left(\frac{3-q}{2\left(1-q\right)}\right)}{\sqrt{q-1}\Gamma\left(\frac{1}{q-1}\right)} & \textrm{if}\ 1 < q < 3.
    \end{cases}
\end{equation}
In 4 dimensions, the final distribution is defined as a product of 4 independent q-Gaussian distributions, for $\hat{x}, \hat{p}_x, \hat{y}, \hat{p}_y$.
In our simulations, the longitudinal distribution is expressed as an exponential of 
the action variable of the single-harmonic RF potential, which is a realistic 
assumption based on profile measurements~\cite{esteban-muller-thesis}.
The parameters of the distribution are chosen such that the r.m.s. bunch length is
equal to as reported in Table~\ref{tab:tab1}.

Fig.~\ref{fig:fig13_5} shows the evolution of the loss rate as a function of time without e-cloud (blue) and in the presence of e-cloud in both MB and MQ magnets (red) for the nominal bunch
intensity and with SEY estimated from heat-load measurements.
The evolution is shown for three different values of the parameter $q$ of the transverse distribution.
It is evident that the sensitivity of the losses to the initial tail population is very strong with and without the presence of e-cloud.

\begin{figure}
    \includegraphics[width=\columnwidth]{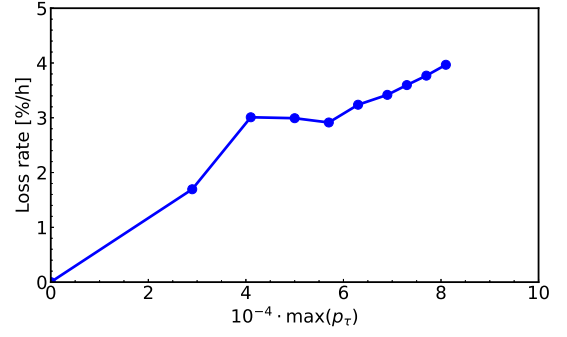}
    \caption{\label{fig:fig12_5}Average loss rate as a function of the synchrotron oscillation amplitude, for the simulation corresponding to the curve $q=1.2$ (with e-cloud) in Fig.~\ref{fig:fig13_5}
    }
\end{figure}

It is interesting to observe the dependence of the losses on the synchrotron oscillation amplitude.
This is illustrated in Fig.~\ref{fig:fig12_5} where we observe that losses are basically absent for on-momentum particles while they become larger as the oscillation amplitude increases.

\begin{figure}
    \includegraphics[width=\columnwidth]{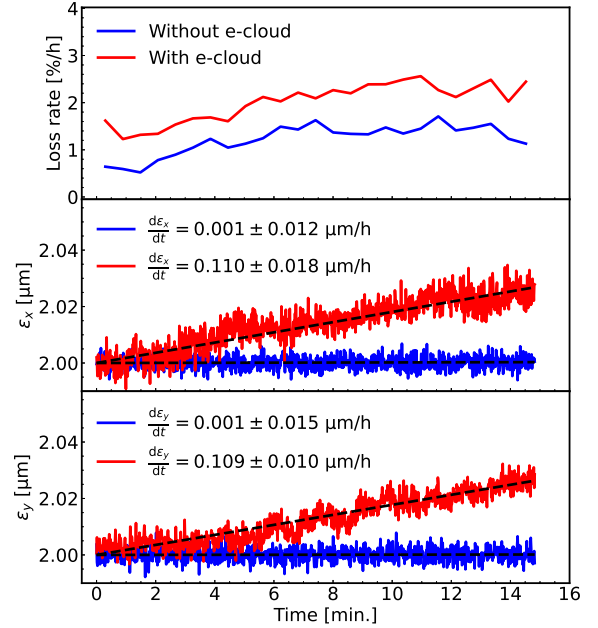}
    \caption{\label{fig:fig13}Long-term tracking simulation with particle distributions for the nominal intensity ($1.2\cdot 10^{11}$~p/bunch) and SEY derived from measurements.
    The relative loss rate (top), the horizontal emittance (middle) and the vertical emittance (bottom) are plotted as a function of time for simulations without e-clouds (blue) and with e-clouds in both the MB and MQ magnets (red).
    The black dashed lines correspond to linear fits
    }
\end{figure}

Figure~\ref{fig:fig13} shows the emittance evolution without e-cloud (blue) and in the presence of e-cloud in both MB and MQ magnets (red).
The emittance is computed by fitting Gaussian functions on the projections of the weighted particle distributions.
The fit is made on the core of the transverse distribution as done on experimental measurements of the transverse emittance in the LHC.
No emittance growth is found in the absence of e-cloud while a weak linear emittance growth of around $0.11$ \textmu m/h is found in the simulations with 
e-cloud effects.
This emittance growth appears to be weaker than what has been 
measured in the LHC~\cite{stefania-evian}.
Effects not included in the present model are likely to influence the emittance growth, for example residual betatron coupling and noise from the transverse feedback system and the magnet power converters.
Including such effects in the simulation will be the subject of future work.

\section{Conclusions and final remarks}

We introduced a general method to simulate slow incoherent effects
related to e-clouds in the presence of the non-linearities of the accelerator lattice.
The method is able to handle arbitrarily complex e-cloud distributions obtained
from PIC simulations, while preserving the
symplecticity of the e-cloud interaction.
This is achieved by applying a tricubic interpolation scheme on a scalar potential, which is
defined on a 3D regular grid, in order to obtain smooth expressions for the e-cloud forces.
The interpolation scheme is observed to introduce numerical artifacts in the forces.
A refinement method based on Poisson's equation was developed, which is shown to successfully mitigate such artifacts.

The method has been applied to the case of the LHC, to
simulate the effect of the e-clouds present in the main dipole and quadrupole magnets of the LHC arcs.
The full non-linear model of the LHC lattice is included in the simulation and computations of dynamic aperture, frequency map analysis and the evolution of beam losses and emittance growth simulated through the long term tracking of particle distributions. 
The simulations show the degradation of single-particle stability, especially for high electron densities in the
vicinity of the beam position.
The non-linear nature of the e-cloud forces results in
a reduction of the DA 
and in an increased chaoticity of the beam particle orbits.
Long-term simulations performed in GPUs clearly show the degradation introduced by the e-cloud on the beam lifetime and emittance evolution.

We underline that the introduced method is not restricted solely to e-cloud effects and that it can be applied to other phenomena introducing forces
that are dependent on both transverse and longitudinal coordinates while preserving the symplecticity.
Examples of such cases are space-charge and beam-beam effects in the presence of complex charge distributions that are difficult to express analytically.


\begin{acknowledgments}
The authors would like to thank
G.~Arduini, R.~De~Maria, H.~Bartosik, L.~Giacomel, M.~Giovannozzi, R.~Jones, N.~Karastathis, S.~Kostoglou, E.~M\'etral, Y.~Papaphilippou, G.~Rumolo, M.~Schwinzerl, S.~E.~Tzamarias and F.~Van~der~Veken
for the fruitful discussions and valuable suggestions on this work. Work supported by the High Luminosity LHC project.
\end{acknowledgments}

\appendix

\section{Shortcomings of a linear interpolation scheme\label{sec:app-linear}}

Typically in PIC codes, the scalar potential $\phi$ is calculated on a regular grid and its derivatives are approximated with central differences:
\begin{align}
e^{(i,j)}_x = -\frac{\partial \phi}{\partial x}^{(i,j)} = -\frac{\phi^{(i+1,j)} - \phi^{(i-1,j)}}{2\Delta x}\label{eq:a1} ,\\
e^{(i,j)}_y = -\frac{\partial \phi}{\partial y}^{(i,j)} = -\frac{\phi^{(i,j+1)} - \phi^{(i,j-1)}}{2\Delta y}\label{eq:a2} ,
\end{align}
where $\Delta x,\Delta y$ are the distances between the grid nodes and $i,j$ are the indices of the grid cells.
The map of the interaction would then have the following form:
\begin{align}
    x&\mapsto x,\label{eq:app-map1}\\
    p_x&\mapsto p_x + A\, e_x(x,y),\label{eq:app-map2}\\
    y&\mapsto y,\label{eq:app-map3}\\
    p_y&\mapsto p_y + A\, e_y(x,y),\label{eq:app-map4}
\end{align}
where $A$ is a constant.
The normalized fields $e_x(x,y),e_y(x,y)$ are interpolated linearly and independently of each other in order to obtain their values at an arbitrary point in the continuous space.
In this case, the interpolating function is explicitly written as:
\begin{multline}\label{eq:a4}
e_{x,y}(x,y) = a^{(i,j)}_{x,y} + b^{(i,j)}_{x,y} \frac{x-x^{(i,j)}}{\Delta x} \\ 
+c^{(i,j)}_{x,y} \frac{y-y^{(i,j)}}{\Delta y} + d^{(i,j)}_{x,y} \frac{(x-x^{(i)})(y-y^{(j)})}{\Delta x \Delta y}  ,
\end{multline}
where $a_{x,y}^{(i,j)},b_{x,y}^{(i,j)},c_{x,y}^{(i,j)},d_{x,y}^{(i,j)}$ are given by:
\begin{align}
    a_{x,y}^{(i,j)} &= e_{x,y}^{(i,j)}\label{eq:a5} ,\\
    b_{x,y}^{(i,j)} &=  - e_{x,y}^{(i,j)} + e_{x,y}^{(i+1,j)}\label{eq:a6} ,\\
    c_{x,y}^{(i,j)} &=  - e_{x,y}^{(i,j)} + e_{x,y}^{(i,j+1)}\label{eq:a7} ,\\
    d_{x,y}^{(i,j)} &=  e_{x,y}^{(i,j)} - e_{x,y}^{(i,j+1)} - e_{x,y}^{(i+1,j)} + e_{x,y}^{(i+1,j+1)}\label{eq:a8} ,
\end{align}
and $x^{(i,j)}$, $y^{(i,j)}$ are the $x^{(i)}$ and $y^{(j)}$ coordinates of the neighbouring grid node.
The Jacobian matrix $J$ of the map in Eqs.~\eqref{eq:app-map1}-\eqref{eq:app-map4} is equal to:
\begin{equation}\label{eq:J}
\mathbf{J}=
    \begin{pmatrix}
     1 & 0 & 0 & 0 \\
     A\partial_x e_x & 1 & A\partial_y e_x & 0 \\
     0 & 0 & 1 & 0 \\
      A\partial_x e_y & 0 & A\partial_y e_y & 1
    \end{pmatrix},
\end{equation}
where we have used $\partial_x \equiv \frac{\partial}{\partial x}$ to ease notation.
The map is symplectic if the following condition is satisfied\cite{wolski}:
\begin{equation}\label{eq:sympapp}
    \mathbf{J}\, \mathbf{S}\, \mathbf{J}^T = \mathbf{S},
\end{equation}
where $\mathbf{S}$ is the antisymmetric matrix:
\begin{equation}\label{eq:smatrix}
    \mathbf{S} = \begin{pmatrix}
     0 & 1 & 0 & 0\\
     -1 & 0 & 0 & 0 \\
     0 & 0 & 0 & 1 \\
     0 & 0 & -1 & 0
    \end{pmatrix}.
\end{equation}
Combining Eqs.~\eqref{eq:J},~\eqref{eq:sympapp} and~\eqref{eq:smatrix}, the following matrix equation follows:
\begin{multline}
    \begin{pmatrix}
     0 & 1 & 0 & 0 \\
     -1 & 0 & 0 & A\left(\partial_y e_x - \partial_x e_y\right) \\
     0 & 0 & 0 & 1\\
     0 & A\left(\partial_x e_y - \partial_y e_x\right) & -1 & 0 \\
    \end{pmatrix} =\\= 
     \begin{pmatrix}
     0 & 1 & 0 & 0\\
     -1 & 0 & 0 & 0 \\
     0 & 0 & 0 & 1 \\
     0 & 0 & -1 & 0
    \end{pmatrix},
\end{multline}
from where we can observe that the symplectic condition of Eq.~\eqref{eq:sympapp} is equivalent to the following condition on the derivatives:
\begin{equation}
\frac{\partial e_y}{\partial x} - 
\frac{\partial e_x}{\partial y} = 0\label{eq:a3} .
\end{equation}
Combining Eqs.~\eqref{eq:a4} and~\eqref{eq:a3}, the condition of symplecticity becomes:
\begin{equation}
\frac{b_y^{(i,j)}}{\Delta x} + \frac{d_y^{(i,j)}}{\Delta x\Delta y}\left(y - y^{(j)}\right) -
\frac{c_x^{(i,j)}}{\Delta y} - \frac{d_x^{(i,j)}}{\Delta x\Delta y}\left(x - x^{(i)}\right) = 0 .
\end{equation}
This equality must be true for any value of $x, y, x^{(i)}, y^{(j)}$, which requires that
\begin{align}
\frac{b_y^{(i,j)}}{\Delta x} -
\frac{c_x^{(i,j)}}{\Delta y} &= 0,\\
d_y^{(i,j)} &=0,\\
d_x^{(i,j)} &= 0 .
\end{align}
By substituting the coefficients from Eqs.~\eqref{eq:a6},~\eqref{eq:a7},~\eqref{eq:a8} and the fields from Eqs.~\eqref{eq:a1} and~\eqref{eq:a2},
we observe that
the symplecticity condition holds in every point of the domain if the following relations hold for the discrete samples of the scalar potential:
\begin{widetext}
\begin{align}
     \phi^{(i,j+1)}
   - \phi^{(i,j-1)}
   - \phi^{(i+1,j+1)}
   + \phi^{(i+1,j-1)}
   - \phi^{(i+1,j)}
   + \phi^{(i-1,j)}
   + \phi^{(i+1,j+1)}
   - \phi^{(i-1,j+1)} &= 0,\\
   - \phi^{(i+1,j)}
   + \phi^{(i-1,j)}
   + \phi^{(i+2,j)}
   - \phi^{(i,j)}
   + \phi^{(i+1,j+1)}
   - \phi^{(i-1,j+1)}
   - \phi^{(i+2,j+1)}
   + \phi^{(i,j+1)} &= 0,\\
   - \phi^{(i,j+1)}
   + \phi^{(i,j-1)}
   + \phi^{(i,j+2)}
   - \phi^{(i,j)}
   + \phi^{(i+1,j+1)}
   - \phi^{(i+1,j-1)}
   - \phi^{(i+1,j+2)}
   + \phi^{(i+1,j)} &=0 .
\end{align}
\end{widetext}
Such relations are not automatically for a potential obtained from the discretized Poisson equation, which means that the map obtained with the scheme defined in Eqs.~\eqref{eq:a1}-\eqref{eq:a7} is in general not symplectic. 

To illustrate the implication using such a non-symplectic map in tracking simulations, we apply it to numerically solve the dynamical system described by the following Hamiltonian:
\begin{equation}
    H = \frac{p_1^2}{2} + \frac{p_2^2}{2} + e^{q_1-q_2} .
\end{equation}

This Hamiltonian has a non-linear potential that cannot be represented exactly by polynomial interpolating functions and is completely integrable.
In addition to the Hamiltonian, the system conserves the following quantities (integrals of motion)\cite{toda}:
\begin{align}
    J_1 &= \left(p_1 - p_2\right)^2 + 4 e^{q_1 - q_2}, \\
    I_1 &= \frac{p_1 - p_2 + \sqrt{J_1}}{p_1 - p_2 - \sqrt{J_1}} \,\exp\left({\sqrt{J_1} \,\frac{q_1 + q_2}{p_1 + p_2}}\right) .
\end{align}
The numerical integration scheme is constructed by splitting the Hamiltonian into its kinetic $\left(H_K\right)$ and potential $\left(\phi\right)$ terms:

\begin{align}
    H_k(p_1, p_2) &= \frac{p_1^2}{2} + \frac{p_1^2}{2},\\
    \phi(q_1, q_2) &= e^{q_1-q_2}.
\end{align}

The system is then integrated by applying Hamilton's equations to the two terms of the Hamiltonian separately.
The scheme is constructed by arranging the solutions in the ``drift-kick-drift'' form:
\begin{align}
    q_1 &\mapsto q_1 + \frac{\Delta t}{2} \,p_1,\\
    q_2 &\mapsto q_2 + \frac{\Delta t}{2} \,p_2,
\end{align}
\begin{align}
    p_1 &\mapsto p_1 - \Delta t \,\frac{\partial \phi}{\partial q_1}\!\left(q_1, q_2\right),\label{eq:appv1}\\
    p_2 &\mapsto p_2 - \Delta t \,\frac{\partial \phi}{\partial q_2}\!\left(q_1, q_2\right),\label{eq:appv2}
\end{align} 
\begin{align}
    q_1 &\mapsto q_1 + \frac{\Delta t}{2} \,p_1,\\
    q_2 &\mapsto q_2 + \frac{\Delta t}{2} \,p_2,
\end{align}

\begin{figure}
    \centering
    \begin{minipage}[t]{0.49\textwidth}
        \includegraphics[width=\textwidth]{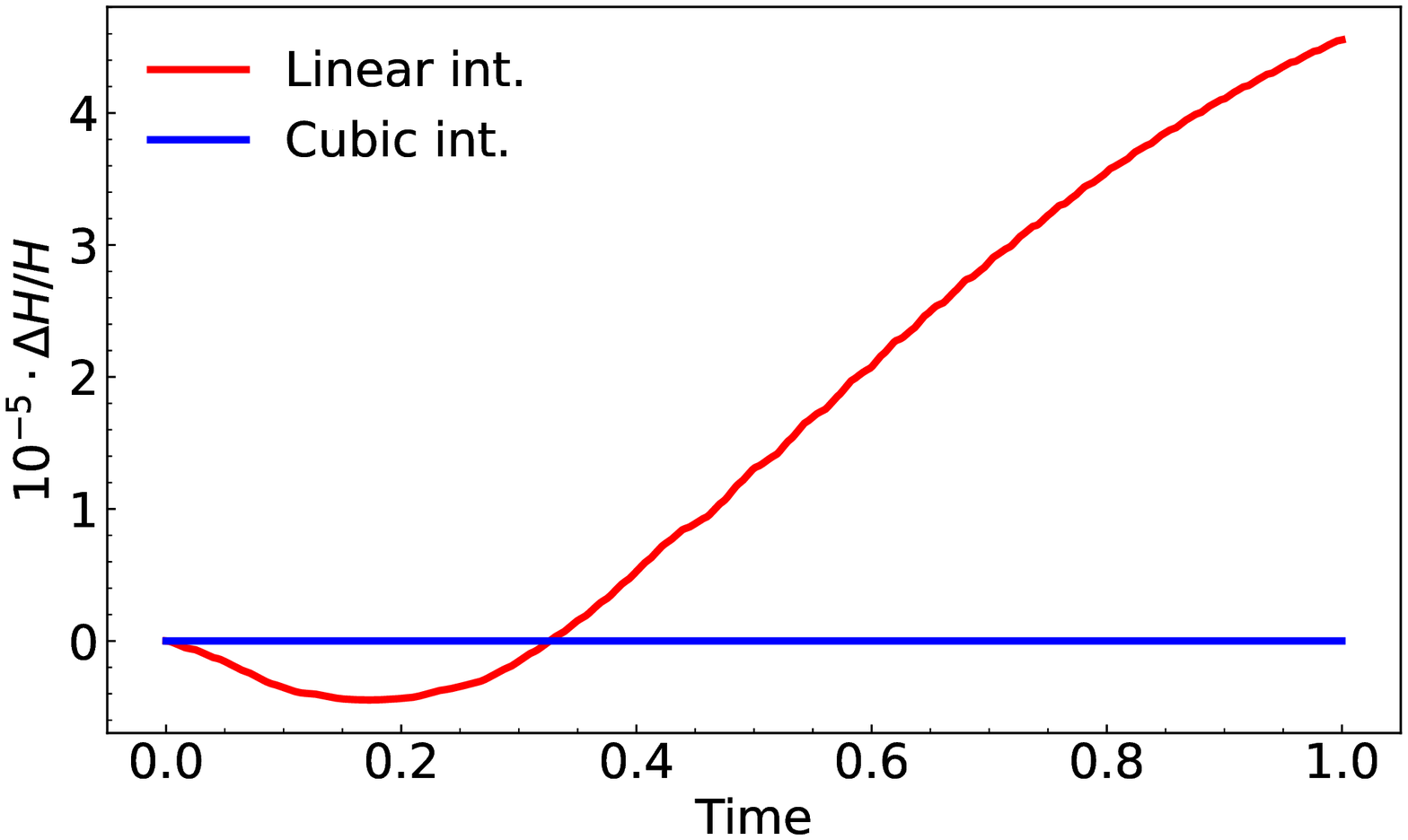}
     \\ (a)
    \end{minipage}
    \begin{minipage}[t]{0.49\textwidth}
        \includegraphics[width=\textwidth]{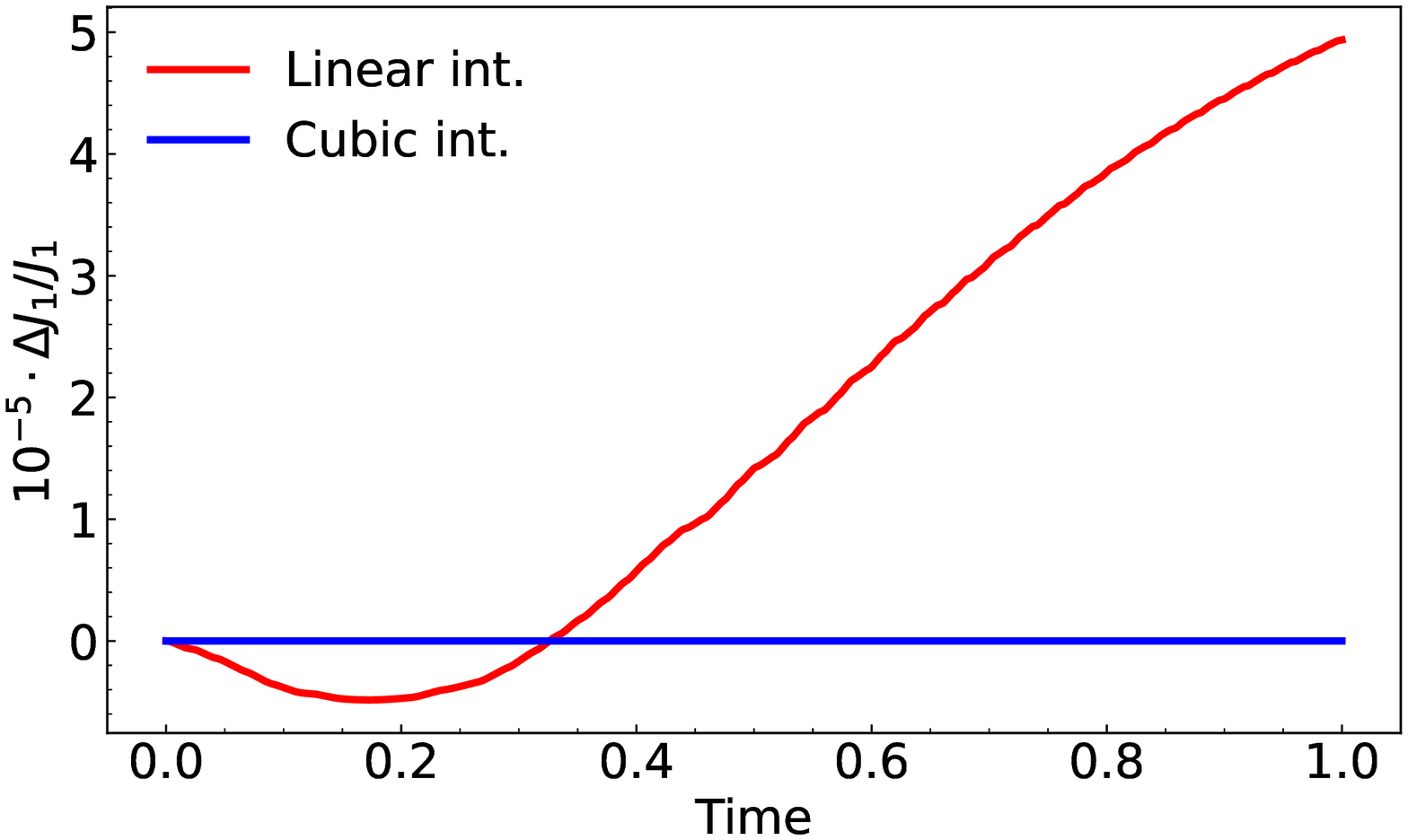}
     \\ (b)
    \end{minipage}
    \begin{minipage}[t]{0.49\textwidth}
        \includegraphics[width=\textwidth]{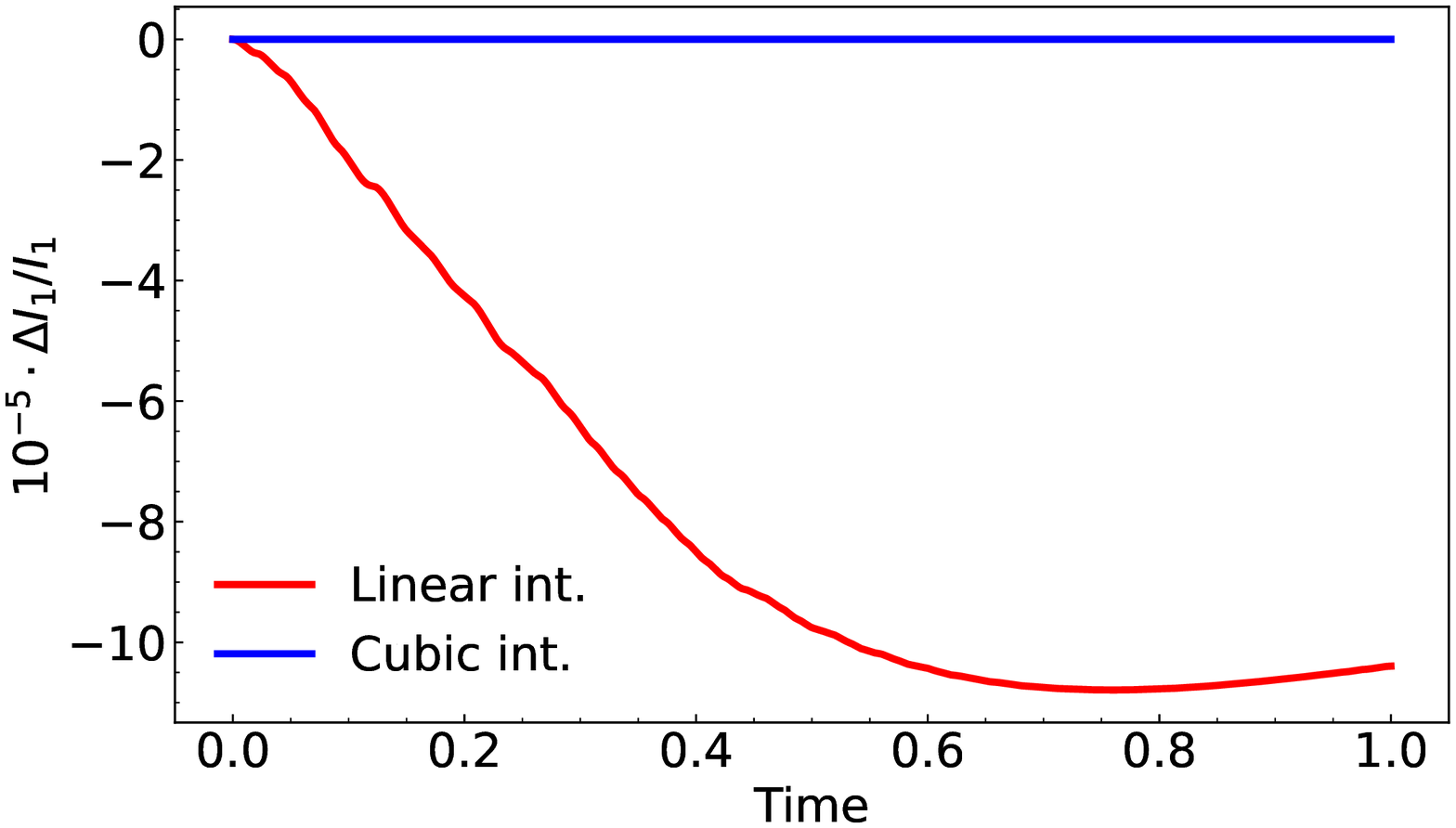}
     \\ (c)
    \end{minipage}
    \caption{\label{fig:app1}Evolution of the error in the integrals of motion with time, using the two interpolation schemes. The error is quoted as a relative absolute difference with respect to the exact value of the integral of motion. The initial conditions used are $q_1 = 0.5,\ p_1=1,\ q_2 = -0.5,\ p_2 = 0$ with a time-step $\Delta t = 10^{-6}$.}
\end{figure}

We compare the performance of two different interpolation schemes in computing the derivatives of the potential in Eqs.~\eqref{eq:appv1} and~\eqref{eq:appv2}. In particular we consider:
1) a linear interpolation scheme on the derivatives $\partial\phi/\partial q_1,\ \partial\phi/\partial q_2$ and 2) a cubic interpolation scheme on $\phi$, as described in Sec. \ref{sec:tricubic}, but in two dimensions.
For both interpolation schemes the parameters used were the same --- a regular two-dimensional interpolation grid of $201\times201$ nodes with a distance of $0.02$ between grid node in both dimensions. 
The evolution of the integrals of motion with respect to time for the different integrals of motion is plotted in Fig.~\ref{fig:app1}. The red lines correspond to the simulations using the linear interpolation while the blue lines to when the cubic interpolation scheme is used.

It is evident that for the same grid spacing the performance of the linear interpolation scheme is worse.
The linear interpolation scheme fails to conserve the integrals of motion which are observed to grow with time.
This is expected as the map produced by this scheme is not symplectic.
On the other hand, the scheme based on the cubic interpolation performs much better, with no observable growth in the integrals of motion.

\section{Formation of e-clouds with different bunch intensities}\label{sec:app-ecloud}

In Sec.~\ref{sec:characterization} it was shown how e-cloud effects become more severe with reduced bunch intensities. In this Appendix we analyze in some more detail the root cause of this effect.

\begin{figure*}[hp!]
    \centering
    \begin{minipage}[h]{0.40\textwidth}
    \centering
    \includegraphics[width=\textwidth]{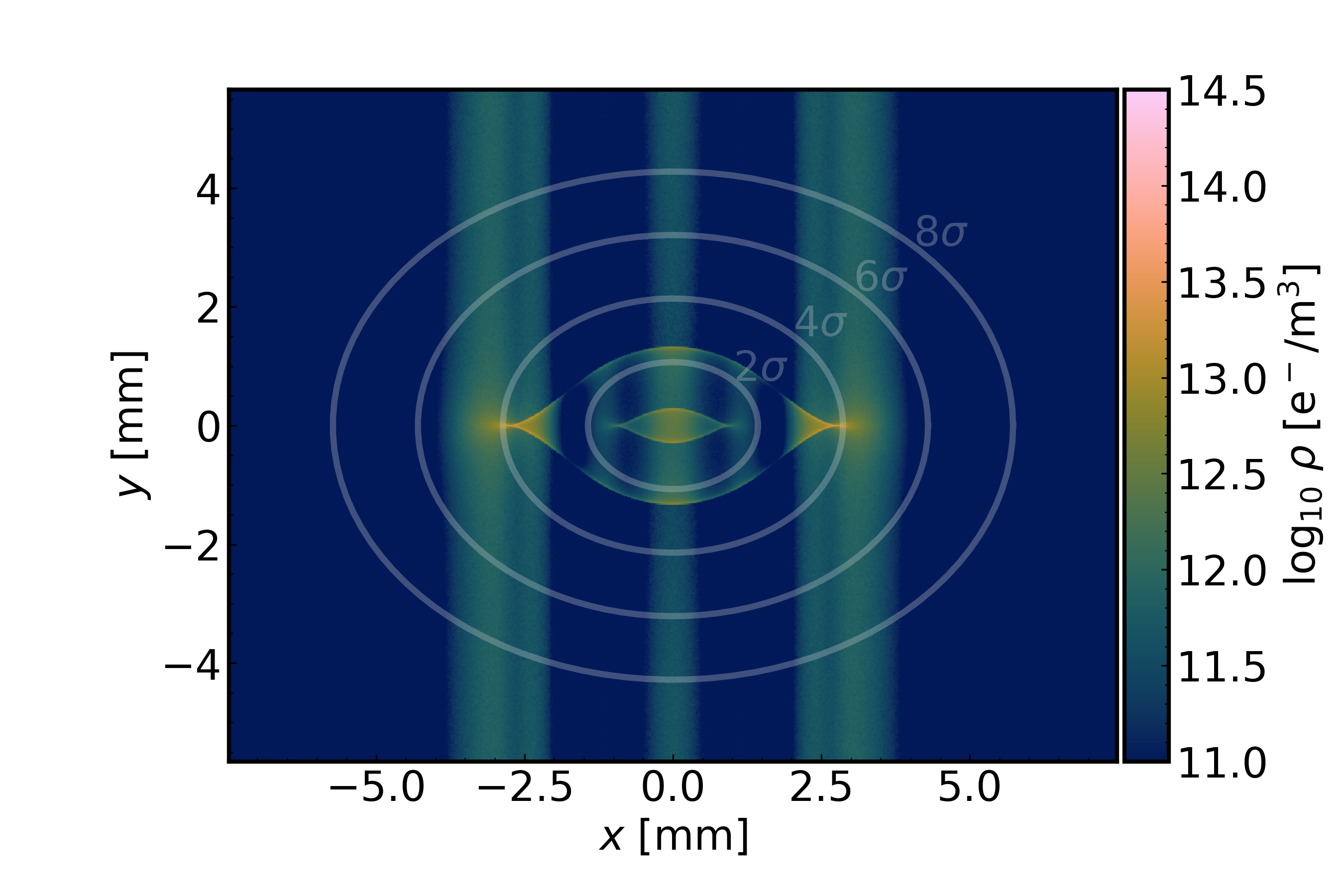}
     \\ (a) 
    \end{minipage}
    \begin{minipage}[h]{0.40\textwidth}
    \centering
    \includegraphics[width=\textwidth]{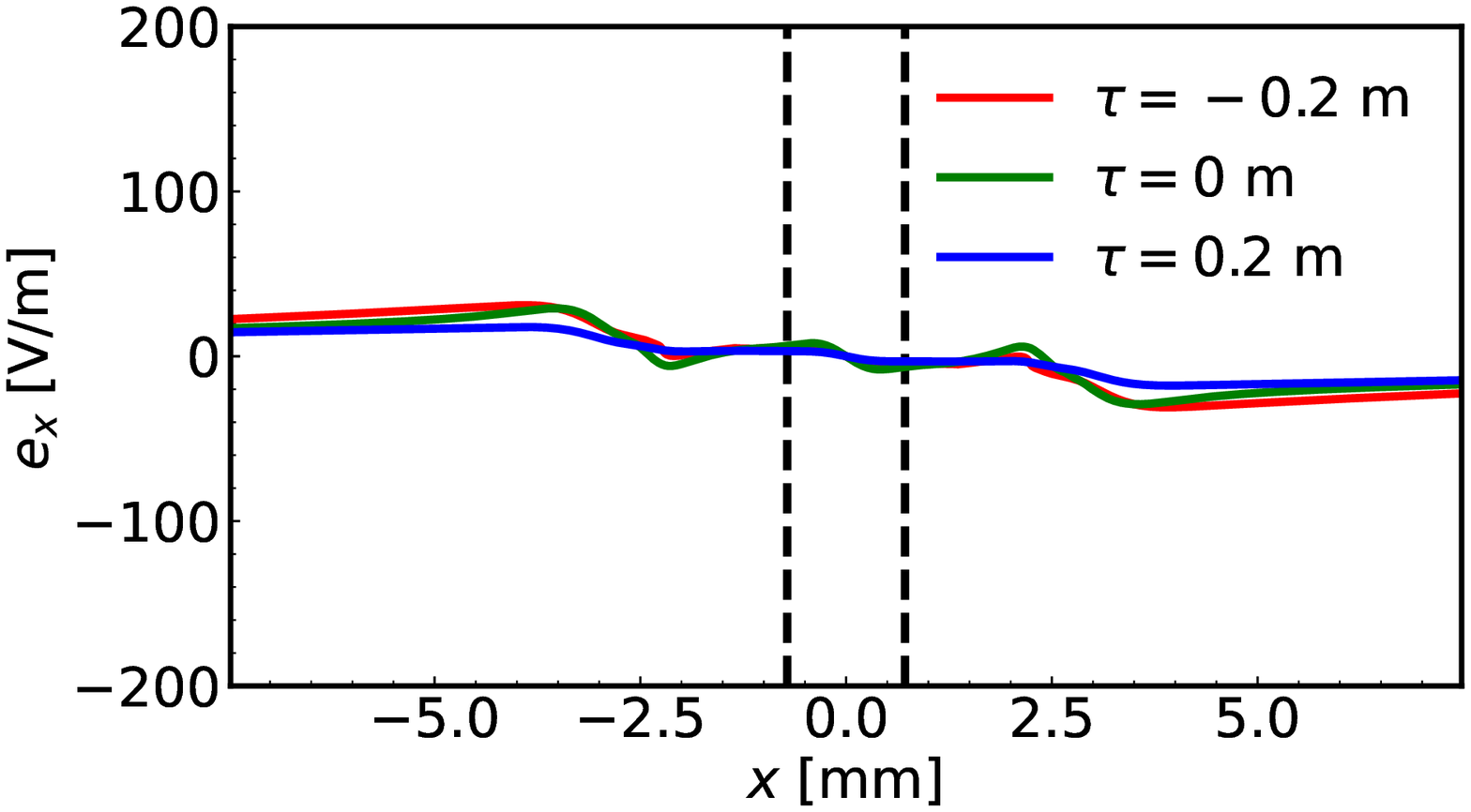}
     \\ (b) 
    \end{minipage}
    \begin{minipage}[h]{0.40\textwidth}
    \centering
    \includegraphics[width=\textwidth]{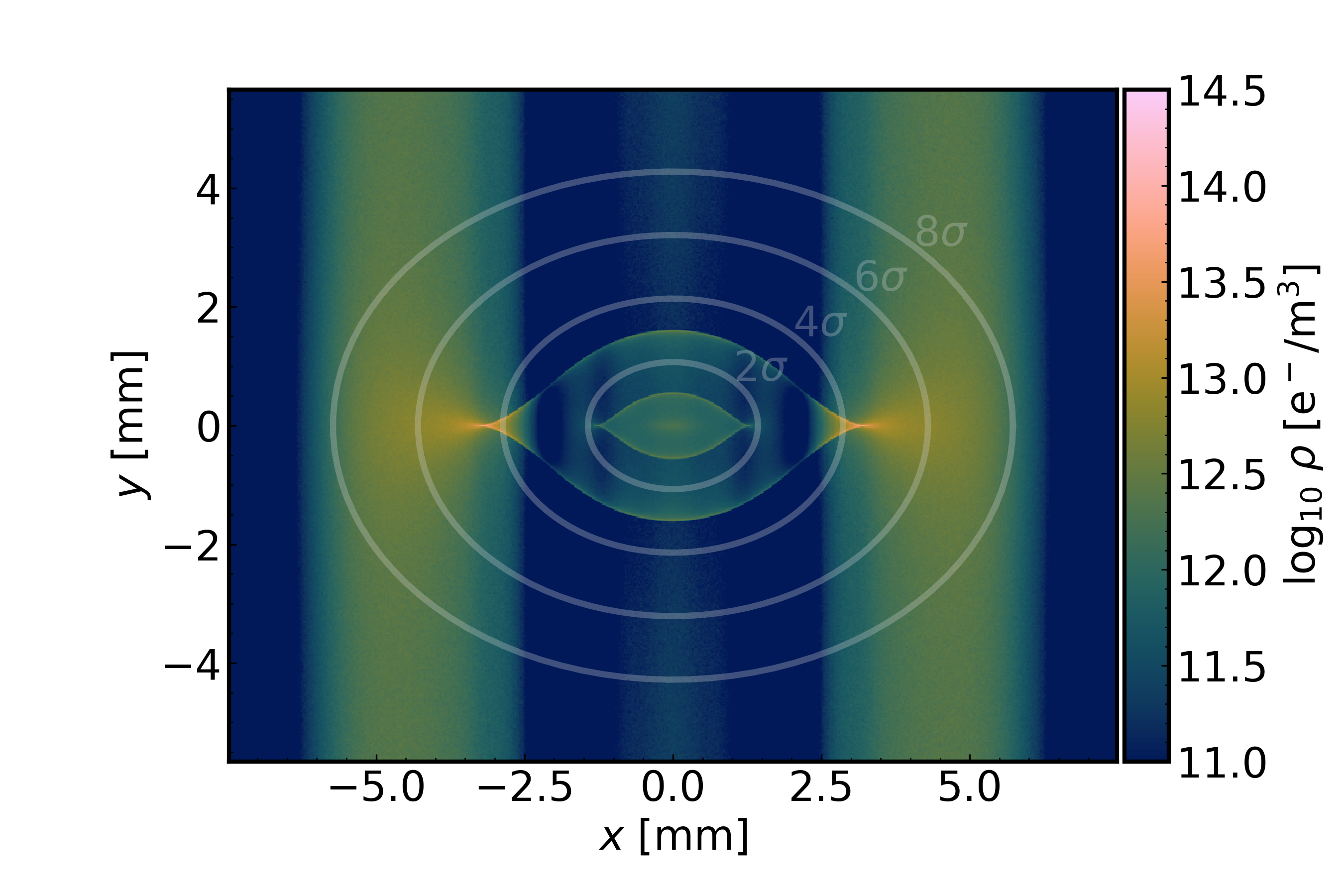}
     \\ (c) 
    \end{minipage}
    \begin{minipage}[h]{0.40\textwidth}
    \centering
    \includegraphics[width=\textwidth]{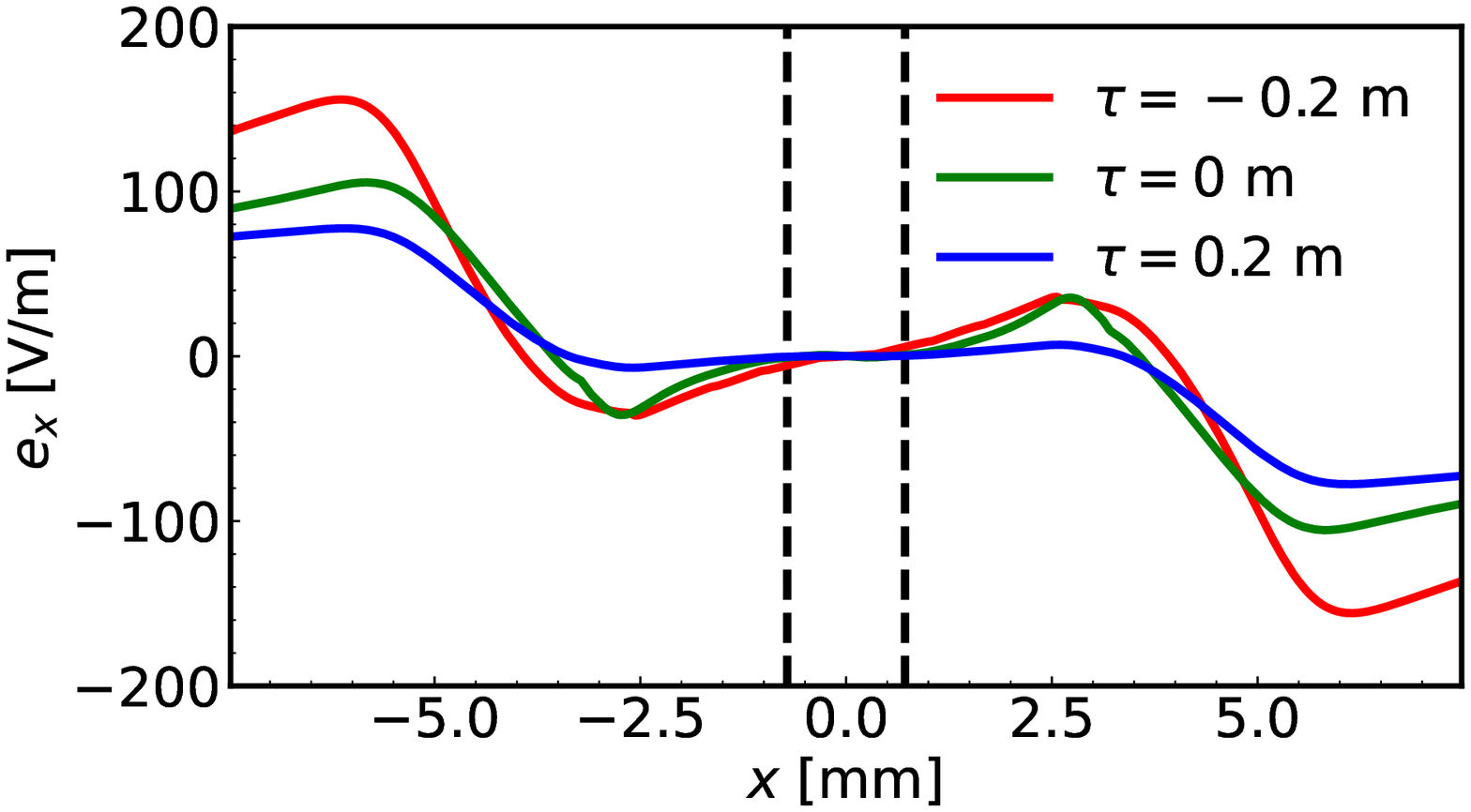}
     \\ (d) 
    \end{minipage}
    \begin{minipage}[h]{0.40\textwidth}
    \centering
    \includegraphics[width=\textwidth]{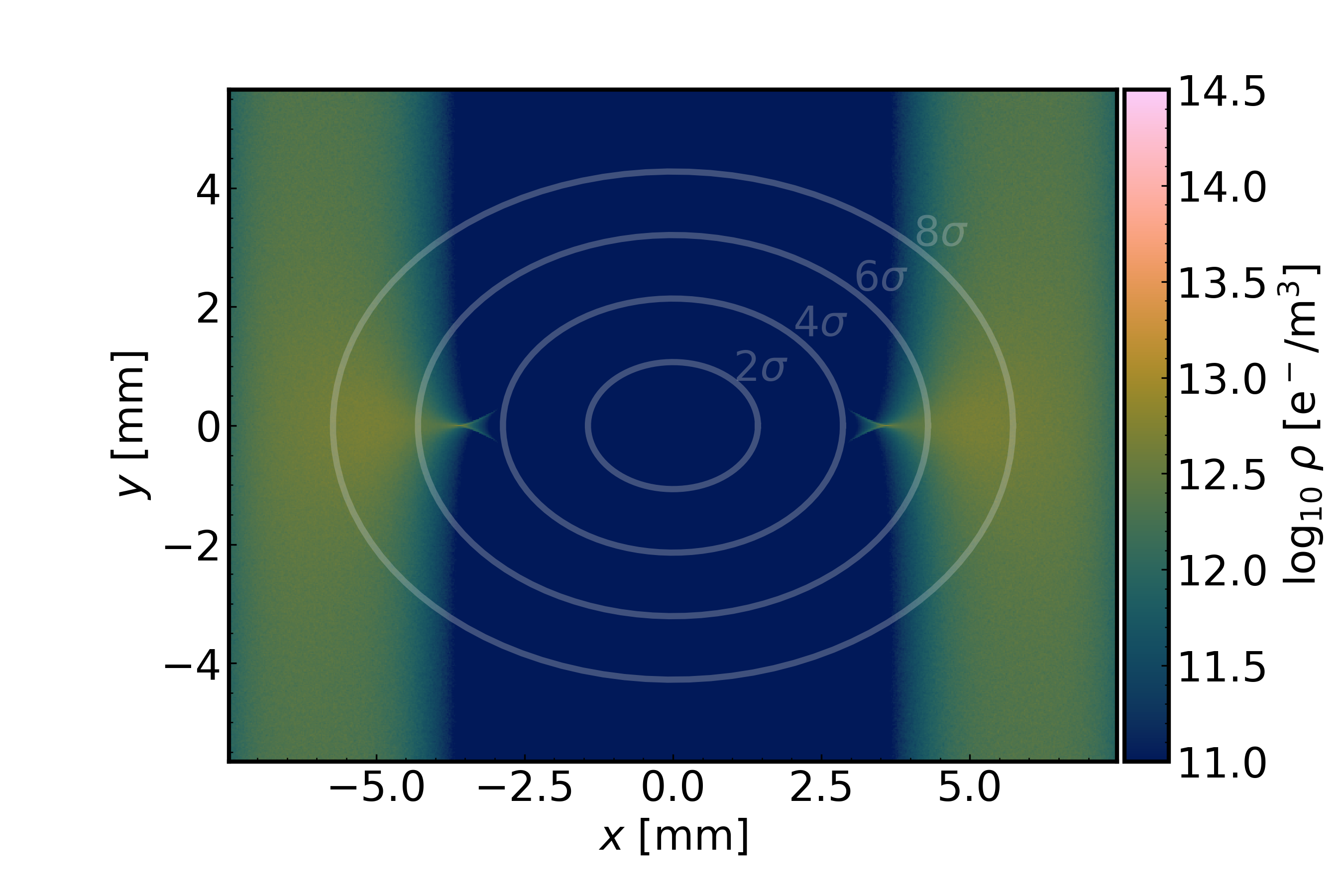}
     \\ (e) 
    \end{minipage}
    \begin{minipage}[h]{0.40\textwidth}
    \centering
    \includegraphics[width=\textwidth]{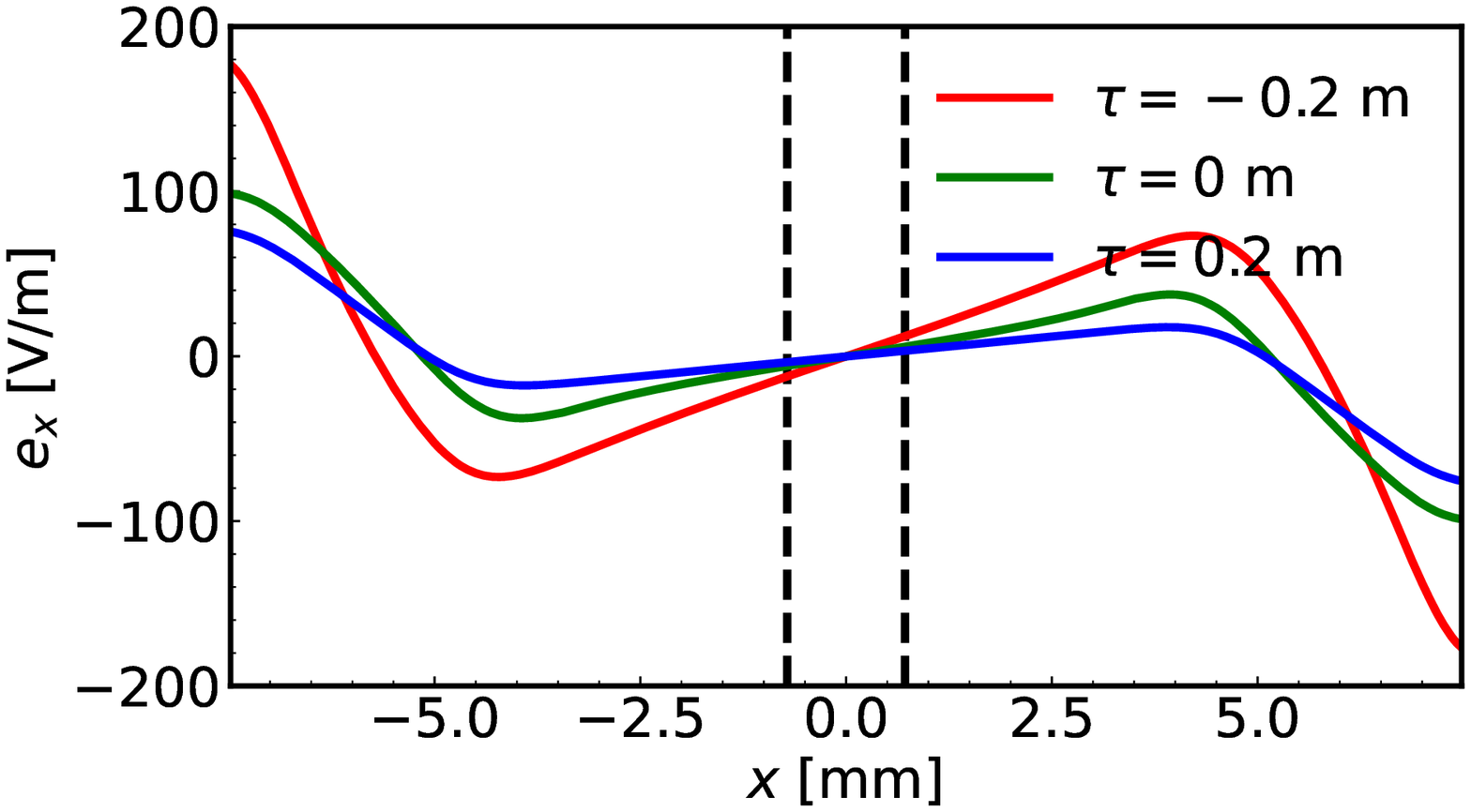}
     \\ (f) 
    \end{minipage}
    \begin{minipage}[h]{0.40\textwidth}
    \centering
    \includegraphics[width=\textwidth]{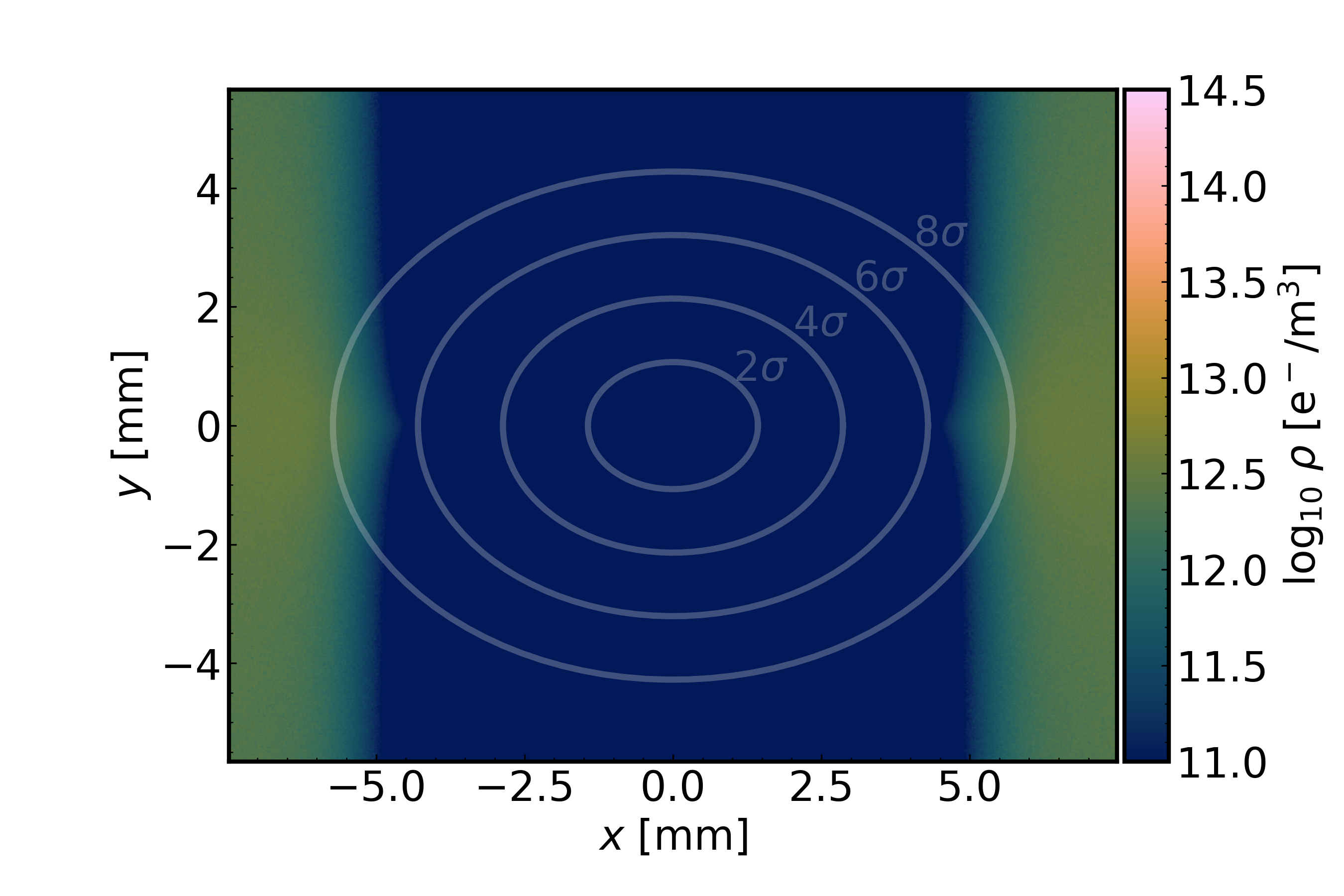}
     \\ (g) 
    \end{minipage}
    \begin{minipage}[h]{0.40\textwidth}
    \centering
    \includegraphics[width=\textwidth]{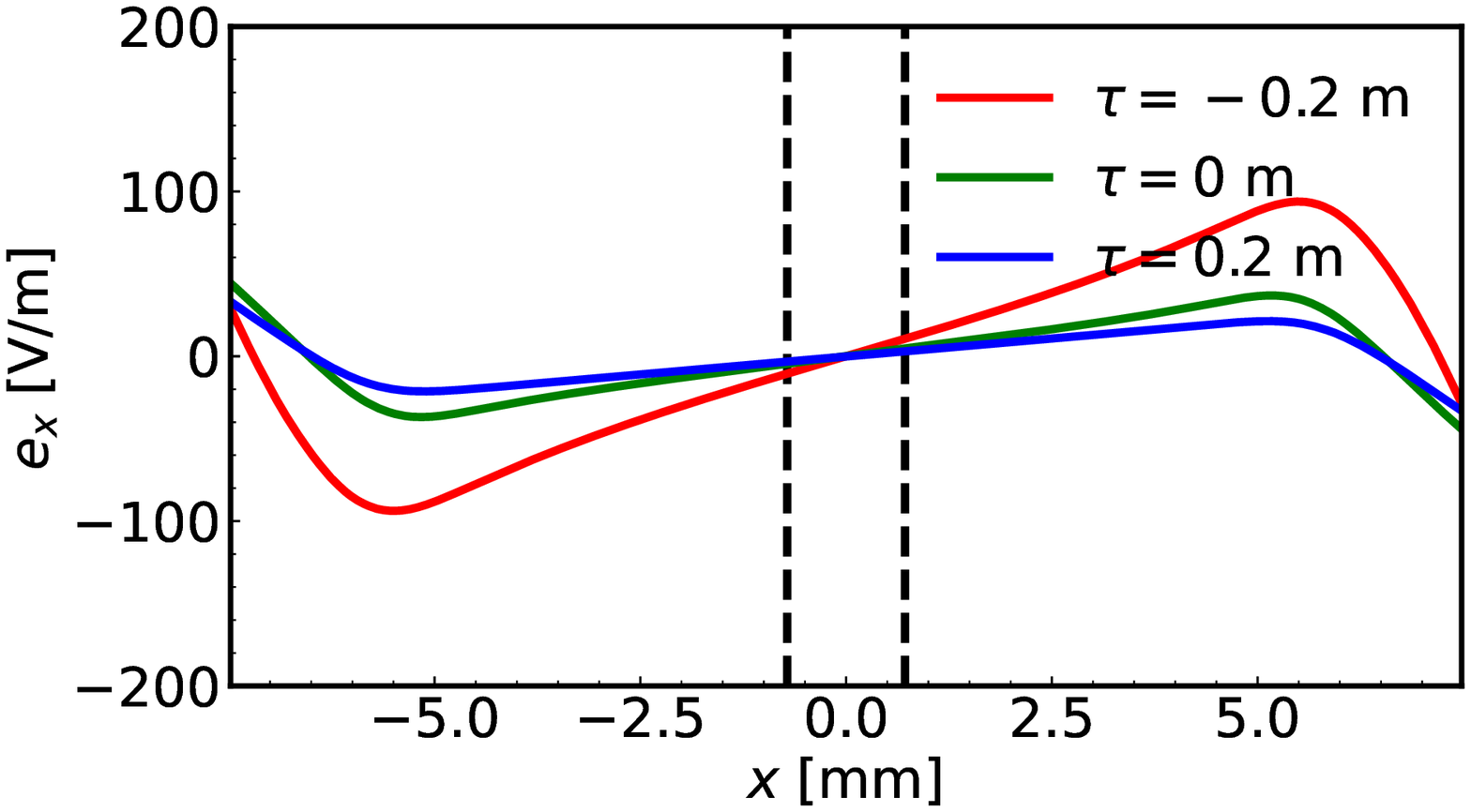}
     \\ (h) 
    \end{minipage}
    \caption{Snapshot of the electron cloud density in an MB magnet (a, c, e, g) and horizontal electric field in the plane $y=0$ at different moments during the bunch passage (b, d, f, h) for a bunch intensity of $0.5\cdot 10^{11}$\,p/bunch (a, b), $0.7\cdot 10^{11}$\,p/bunch (c, d), $0.9\cdot 10^{11}$\,p/bunch (e, f), $1.1\cdot 10^{11}$\,p/bunch (g, h),
    and $\mathrm{SEY}=1.3$.
    }
    \label{fig:app21}
\end{figure*}

Figure~\ref{fig:app21} shows the e-clouds forming in the dipole magnets (MB).
The two stripes of electrons forming left and right of the beam location come closer together as the bunch intensity decreases.
This results in stronger non-linearities in the fields in the region closer to the beam location.

\begin{figure*}[hp!]
    \centering
    \begin{minipage}[h]{0.40\textwidth}
    \centering
    \includegraphics[width=\textwidth]{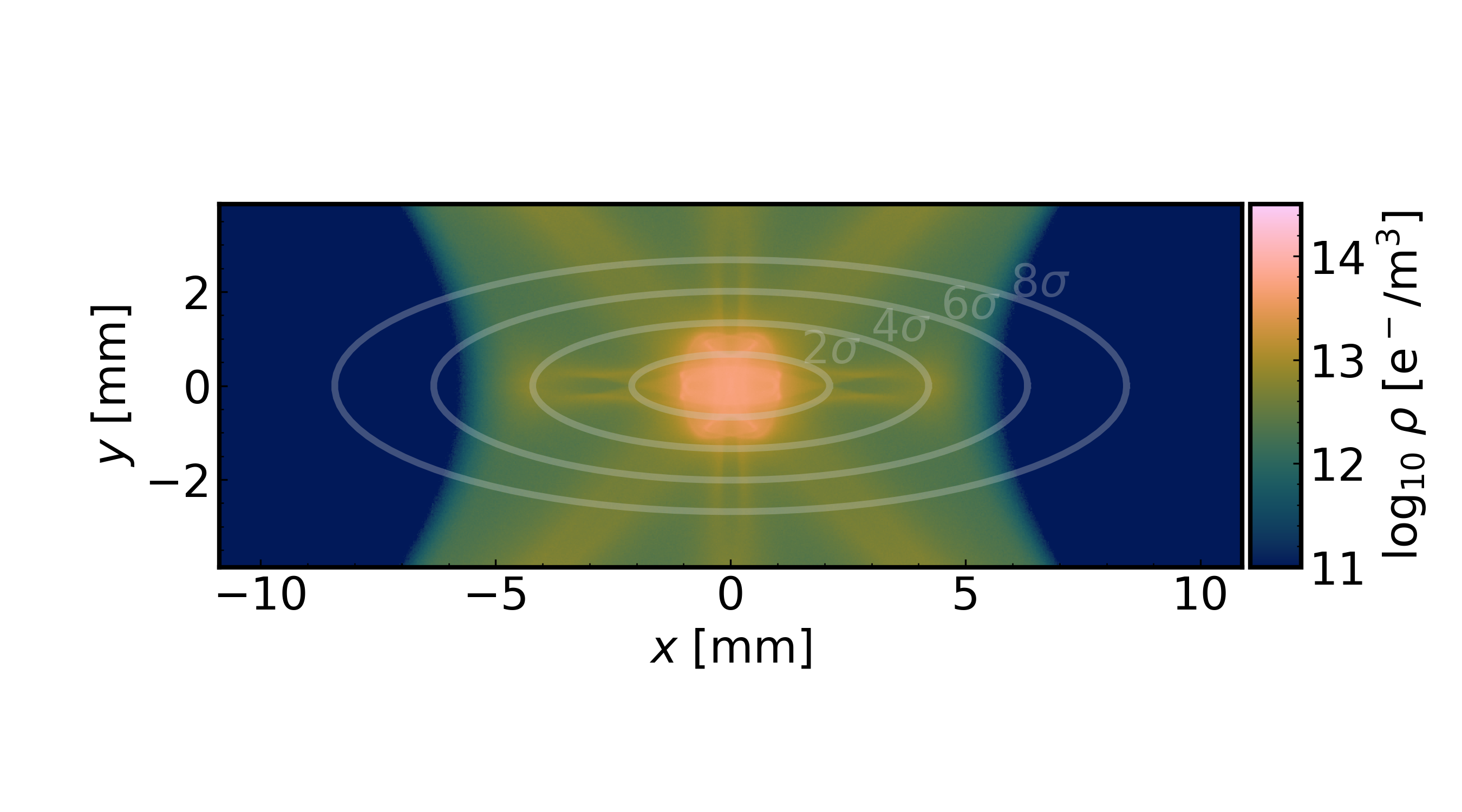}
     \\ (a) 
    \end{minipage}
    \begin{minipage}[h]{0.40\textwidth}
    \centering
    \includegraphics[width=\textwidth]{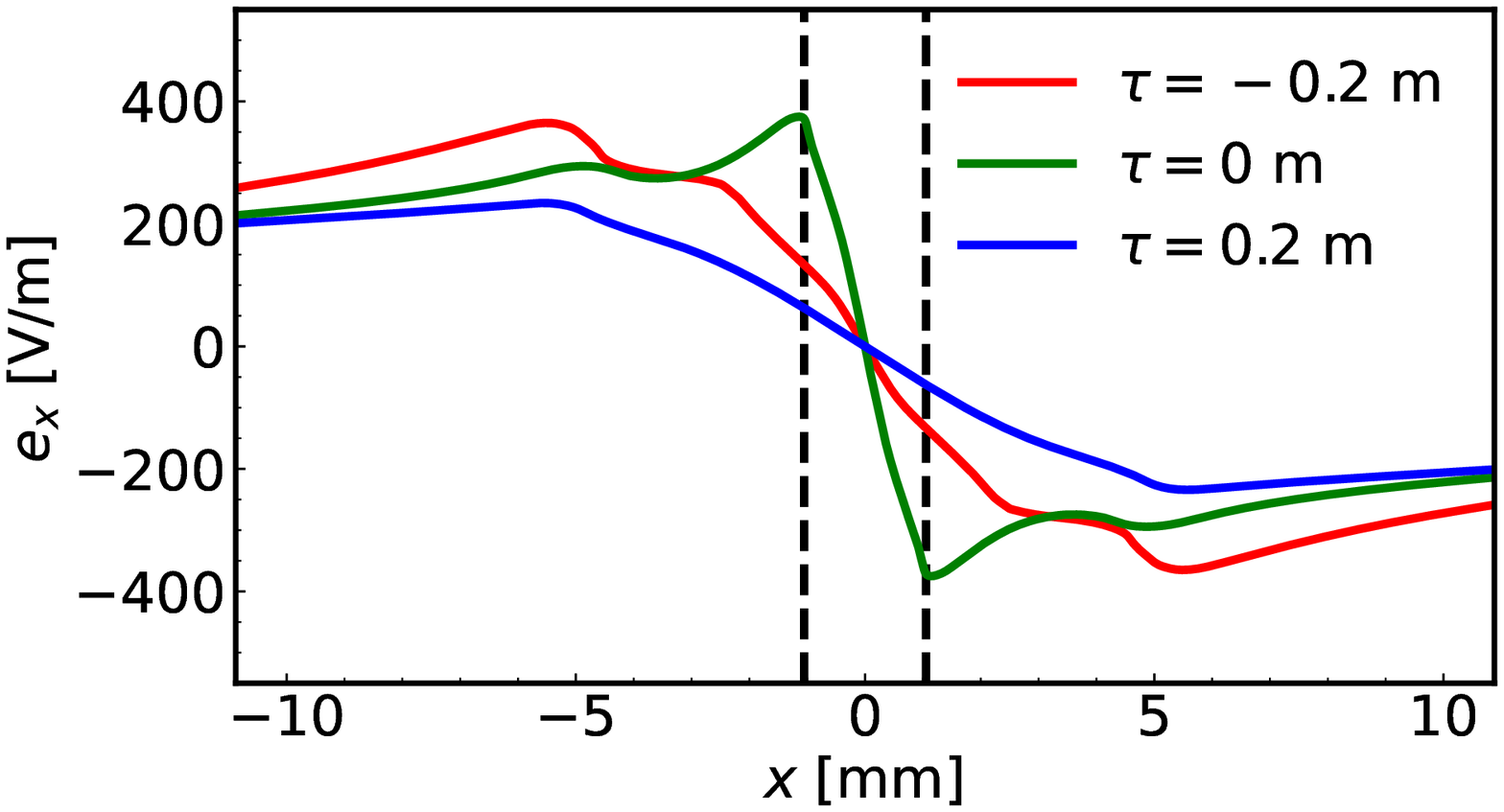}
     \\ (b) 
    \end{minipage}
    \begin{minipage}[h]{0.40\textwidth}
    \centering
    \includegraphics[width=\textwidth]{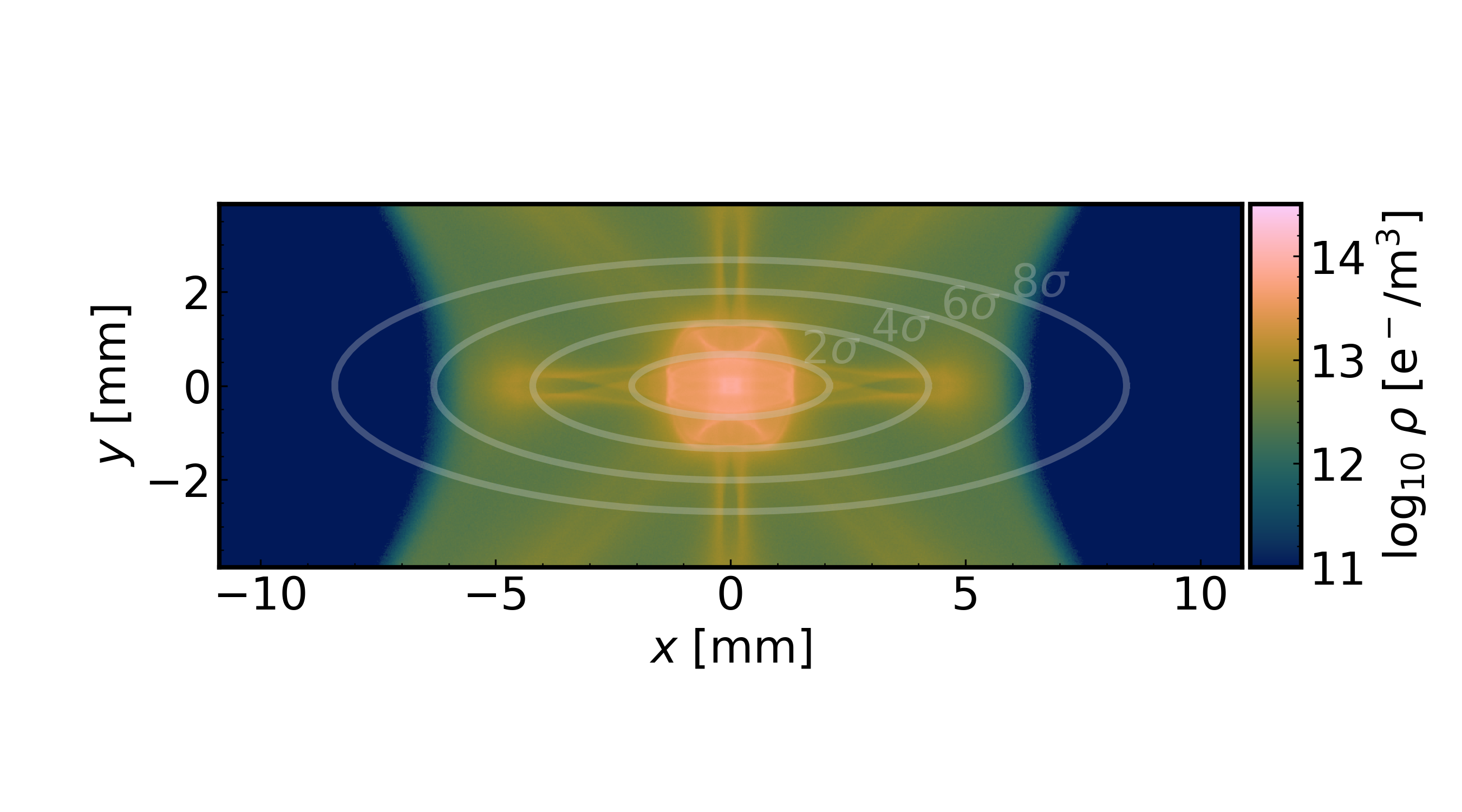}
     \\ (c) 
    \end{minipage}
    \begin{minipage}[h]{0.40\textwidth}
    \centering
    \includegraphics[width=\textwidth]{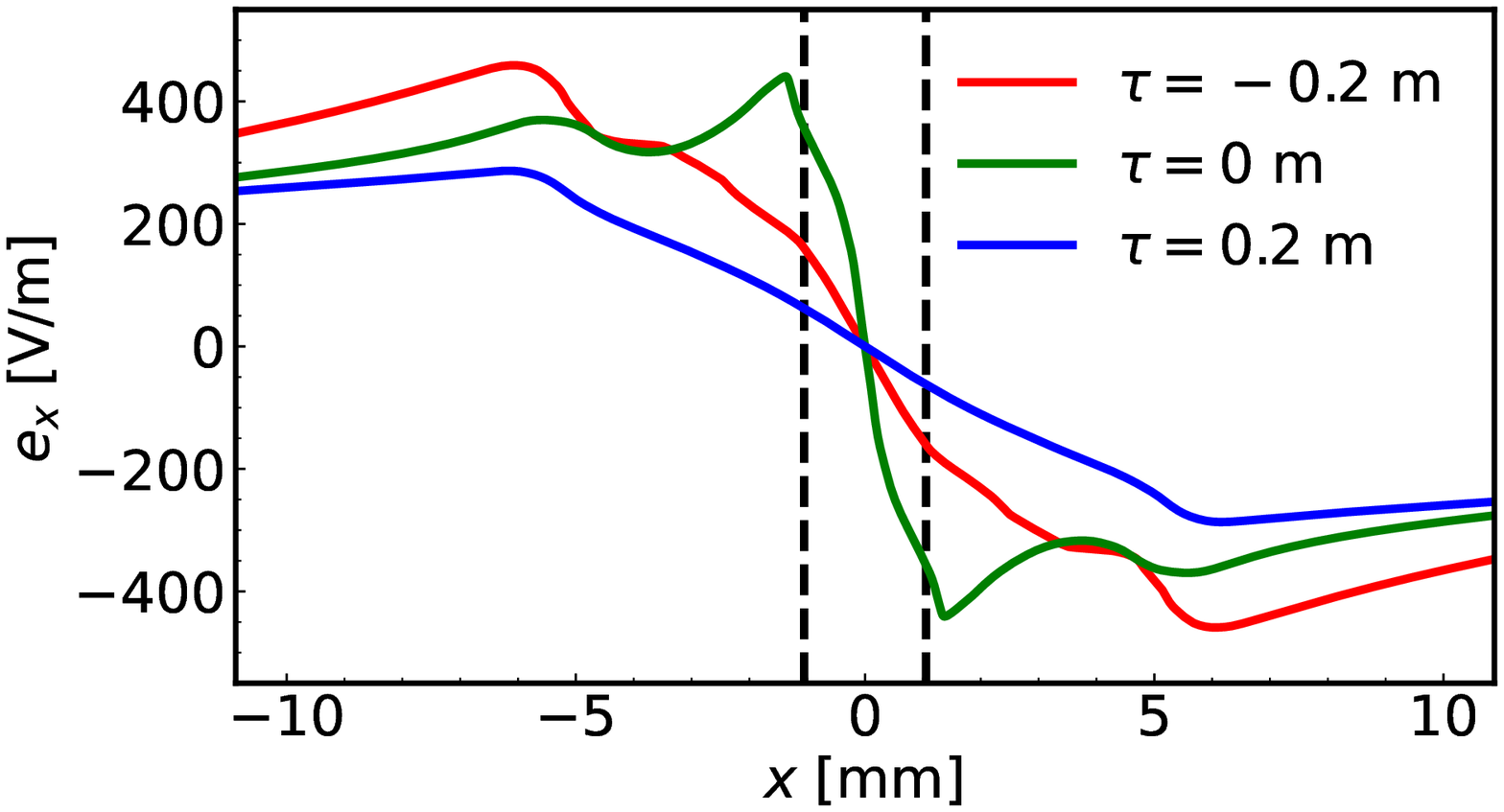}
     \\ (d) 
    \end{minipage}
    \begin{minipage}[h]{0.40\textwidth}
    \centering
    \includegraphics[width=\textwidth]{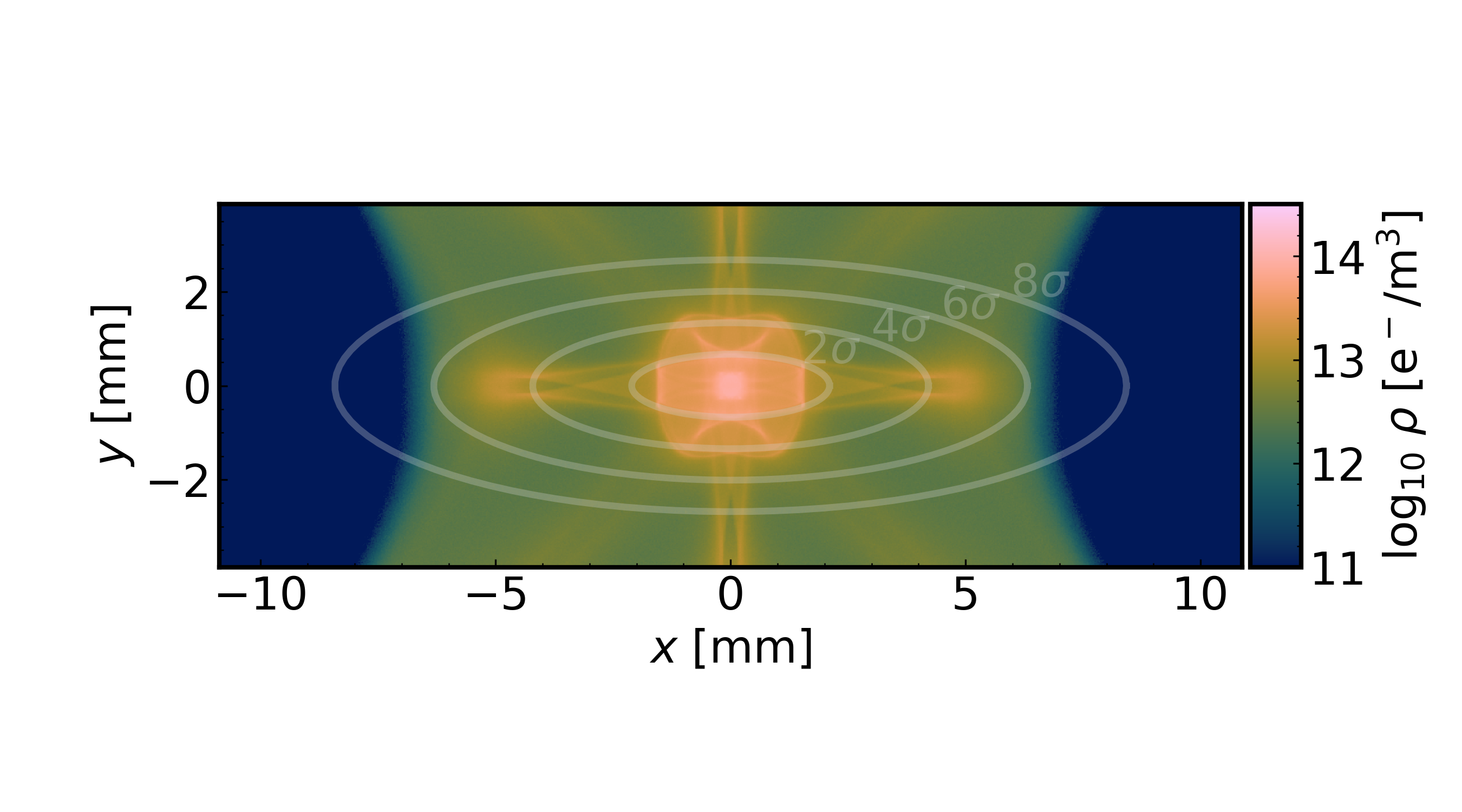}
     \\ (e) 
    \end{minipage}
    \begin{minipage}[h]{0.40\textwidth}
    \centering
    \includegraphics[width=\textwidth]{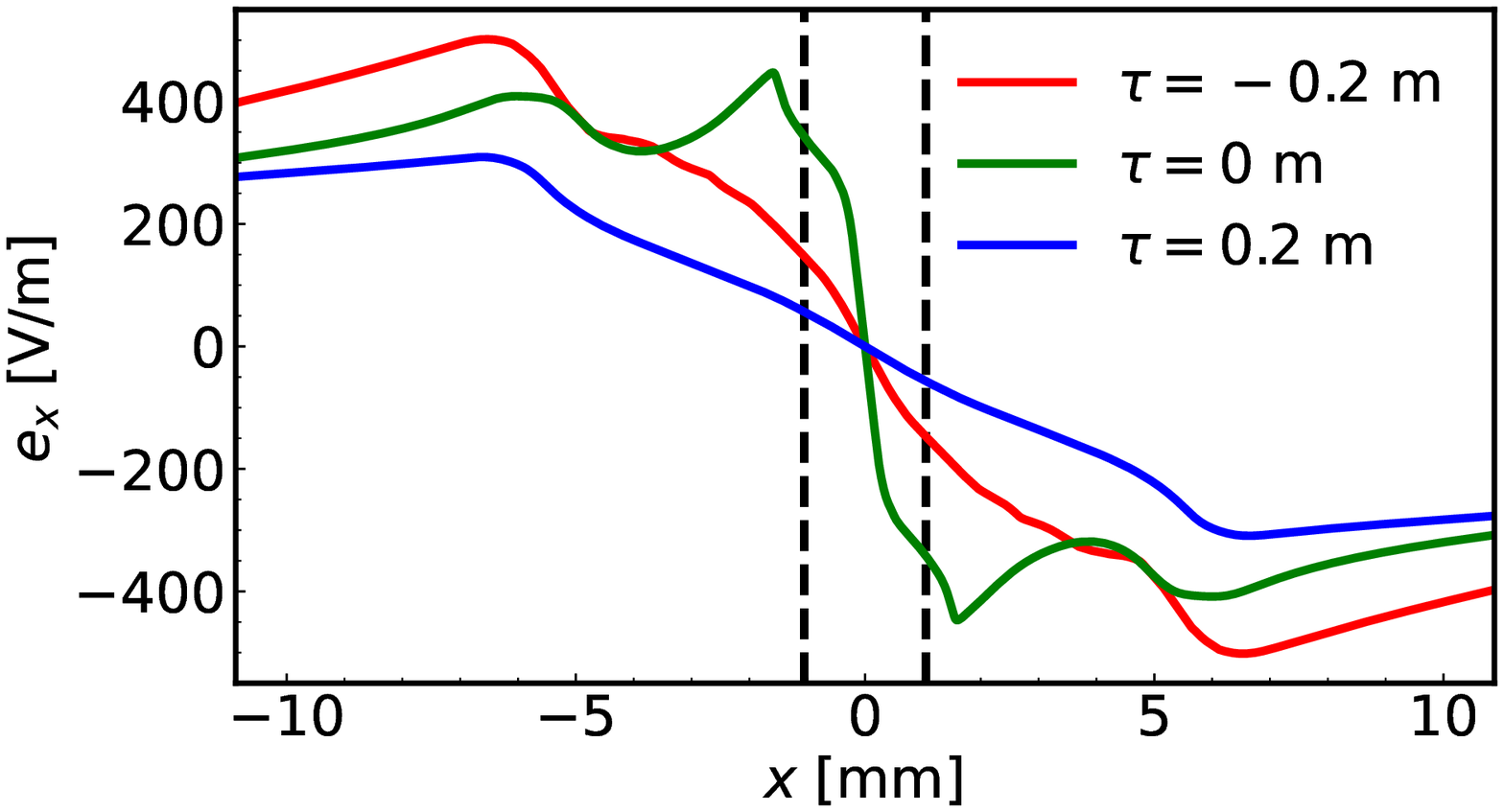}
     \\ (f) 
    \end{minipage}
    \begin{minipage}[h]{0.40\textwidth}
    \centering
    \includegraphics[width=\textwidth]{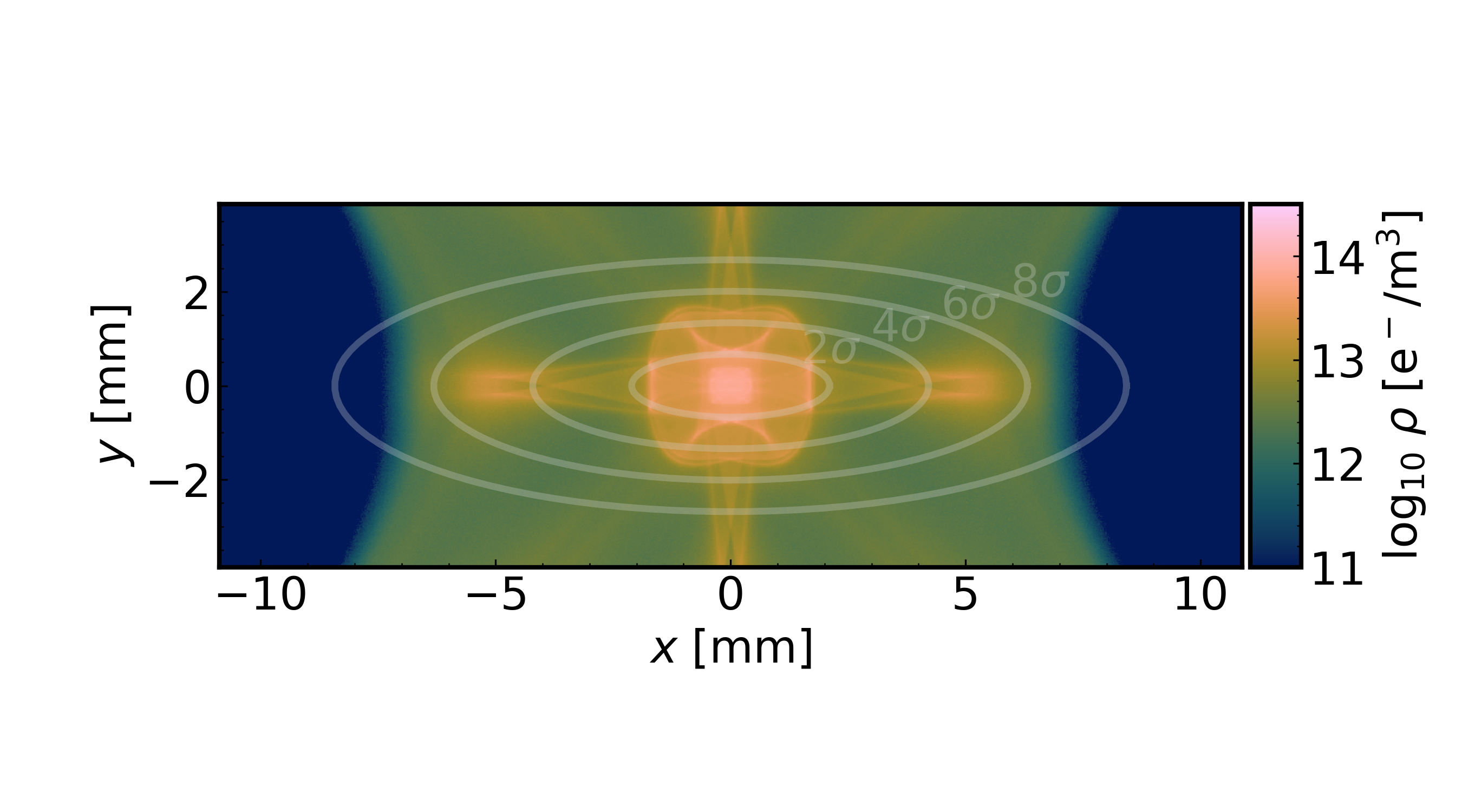}
     \\ (g) 
    \end{minipage}
    \begin{minipage}[h]{0.40\textwidth}
    \centering
    \includegraphics[width=\textwidth]{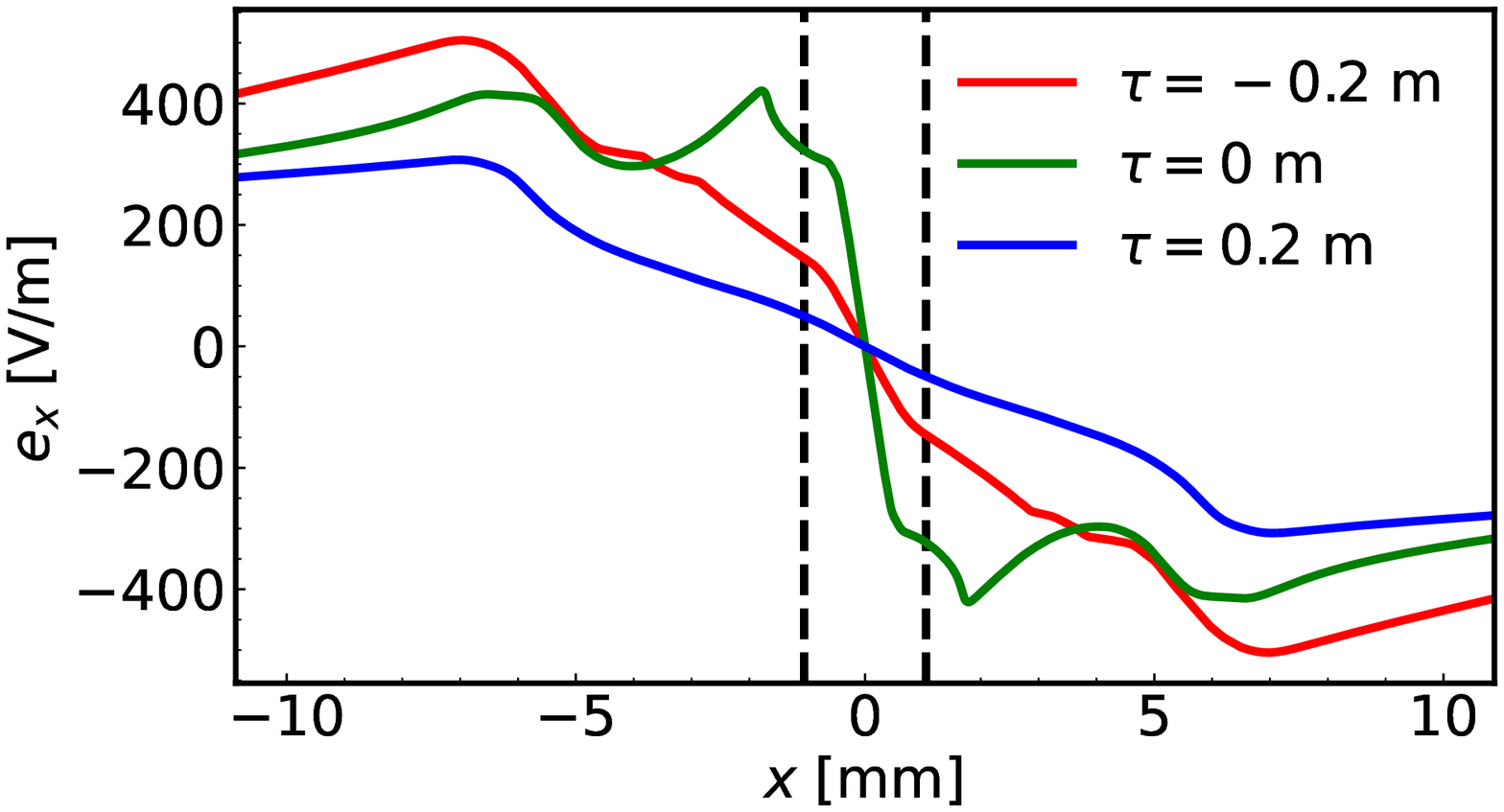}
     \\ (h) 
    \end{minipage}
    \caption{Snapshot of the electron cloud density in an MQ magnet (a, c, e, g) and horizontal electric field in the plane $y=0$ at different moments during the bunch passage (b, d, f, h) for a bunch intensity of $0.5\cdot 10^{11}$\,p/bunch (a, b), $0.7\cdot 10^{11}$\,p/bunch (c, d), $0.9\cdot 10^{11}$\,p/bunch (e, f), $1.1\cdot 10^{11}$\,p/bunch (g, h),
    and $\mathrm{SEY}=1.3$.
    }
    \label{fig:app22}
\end{figure*}

Similarly, Fig.~\ref{fig:app22} shows the electron densities and the corresponding electric fields for e-clouds forming in the main quadrupole magnets (MQ).
The effect of the bunch intensity is less visible in this case.
In this case, the region of strong electron densities becomes only slightly smaller with decreasing bunch intensities.

\clearpage

%

\end{document}